\input{aipcheck}

\documentclass[
draft            
,numberedheadings 
,sort&compress]{aipproc}

\layoutstyle{6x9}

\usepackage{amsmath}
\usepackage{amsfonts}
\usepackage{amssymb}
\usepackage{bm}

\begin{document}

\title[]{Dynamical Mean-Field Theory of Electronic Correlations in Models and Materials}


\date{~}

\author{Dieter Vollhardt}{
  address={Theoretical Physics III,
Center for Electronic Correlations and Magnetism,
Institute of Physics,
University of Augsburg,
86135 Augsburg,
Germany
}
}



\begin{abstract}
The concept of electronic correlations plays an important role in modern condensed matter physics. It refers to interaction effects which cannot be explained within a static mean-field picture as provided by Hartree-Fock theory. Electronic correlations can have a very strong influence on the properties of materials. For example, they may turn a metal into an insulator (Mott-Hubbard metal-insulator transition). In these lecture notes I (i) introduce basic notions of the physics of correlated electronic systems, (ii) discuss the construction of mean-field theories by taking the limit of high lattice dimensions, (iii) explain the simplifications of the many-body perturbation theory in this limit which provide the basis for the formulation of a comprehensive mean-field theory for correlated  fermions, the dynamical mean-field theory (DMFT), (v) derive the DMFT self-consistency equations, and (vi) apply the DMFT to investigate electronic correlations in models and materials.
%

\end{abstract}

\maketitle

\tableofcontents


\section{Introduction}

\subsection{What are electronic correlations and where do they show up?}

The term \emph{correlations} (which means ``with relation'', from Latin {\it con + relatio}) is used not
only in physics, but also in many other fields. For example, in grammar the
two words \emph{either ... or} are called a ``correlate''. Indeed, in a
grammatically correct sentence the word \emph{either} always has to be
followed by the word \emph{or}. Obviously the two words are
correlated. In mathematics and natural sciences the term correlation
is used to express the fact that the average or expectation value of a product of quantities is usually not
equal to the product of the averages of the individual quantities:
\begin{equation}
\langle AB \rangle \neq \langle A \rangle \langle B \rangle.
\end{equation}
For example, the
mass or charge density $n(\bm{r})$ of a many-body system at position $\bm{r}$ is influenced, in general, by the density of the particles at other positions $\bm{r^{\prime}}$. Therefore the density-density correlation function is not simply given
by the product of the \emph{average} density $n$:
\begin{equation}
\langle n(\bm{r}) n(\bm{r^{\prime}}) \rangle \neq \langle n(\bm{r}) \rangle \langle n(\bm{r^{\prime}}) \rangle = n^2.
\end{equation}
Indeed, this very property defines correlations: they express effects which go
beyond approximations obtained by the factorization of correlation functions, i.e., beyond static mean-field theories such as the Weiss mean-field
theory for the Ising model or the Hartree approximation for the Hubbard
model.

Correlations in space and time are by no means abstract notions, but occur frequently in everyday life. Persons in an elevator or in a
car are strongly correlated both in space and time, and it would be
quite inadequate to describe the situation of a person in such a
case within a factorization approximation where the influence of
the other person(s) is described only by a static mean-field, i.e., a structureless cloud.

As in the case of two persons riding together on an elevator,
two electrons with different spin direction occupying the same narrow $d$ or $f$ orbital in a real
material are also correlated. Here the degree of correlation can be estimated in a
 very simplified picture as follows. Assuming the correlated electrons  (or rather the quasiparticles, i.e., excitations) to
have a well-defined dispersion $\epsilon_{\bm{k}}$, their velocity is given by $v_{\bm{k}} = \frac{1}{ \hbar} |\nabla _{\bm{k}} \epsilon_{\bm{k}}|$.
The typical velocity is given by $v_{\bm{k}} \sim \frac{a}{ \tau}$, where $a$ is  the lattice spacing and $\tau$ is the
average time spent on an atom. The derivative
can be estimated as $\frac{1}{ \hbar}|\nabla_{\bm{k}} \epsilon_{\bm{k}}|\sim \frac{1}{ \hbar} a W$ since $|\nabla_{\bm{k}}|\sim 1/k \sim a$ and $|\epsilon_{\bm{k}}|$ corresponds to the band overlap $t$ and
hence to the band width $W$. Altogether this means that
\begin{equation}
\tau \sim \frac{\hbar}{W}.
\end{equation}
The narrower an orbital, the longer an electron therefore resides on an atom and thereby feels the presence of other electrons.
Hence a narrow band width implies strong electronic correlations.

Indeed this is the case for many elements in the periodic table.
Namely, in many materials with partially filled $d$ and $f$ electron
shells, such as the transition metals V, Fe, and Ni and their oxides,
or rare--earth metals such as Ce, electrons occupy narrow
orbitals. This spatial confinement enhances the effect of the
Coulomb interaction between the electrons, making them ``strongly
correlated''.
Correlation effects lead to profound quantitative and qualitative changes of the physical properties of electronic systems as compared to non-interacting particles. In particular, they often respond very strongly to changes in external parameters. This is expressed by large renormalizations of the response functions of the system, e.g., of the spin susceptibility and the charge compressibility. Electronic correlations also play an essential role in high temperature superconductivity. In particular, the interplay between the spin, charge and orbital degrees of freedom of the correlated $d$ and $f$ electrons and with the lattice degrees of freedom leads to a wealth of unusual phenomena at low temperatures \cite{tokura}.
These
properties cannot be explained within conventional mean-field theories,
e.g., Hartree-Fock theory, since they describe the
interaction only in an average way.

\subsection{Electrons \emph{vs.} Landau quasiparticles}

Electrons are fermions which obey the Fermi-Dirac statistics. The Pauli
exclusion principle implies the existence of a Fermi body of
occupied states and thereby of a Fermi {\em surface} which distinguishes between occupied states (inside the Fermi body) and empty
states (outside the Fermi body). The existence of a Fermi surface is something quite extraordinary. It
allows for the formulation of Landau Fermi-liquid theory and thereby for
a deep understanding of interacting fermionic systems \cite{Pines+Nozieres}.

In the Landau Fermi-liquid theory  a one-to-one correspondence between
$\bm{k}$-states of the non-interacting and the interacting system is assumed. Therefore
there exist well-defined $\bm{k}$-states, called \emph{quasiparticles}, which have a
finite lifetime (the closer they are to the Fermi surface the more
well-defined they are, i.e., the longer they live), an effective mass $m^*$ and an effective interaction.
Quasiparticles are the \emph{elementary excitations} of a Fermi liquid and determine the entire low temperature
thermodynamics of a Landau Fermi liquid. They are rather
abstract objects which should not be confused with the particles of the
non-interacting system.  The Fermi-liquid concept is independent of the strength of the bare
interaction between particles. Hence it can not only describe simple metals
such as potassium where $m^*$ is not much different from
the bare electronic mass, but even ``heavy fermion systems'' such as UBe$_{13}$ where $m^*$ can be a
factor 1000 larger than the bare electronic mass \cite{Stewart}.

\subsection{The simplest model for correlated electrons}

The simplest model describing interacting electrons in a solid is the one-band, spin-1/2
Hubbard model \cite{Gutzwiller,HubbardI,Kanamori} where the interaction between the electrons is assumed to be so strongly screened that it is taken as purely local. The Hamiltonian consists of two terms, the kinetic energy $\hat{H}_0$ and the interaction energy  $\hat{H}_I$ (here and in the following operators are denoted by a hat):
\begin{subequations}
\label{G11.7}
\begin{eqnarray}
\hat{H} & = & \hat{H}_0 + \hat{H}_I \\[10pt]
\label{G11.7a}
\hat{H}_0 & = & \sum_{\bm{R}_i , \bm{R}_j} \sum_{\sigma}
t_{ij}  \hat{c}_{i \sigma}^{+} \hat{c}_{j \sigma}^{} = \sum_{\bm{k} , \sigma}
\epsilon_{\bm{k}} \hat{n}_{\bm{k} \sigma}^{} \label{G11.7b} \\[10pt]
\hat{H}_{\rm I} & = & U \sum_{\bm{R}_i} \hat{n}_{i \uparrow} \hat{n}_{i \downarrow},
\label{G11.7c}
\end{eqnarray}
\end{subequations}%
where $\hat{c}_{i \sigma}^{+} (\hat{c}_{i \sigma}^{})$ are creation (annihilation)
operators of electrons with spin $\sigma$ at site $\bm{R}_i$, and
$\hat{n}_{i \sigma}^{} = \hat{c}_{i \sigma}^{+} \hat{c}_{i \sigma}^{}$. The Fourier transform
of the kinetic energy in \eqref{G11.7b}, where $t_{ij}$ is the hopping amplitude, involves the dispersion
$\epsilon_{\bm{k}}$ and the momentum distribution operator
$\hat{n}_{\bm{k} \sigma}^{}$ .

\begin{figure}
\includegraphics[width=1.0\textwidth]{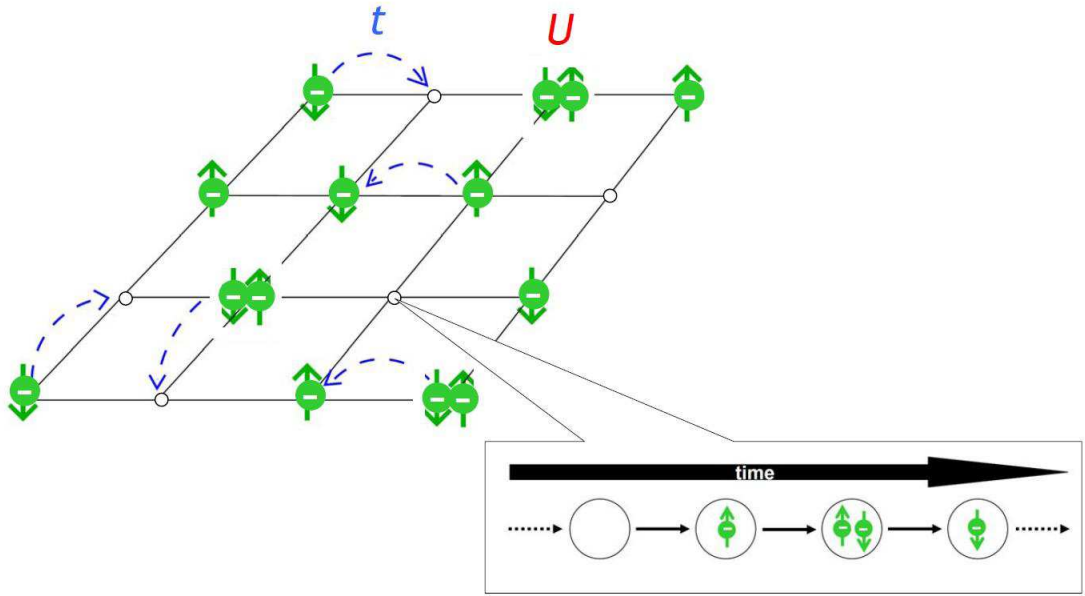}
\caption{Schematic illustration of interacting electrons in a solid in terms of the Hubbard model. The ions appear only as a rigid lattice (here represented as a square lattice). The electrons, which have a mass, a negative charge, and a spin ($\uparrow$   or $\downarrow$), move from one lattice site to the next with a hopping amplitude $t$. The quantum dynamics thus leads to fluctuations in the occupation of the lattice sites as indicated by the time sequence. When two electrons meet on a lattice site (which is only possible if they have opposite spin because of the Pauli exclusion principle) they encounter an interaction $U$. A lattice site can either be unoccupied, singly occupied ($\uparrow$    or   $\downarrow$), or doubly occupied. }
\label{Hubbard_model}
\end{figure}

A schematic picture of the Hubbard model is shown in Fig.~\ref{Hubbard_model}. When we look only at a single site of this lattice model, this site will sometimes
be empty, singly occupied or doubly occupied. In particular, for strong
repulsion $U$ double occupations are energetically very unfavorable and are therefore strongly suppressed. In this situation $\langle \hat{n}_{i \uparrow} \hat{n}_{i \downarrow} \rangle$ must not be factorized since   $\langle \hat{n}_{i \uparrow} \hat{n}_{i \downarrow} \rangle \neq \langle \hat{n}_{i\uparrow} \rangle \langle \hat{n}_{i\downarrow} \rangle $. Otherwise, correlation phenomena such as the Mott-Hubbard metal-insulator
transition are eliminated from the very beginning. This explains why
Hartree-Fock-type mean-field theories are generally insufficient to
explain the physics of electrons in the paramagnetic phase for strong interactions.


The Hubbard model looks very simple. However, the competition between the kinetic energy and the interaction leads to a complicated many-body problem. which is impossible to solve analytically, except in dimension $d=1$ \cite{Lieb+Wu}. This model provides the basis for most of the theoretical research on correlated electrons during the last decades.


\section{Mean-field theories for many-body systems
}

\subsection{Construction of mean-field theories}
\label{sec:General}
It is well known that theoretical investigations of quantum-mechanical
many-body systems are faced with severe technical problems,
particularly in those dimensions which are most interesting to us,
i.~e., $d = 2,3$. This is due to the complicated
dynamics and, in the case of fermions, the non-trivial
algebra introduced by the Pauli exclusion principle. In the absence of exact methods there is clearly
a great need for reliable, controlled approximation schemes.
Their construction is
not straightforward.


In the statistical theory of classical and quantum-mechanical systems a rough,
overall description of the properties of a model is often obtained
within a so-called \emph{mean-field theory}.
Although the term is frequently used, the actual meaning
of what a mean-field theory is
or should be is rather vague, because there is no unique
prescription of how to construct such a theory (and sometimes
 it is only a respectable name for a
questionable approximation \ldots). Hence every time one encounters
the term ``mean-field theory'', one should ask about the reliability of this approximation,
i.~e., about its range of validity with respect to the input parameters and its
thermodynamic consistency.

There exists a well-established branch of
approximation techniques which makes use of the simplifications that occur when
some parameter is taken to be large (in fact, infinite), e.g., the length of the spins $S$, the spin degeneracy $N$, the coordination number  $Z$ (the number of nearest neighbors of a lattice site).
Investigations in this limit,
supplemented, if at all possible, by an expansion in the inverse of the large
parameter, often provide valuable insight into the fundamental properties of a system even when this parameter is not large.

One of the best-known mean-field theories is the Weiss molecular-field theory
for the Ising model~\cite{Baxter}. It is a prototypical
{\em single-site mean-field theory} which becomes exact for infinite-range interaction, as well as in the limit of the coordination number $Z \to \infty$ or\footnote{For regular lattices, e.g., Bravais-lattices, both a dimension $d$ and a coordination number $Z$ can be defined. In this case either $d$ or $Z$ can be used alternatively as an expansion parameter. However, there exist other lattices (or rather graphs) which cannot be associated with a physical dimension $d$ although a coordination number $Z$ is well-defined. The best-known example is the Bethe lattice, an infinitely extended Cayley tree \cite{Baxter,Bethe-Eckstein}, which is not a regular lattice because it does not have loops. The coordination number $Z$ is therefore a very useful parameter for theoretical investigations of lattice models, although the dimension $d$ is the more general physical parameter. In the following discussion we mostly use both $d$ and $Z$ in parallel.\label{d_vs_Z}}
the dimension
$d \to \infty$. In the latter case $1/Z$ or $1/d$ is a small parameter which can sometimes be used to improve the mean-field theory
systematically. This mean-field theory contains no unphysical singularities and is applicable for
all values of the input parameters, i.e., coupling parameters, magnetic field, and temperature. It is also diagrammatically
controlled \cite{Itzykson89}. Insofar it is a very respectable approximation which
sets very high standards for other mean-field theories.

\subsubsection{Motivation for using the limit of high dimensions to construct mean-field theories}

In a perfectly crystalline system every lattice site has the same number of
nearest neighbors  $Z$. In three dimensions ($d = 3$)
one has $Z = 6$ for a simple cubic lattice ($Z = 2d$ for
a hypercubic lattice in general dimensions
$d$), $Z = 8$ for a bcc lattice and $Z = 12$ for an fcc-lattice.
The dimensionality of a lattice system is directly described by the number
$Z$ (see footnote \ref{d_vs_Z}). Since $Z \sim {\cal O} (10)$ is already quite large
in $d = 3$, such that $1/Z$ is rather small, it is only natural and in the general
spirit of theoretical physics to consider the extreme limit $Z \to \infty$ to simplify the problem. Later, if possible, one can try to improve the result by expanding in the small parameter $1/Z$.
The limit $d \to \infty$ is not as
academic as it might seem. In fact, it turns out that
several standard approximation schemes which are commonly used to
explain experimental results in dimension $d = 3$, are exact only in $d = \infty$ \cite{Jerusalem}.

\subsection{A prototypical example: The Weiss mean-field theory for the Ising model}
\label{sec:ising}
In the case of classical spin models (e.g., Ising, Heisenberg) the $Z \to \infty$
limit is well-known \cite{Baxter,Itzykson89}. It leads
to the results of the Weiss molecular-field theory which may be
viewed as the prototypical method for constructing a mean-field theory.
The Hamiltonian for the Ising model with nearest-neighbor (NN) coupling is given by
\begin{equation}
 H  = - \frac{1}{2}
J \sum_{\langle \bm{R}_{i}^{} , \bm{R}_{j}^{} \rangle}
\; S_i S_j,
\label{G11.1}
\end{equation}
where we assume ferromagnetic coupling ($J > 0$).
Every spin $S_i$ interacts with a local field $h_i$,
produced by its nearest neighbors at site $\bm{R}_i$.
In the Weiss mean-field approach the two-spin interaction in \eqref{G11.1} is decoupled,
i.~e., $H$ is replaced by a mean-field Hamiltonian
\begin{subequations}
\label{G11.2}
\begin{equation}
H^{\rm MF} = - h_{\rm MF} \sum_{\bm{R}_i} \; S_i + E_{\rm shift}.
\label{G11.2a}
\end{equation}
Now a spin $S_i$ interacts only with a global (``molecular'') field
\begin{eqnarray}
h_{\rm MF} & = & J \sum_{\bm{R}_j}^{(i)} \langle S_j \rangle
\label{G11.2b} \\[10pt]
& \equiv & J \langle S \rangle.
\label{G11.2c}
\end{eqnarray}
\end{subequations}
Here $\langle  \; \rangle$ indicates the thermal average, $E_{\rm shift} = \frac{1}{2}
\; L J Z \langle S \rangle^2$ is a constant energy shift
with $L$ as the number of lattice sites and the superscript
$(i)$ implies summation over only
NN-sites of $\bm{R}_i$. This corresponds to the factorization
\begin{equation}
\left\langle [ S_i - \langle S \rangle ] [ S_j - \langle S \rangle ]
\right\rangle \equiv 0,
\label{G11.2d}
\end{equation}
whereby correlated fluctuations of spins at sites $\bm{R}_i$ and $\bm{R}_j$
are neglected. In the limit $Z \to \infty$ the coupling constant $J$ has to be rescaled as
\begin{equation}
J \to  \frac{J^*}{Z} \; , \; J^* = {\rm const}
\label{G11.3}
\end{equation}
for $h_{\rm MF}$ to remain finite. In this limit the factorization
procedure \eqref{G11.2d}, and hence the replacement of \eqref{G11.1}, by the mean-field Hamiltonian \eqref{G11.2a},
 becomes
exact \cite{Brout60,Thompson74}.

\begin{figure}
\includegraphics[width=0.8\textwidth]{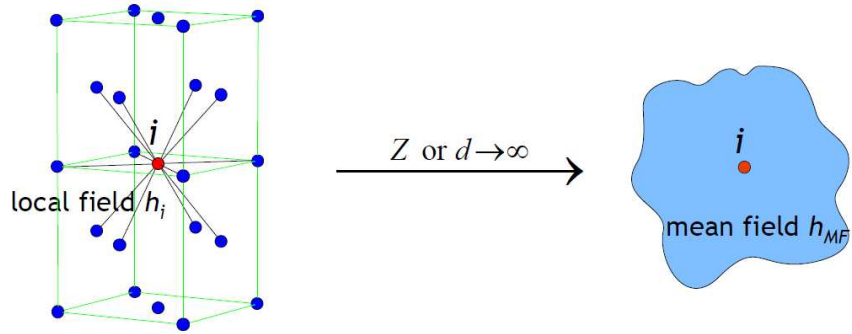}
\caption{Already in three dimensions ($d=3$) the coordination number $Z$ of a lattice can be quite high, as in the face-centered cubic lattice where $Z=12$. In the limit $Z \to \infty$, or equivalently $d \to \infty$, the Ising model effectively
reduces to a single-site problem where the local field $h_i$ is replaced
by a global mean (``molecular'') field $h_{\rm MF}$.}
\label{Ising_MF}
\end{figure}

 Eq.~\eqref{G11.2a} implies that in the
limit $Z \to \infty$ fluctuations in the ``bath'' of surrounding neighbors become
unimportant, such that the surrounding of any site
is completely described by a single mean-field parameter $h_{\rm MF}$
(see Fig.~\ref{Ising_MF}).
Hence the
Hamiltonian becomes purely local
\begin{equation}
H^{\rm MF}  = \sum_{\bm{R}_i} H_i + E_{\rm shift}
\label{G11.4}
\end{equation}
\begin{equation}
H_i = - h_{\rm MF} S_i.
\label{G11.5}
\end{equation}
Thereby the problem reduces to an effective {\em single-site problem}. The value of $\langle S \rangle$
is determined by the Curie-Weiss (or Bragg-Williams) self-consistent
equation
\begin{equation}
\langle S  \rangle = \tanh (\beta J^{*} \langle S \rangle ),
\label{G11.6}
\end{equation}
where $\beta = 1/T$ (here $k_B = 1)$.

It should be noted that the scaling \eqref{G11.3} is typical for localized
spin models with isotropic coupling, i.~e.,
when the spatial average $\overline{J_{ij}} \equiv J $
is non-zero. On the
other hand, in the classical spin glass problem with \emph{random} coupling one has
 $\overline{J}_{ij} = 0$, but $\overline{J_{ij}^2} \neq 0$. In this case a different kind
of scaling, namely $J_{ij} \to J_{ij}^* /\sqrt{Z},$ has to be
used \cite{Sherrington75}.


\subsection{Hartree approximation for the Hubbard model}

To elucidate some of the shortcomings of conventional mean-field theories for Hubbard-type
models, we consider the Hartree approximation\footnote{The Hubbard interaction acts only between electrons with opposite spin; therefore the Hartree-Fock approximation does not lead to an exchange term.}, which corresponds to a factorization of the interaction term in direct analogy to the factorization of the spins in the Ising model. To this end let us consider a generalization of Hubbard model
\begin{equation}
\hat{H} =  \hat{H}_{\rm kin} +
\frac{1}{2}  \sum\limits_{\bm{R}_i , \bm{R}_j  }
\sum\limits_{\sigma \sigma'} \; V_{ij}^{\sigma \sigma'}
\hat{n}_{i \sigma}^{} \hat{n}_{j \sigma'},
\label{G14.1}
\end{equation}
where the interaction $V_{ij}^{\sigma \sigma'}$ describes a local part,
$U  \delta_{\sigma, - \sigma'}$, i.~e.,
the Hubbard interaction, as well as a nearest-neighbor contribution $V^{\sigma \sigma'}$.
The form of the interaction term is similar to \eqref{G11.1} for the Ising model.
In the spirit of a conventional mean-field  approximation
the two-particle interaction in \eqref{G14.1} is factorized, i.~e.,
$\hat{H}_{\rm ext}$ is replaced by
\begin{equation}
\hat{H}^{\rm MF} = \hat{H}_{\rm kin} + \sum_{\bm{R}_i , \sigma}
\hat{n}_{i \sigma}^{} \langle \hat{h}_{i \sigma} \rangle + E_{\rm shift}
\label{G14.2}
\end{equation}
in complete analogy with the Weiss molecular field theory for the
Ising model in Sec.~\ref{sec:ising}. In \eqref{G14.2} a $\sigma$-electron at site
$\bm{R}_i$ interacts only with a {\em local} field (a c-number)
\begin{equation}
\langle \hat{h}_{i \sigma} \rangle
= \sum_{\bm{R}_j , \sigma}^{(i)} V_{ij}^{\sigma \sigma'} \langle
\hat{n}_{j \sigma} \rangle.
\label{G14.3}
\end{equation}
 The above decoupling of the operators
is equivalent to the Hartree approximation which sets
\begin{equation}
\left\langle [ \hat{n}_{i \sigma}^{} - \langle \hat{n}_{i \sigma}^{} \rangle ] [\hat{n}_{j \sigma'} -
\langle \hat{n}_{j \sigma'} \rangle ]  \right\rangle \equiv 0
\label{G14.4}
\end{equation}
for all $i, j, \sigma, \sigma'$. Thereby correlated fluctuations on the sites
$\bm{R}_i$ and $\bm{R}_j$ are neglected.
 Although \eqref{G14.2} is a one-particle
problem it cannot be solved exactly in any systematic way,
since, in principle, the potential, i.~e., the mean field
$\langle \hat{h}_{i \sigma} \rangle$, may be an arbitrarily complicated function of
position.

An obvious question concerning \eqref{G14.2} is, whether the mean-field decoupling ever
becomes exact for all values of $U, V^{\sigma \sigma'}$ and $n$,
 i.~e., beyond the weak-coupling or low-density limit. How about the limit $d \to \infty$?
We will come back to this question in Sec.~3.3.1 once we understood how this limit can be employed in the case of lattice fermion models.

\section{Lattice fermions in the limit of high dimensions}

\subsection{Scaling of the hopping amplitude}

\begin{figure}
\includegraphics[width=0.8\textwidth]{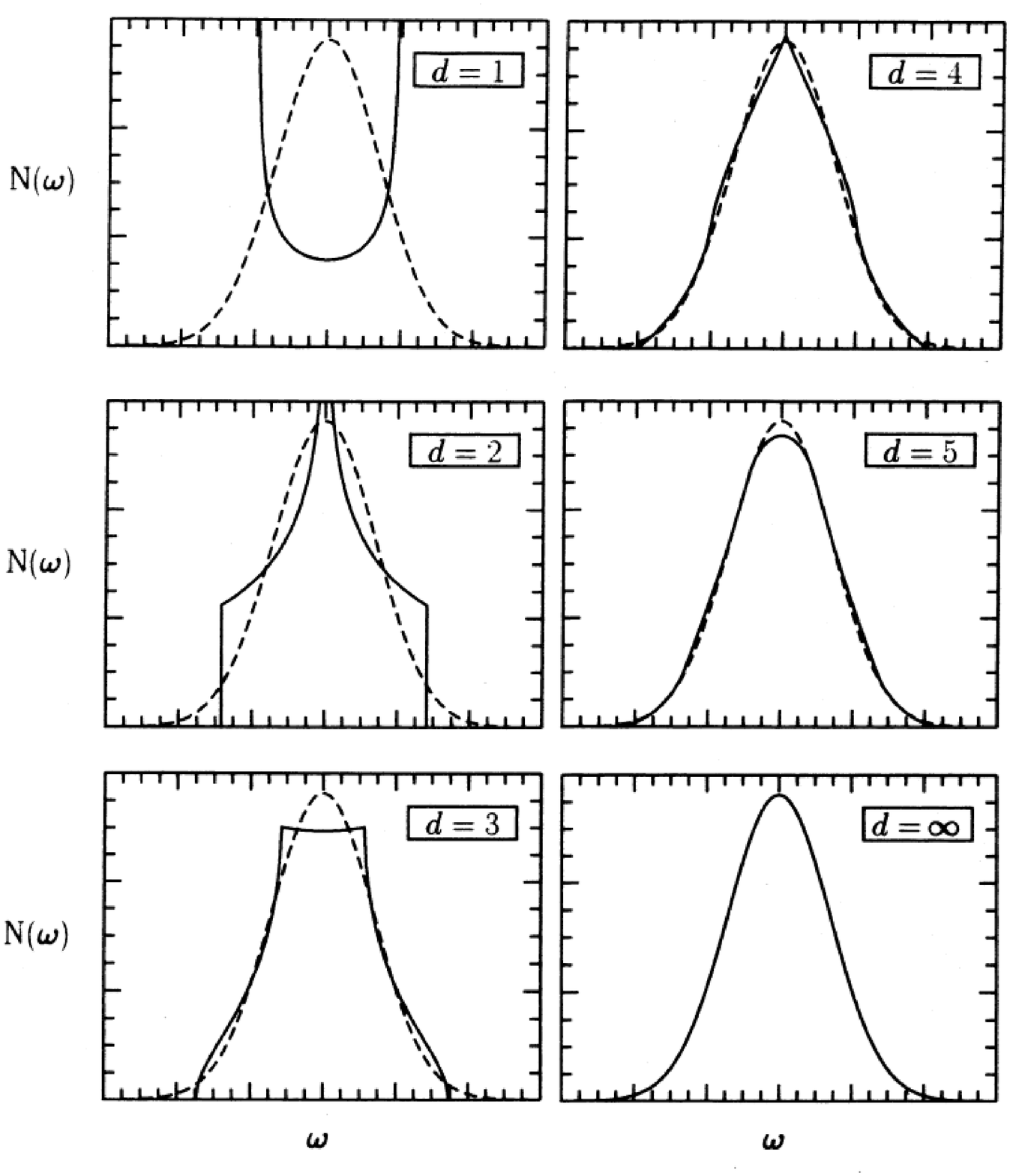}
\caption{Density of states of a tight-binding dispersion relation $\epsilon_{\bm{k}}$ on a hypercubic lattice in $d = 1, 2, 3, 4, 5$ (full lines) as compared
to the result for $d = \infty$ (dashed line); from Ref. \cite{Jerusalem}.}
\label{fig1.2}
\end{figure}

It is  natural  to ask whether the limit $d \to \infty$ may
also be useful in the investigation of lattice models with itinerant
quantum-mechanical degrees of freedom  and, in particular, in the case of the Hubbard model.
Following Metzner and Vollhardt \cite{metzner89} we take a look at the kinetic energy term (\ref{G11.7a}), since the interaction term is purely local and is thereby completely independent of the lattice structure and the dimension. For nearest-neighbor hopping on a d-dimensional hypercubic
lattice (where $Z=2d$) $\epsilon_{\bm{k}}$ is given by\footnote{In the following we set Planck's constant $\hbar $, Boltzmann's constant $k_{B}$, and the lattice spacing
equal to unity.}
\begin{equation}
\epsilon_{\bm{k}} = - 2t \sum_{i = 1}^{d} \; \cos k_i.
\label{G11.8}
\end{equation}
The density of states (DOS) corresponding to $\epsilon_{\bm{k}}$ is
\begin{equation}
N_d (\omega) = \sum_{\bm{k}} \delta (\hbar\omega - \epsilon_{\bm{k}}).
\label{G11.9}
\end{equation}
This is simply the probability density for finding $\omega = \epsilon_{\bm{k}}$
for a random choice of $\bm{k} = (k_1, \ldots, k_d)$. If the $k_i$ are
chosen randomly, $\epsilon_{\bm{k}}$ in \eqref{G11.8} is the sum of
(independent) random numbers $-2t \cos k_i$. The central limit theorem then implies that in the limit $d \to \infty$ the DOS is given by a Gaussian
\begin{equation}
N_d (\omega) \stackrel{d \to \infty}{\longrightarrow} \; \frac{1}{2t \sqrt{\pi d}}
 \exp \Bigg[ - \Big( \frac{\omega}{2t \sqrt{d}} \Big)^2 \Bigg].
\label{G11.10}
\end{equation}
Unless $t$ is scaled properly with $d$  this DOS  will become arbitrarily broad and featureless for
\hbox{$d \to \infty$}. Clearly only the  ``quantum'' scaling
\begin{equation}
t \to \frac{t^*}{\sqrt{d}}, \; t^* = {\rm const.},
\label{G11.11}
\end{equation}
yields a non-trivial DOS \cite{Wolff83},\cite{metzner89}:
\begin{equation}
N_{\infty} (\omega) = \frac{1}{\sqrt{2 \pi} t^*} \; \exp \Bigg[ - \frac{1}{2}
\Big( \frac{\omega}{t^*} \Big)^2 \Bigg].
\label{G11.12}
\end{equation}
This DOS does not have any van Hove singularities.
The reason for this can  be seen when $N_d (\omega)$ is calculated explicitly
from \eqref{G11.9} \cite{Wolff83,MH89a}.
Expressing the $\delta$-function as a Fourier series one has
\begin{subequations}
\label{G11.13}
\begin{equation}
N_d (\omega) = \prod_{i = 1}^{d} \; \int_{-\pi}^{\pi} \;
\frac{dk_i}{2 \pi} \; \int_{-\infty}^{\infty} \; d \tau
e^{i \tau (\omega- \epsilon_{\bm{k}})}
\label{G11.13a}
\end{equation}
\begin{equation}
= \int_{- \infty}^{\infty} d \tau \; e^{i \omega \tau} [ \; J_0 (2 \tau t) \; ]^d,
\label{G11.13b}
\end{equation}
\end{subequations}%
where $J_0 (x) = 1 - x^2 + {\cal O} (x^4)$, $x \ll 1$, is the zero-order Bessel-function.
For $d \gg 1$ the main contribution to the integral comes from the first
extremum of $J_0(x)$, i.~e., $\mid \tau \mid \stackrel{<}{\sim} 1/2t \sqrt{d}$,
while van Hove
singularities are due to higher extrema, yielding exponentially small contributions to
$N_d (\omega)$. Hence, using the scaling \eqref{G11.11}, one finds for $d \gg 1$
\begin{equation}
N_d (\omega) = \frac{1}{\sqrt{2 \pi}\;  t^*} \; e^{- \frac{1}{2} (\omega/t^*)^2}
\Bigg\{ 1 - \frac{1}{16d} \Bigg[ \Big( \frac{\omega}{t^*} \Big)^4
- 6 \Big( \frac{\omega}{t^*} \Big)^2 + 3 \Bigg] + {\cal O} \Big( \frac{1}{d^2} \Big)
\Bigg\}.
\label{G11.14}
\end{equation}
It is interesting to compare $N_d (\omega)$ for different $d$ as shown in Fig.~\ref{fig1.2}.
For $d \geq 3$ the shapes rapidly approach the $d = \infty$
result; the main difference is that for $d < \infty$
the band has a finite width, while in $ d = \infty $ there exist exponentially small
tails for all $\omega$.

The scaling \eqref{G11.11} expresses the fact that, for some
randomly chosen $\bm{k}$,
\begin{equation}
\epsilon_{\bm{k}}/t \sim {\cal O} (\sqrt{d}) \; ,  \; d \to \infty
\label{G11.15}
\end{equation}
since $\epsilon_{\bm{k}}/t$ is the sum of $d \to \infty$ many
random numbers from the interval $[-1,1]$. Clearly, $\bm{k} = 0$ and
$\bm{k} = (\pi , \ldots , \pi)$ are special values,
for which \eqref{G11.15} does not hold. However, as long as $\epsilon_{\bm{k}}$
appears under an integral these points have zero measure.

Using the relation
\begin{equation}
\frac{1}{L} \sum_{\bm{k}} F(\epsilon_{\bm{k}}) = \int_{- \infty}^{\infty} d \omega N(\omega) F(\omega)
\label{G11.16}
\end{equation}
where $L$ is the number of lattice sites, the kinetic energy of the non-interacting
electrons is found as
\begin{equation}
E_{\rm kin}^{0} = \sum_{\bm{k}, \sigma} \epsilon_{\bm{k}} n_{\bm{k}\sigma}
= - 2Lt^{*2} N_{\infty} (E_F).
\label{G11.17}
\end{equation}
We see that only the quantum scaling \eqref{G11.11} leads
to a  finite kinetic energy in $d = \infty$. Without this scaling $E_{\rm kin}^0$ would diverge, while the classical
scaling $t \to t^*/Z$ would reduce $E_{\rm kin}^0 $ to zero\footnote{To obtain a physically meaningful mean-field
theory for a model its internal or free energy  has to remain finite in the
limit $d$ or $Z\rightarrow \infty $.
While for the Ising model
the scaling $J\rightarrow {\tilde{J}}/Z$, ${\tilde{ J}}$= const., was rather obvious this is not so for more complicated
models. Namely, fermionic or bosonic many-particle systems are
usually described by a Hamiltonian consisting of several
non-commuting terms, e.g., a kinetic energy and an interaction, each of which is associated with a coupling
parameter, usually a hopping amplitude and an interaction, respectively. In such a case
the question of how to scale these parameters has no unique answer
since this depends on the physical effects one wishes to
explore. In any case, the scaling should be
performed such that the model remains non-trivial and that its internal or free
energy stays finite in the $Z\rightarrow \infty $ limit. By
``non-trivial'' we mean that not only $\langle \hat{H}_{0}\rangle $ and
$\langle \hat{H}_{\mathrm{int}}\rangle $, but also the \emph{competition}
between these terms, expressed by $\langle \lbrack
\hat{H}_{0},\hat{H}_{\mathrm{int}}]\rangle $, should remain finite. In the case of the Hubbard model it would be possible to employ classical scaling for the hopping amplitude, i.e., $t \to t^*/Z, \; t^* = {\rm const.}$,  but then the kinetic energy would be reduced to zero in the limit $d \to \infty$, making the resulting model uninteresting (but not unphysical) for most purposes. For the bosonic Hubbard model the situation is more subtle due to the occurrence of Bose-Einstein condensation; for a discussion see Ref. \cite{B-DMFT}.\label{Competition}}.

The interaction term in \eqref{G11.7} is seen to be purely local and independent
of the surrounding; hence it is independent of the spatial dimension of the system.
Consequently, the on-site interaction $U$ need not be scaled.
So we see that the scaled Hubbard Hamiltonian
\begin{equation}
\hat{H} = - \frac{t^*}{\sqrt{Z}} \; \sum_{\langle \bm{R}_i, \bm{R}_j \rangle}
\sum\limits_\sigma
 \; \hat{c}_{i \sigma}^{+} \hat{c}_{j\sigma}
+ U \sum_{\bm{R}_i} \hat{n}_{i \uparrow} \hat{n}_{i \downarrow}
\label{G11.18}
\end{equation}
 has a nontrivial $Z \to \infty$ limit, where
 both terms, the kinetic energy and the interaction, are of
the same order of magnitude and are thereby able to compete. It is this competition between the two terms which leads to interesting many-body physics (see footnote \ref{Competition}).

The quantum scaling \eqref{G11.11} was determined within a $\bm{k}$- space formulation. We will now derive the same result within a position-space formulation.

\subsection{Simplifications of the many-body perturbation theory}
\label{sec:scaling}

The most important consequence of the scaling \eqref{G11.11} is the fact that it leads to significant \emph{simplifications} in the investigation
of Hubbard-type lattice models \cite{metzner89,MH89a,metzner89a,MH89b,Czycholl,Brandt}. To understand this point better we take a look at the perturbation theory in terms of $U$.
At $T = 0 $ and $U = 0$ the kinetic energy \eqref{G11.17} may be written
as
\begin{equation}
E_{\rm kin}^0 = - t \sum_{\langle \bm{R}_i, \bm{R}_j \rangle} \sum_{\sigma}
 \; g_{ij, \sigma}^{0},
\label{G11.19}
\end{equation}
where $ g_{ij , \sigma}^0 = \langle \hat{c}_{i \sigma}^{+} \hat{c}_{j \sigma}^{} \rangle_0$
is the one-particle density matrix. This quantity can also be interpreted as the amplitude for transitions
between site $\bm{R}_i$ and $\bm{R}_j$, whose square  is proportional to the \emph{probability} for a particle to hop from $\bm{R}_j$ to $\bm{R}_i$, i.e., $\mid g_{ij , \sigma}^0 \mid^2 \sim 1/Z \sim 1/d$ since $\bm{R}_j$ has ${\cal O}(d)$ nearest neighbors $\bm{R}_i$. Thus the sum of $\mid g_{ij , \sigma}^0 \mid^2$
over all nearest neighbors must yield a constant.
In the limit $d \to \infty$ we then have
\begin{equation}
g_{ij, \sigma}^0 \sim {\cal O} \Big( \frac{1}{\sqrt{d}} \Big) \; , \;
\bm{R}_i \; NN \; {\rm of} \; \bm{R}_j.
\label{G11.20}
\end{equation}
Since the sum over the NN-sites $\bm{R}_i$ in \eqref{G11.19} is of ${\cal O}(d)$ the NN-hopping amplitude $t$ must obviously be scaled according to \eqref{G11.11} for $E_{\rm kin}^0$ to remain finite in the limit $d, Z \to \infty$. Hence, as expected,  a real-space formulation yields the same results for the required scaling of the hopping amplitude.

The one-particle Green function
(``propagator'') $G_{ij, \sigma}^0 (\omega)$ of the non-interacting system
obeys the same scaling as $ g_{ij , \sigma}^0$.
%
This  follows directly from its definition
\begin{equation}
G_{ij , \sigma}^0(t)\equiv -\langle T \hat{c}_{i\sigma}(t) \hat{c}^+_{j\sigma}(0) \rangle_0,
\label{G11.21}
\end{equation}
where $T$ is the time ordering operator, and the time evolution of the operators is provided by the Heisenberg representation. The one-particle density matrix is obtained as $g_{ij,\sigma}^0 =\lim_{t\rightarrow 0^-} G^0_{ij,\sigma}(t)$.  If $g_{ij,\sigma}^0$ obeys (\ref{G11.20}) the one-particle Green function must follow the same scaling at all times since this property does not dependent on the time evolution and the quantum mechanical representation. The Fourier transform $G_{ij , \sigma}^0 (\omega)$ also preserves this property.

It is important to realize that, although the propagator $G_{ij , \sigma}^0 \sim 1/\sqrt{d}$ vanishes
for $d \to \infty$, the particles are {\em not} localized, but are still mobile. Indeed,
 even in the limit
$d \to \infty$ the off-diagonal elements of $G_{ij , \sigma}^0$ contribute, since a particle may
hop to $d$ nearest neighbors with reduced amplitude $t^*/\sqrt{2d}$. For general $i, j$ one finds
\cite{vanDongen89,metzner89a}
\begin{equation}
G_{ij , \sigma}^0 \sim {\cal O}
\Big( 1/d^{\parallel \bm{R}_i - \bm{R}_j \parallel/2 } \Big),
\label{G11.22}
\end{equation}
where $\parallel \bm{R} \parallel = \sum_{n = 1}^{d} \mid R_n \mid $ is the length of $\bm{R}$ in
the so-called ``New York metric'' (also called ``taxi cab metric'', since particles
only hop along horizontal or vertical lines, never along a diagonal).

\begin{figure}
\includegraphics[width=0.8\textwidth]{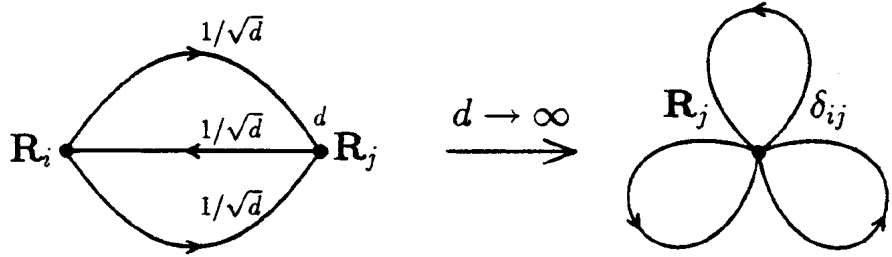}
\caption{Contribution to the irreducible self-energy for the Hubbard model in second-order perturbation theory
in $U$, and its collapse
in the limit $d \to \infty$.}
\label{fig1.3}
\end{figure}

It is the property \eqref{G11.22}
which
is the origin of all simplifications arising in the limit
$d \to \infty$. In particular, it implies the collapse of all
connected, irreducible perturbation theory diagrams in position space
\cite{metzner89,MH89a, metzner89a}. This
is illustrated in Fig.~\ref{fig1.3}, where a contribution in second-order perturbation
theory to the irreducible self-energy, $\Sigma_{ij}^{(2)}$,
is shown.

\noindent
In all physically interesting circumstances $\Sigma_{ij}^{(2)}$ will
only enter in a sum over $\bm{R}_i$ and $\bm{R}_j$. Then it becomes
apparent that $\Sigma_{ij}^{(2)}$ is only of order $1/\sqrt{d}$ small,
unless $i = j.$
Namely, for $j \neq i$ the three lines, corresponding to $G_{ij , \sigma}^0$, contribute a
factor $1/d^{3/2}$, while the sum over
NN- sites $\bm{R}_j$ of $\bm{R}_i$ contributes a factor
$d$. Only for $i = j$ is the value of $\Sigma_{ij}^{(2)}$
independent of $d$. Hence in the limit $d \to \infty$ the
diagram on the left-hand side of Fig.~\ref{fig1.3} is
equivalent to the ``collapsed'', petal-shaped diagram on the r.h.s., provided $i = j$; otherwise
it is zero. More generally, any two vertices
which are connected by more than two separate paths
will collapse onto the same site \footnote{Here a ``path''
is any sequence of lines in a diagram; they are ``separate'' when they have no lines
in common.}. In particular, the
external vertices of any irreducible self-energy diagram are
always connected by three separate paths and hence always
collapse. As a consequence the full, irreducible self-energy becomes a purely local
quantity \cite{metzner89,MH89a}:
\begin{subequations}
\label{G11.23}
\begin{equation}
\Sigma_{ij, \sigma} (\omega) \stackrel{d \to \infty}{=}
\Sigma_{ii, \sigma}(\omega) \delta_{ij}.
\label{G11.23a}
\end{equation}
In the paramagnetic phase we may write $\Sigma_{ii, \sigma}(\omega) \equiv \Sigma (\omega)$.
The Fourier transform of $\Sigma_{ij , \sigma}$ is seen to become
momentum-independent
\begin{equation}
\Sigma_{\sigma} (\bm{k}, \omega) \stackrel{d \to \infty}{\equiv}
\Sigma_{\sigma} (\omega).
\label{G11.23b}
\end{equation}
\end{subequations}%
This leads to tremendous simplifications in all many-body
calculations for the Hubbard model and related models. It
should be noted that a $\bm{k}$-independence of $\Sigma$ is sometimes
\emph{assumed} as a convenient approximation (``local approximation'') \cite{Kajzar78,Treglia80,Bulk90}. Here we identified the
limit where this is indeed exact.

\begin{figure}
\includegraphics[width=0.5\textwidth]{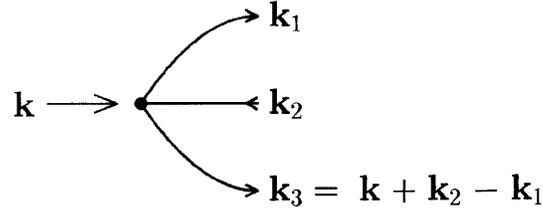}
\caption{Interaction vertex for the Hubbard model with external
momentum and frequency $\textbf{k} = (\bm{k}, \omega)$.}
\label{fig1.4}
\end{figure}

The result expressed in \eqref{G11.23b} may equally be obtained
by working in $\bm{k}$-space from the beginning \cite{MH89a}. For this
we consider an external vertex where a momentum $\bm{k}$ enters from
outside (see Fig.~\ref{fig1.4}).

Making use of the fact that

\noindent (i) the Hubbard interaction is momentum
independent,

\noindent (ii) the $\bm{k}$-dependent, free propagator is given by
\begin{equation}
G_{\bm{k}}^0 (\omega) = \frac{1}{\omega - \epsilon_{\bm{k}} + \mu}
\equiv G^0 (\epsilon_{\bm{k}}, \omega)
\label{G11.24}
\end{equation}
and only depends on $\bm{k}$ via $\epsilon_{\bm{k}}$ (we do not write the
spin index explicitly), and

\noindent (iii) that momenta only enter
explicitly in the conservation of momentum at the vertex, we see that the
evaluation of the vertex involves the momentum summation
\begin{subequations}
\label{G11.25}
\begin{equation}
\frac{1}{L^3} \sum_{\bm{k}_1 , \bm{k}_2, \bm{k}_3} G_{\bm{k}_1}^0
(\omega_1) G_{\bm{k}_2}^0 (\omega_2) G_{\bm{k}_3}^0 (\omega_3)
\delta^* ( \bm{k} - \bm{k}_1 +\bm{k}_2 - \bm{k}_3 )
\label{G11.25a}
\end{equation}
\begin{equation}
\equiv \prod_{ i = 1}^{3} \Big[ \int d \epsilon_i G^0 (\epsilon_i , \omega_i)
\Big] N_{\bm{k}} (\epsilon_1 , \epsilon_2 , \epsilon_3 ),
\label{G11.25b}
\end{equation}
\end{subequations}%
where $\omega_1 - \omega_2 + \omega_3 = \omega$ and
\begin{equation}
\delta^* (\bm{q}) = \sum_{\bm{K}} \ \delta (\bm{q} + \bm{K} )
= \frac{1}{(2\pi)^d}
\sum_{\bm{R}} e^{i \bm{q} \cdot (\bm{R} - \bm{R}_0)}
\label{G11.26}
\end{equation}
is the ``Laue-function'' which guarantees momentum
conservation up to a reciprocal lattice vector $\bm{K}$. The lattice
summation extends over all sites $\bm{R}$ relative to some
origin $\bm{R}_0$ (without loss of generality we may put $\bm{R}_0 = 0$)
 and couples momenta explicitly. Without this coupling the momentum-summation would be
simple because we would be able to use \eqref{G11.16}. In \eqref{G11.25b} we
therefore  introduced a generalized density of states
\begin{equation}
N_{\bm{k}} (\epsilon_1, \epsilon_2, \epsilon_3) =
\frac{1}{L^3} \sum_{\bm{k}_1,\bm{k}_2, \bm{k}_3} \delta
(\epsilon_1 - \epsilon_{\bm{k}_1} ) \delta (\epsilon_2 - \epsilon_{\bm{k}_2} )
 \delta (\epsilon_3 - \epsilon_{\bm{k}_3} )
 \delta^* (\bm{k} - \bm{k}_1 + \bm{k}_2 - \bm{k}_3 ),
\label{G11.27}
\end{equation}
which is the probability density for $\epsilon_i = \epsilon_{\bm{k}_i}
(i = 1,2,3)$ for given $\bm{k}$.
Writing the $\delta$-functions in \eqref{G11.27} as a Fourier
series (see \eqref{G11.13a}) and using \eqref{G11.26}, one
finds in the limit $d = \infty$
\begin{equation}
N_{\bm{k}} (\epsilon_1, \epsilon_2 , \epsilon_3) = N_{\infty} (\epsilon_1)
N_{\infty} (\epsilon_2) N_{\infty} (\epsilon_3 ) , \quad d = \infty
\label{G11.28}
\end{equation}
i.~e., for all $\bm{k}$ the quantity $N_{\bm{k}}$ factorizes into
a product of one-particle DOS's. This
is equivalent to replacing the Laue-function in \eqref{G11.27} by  unity
\begin{equation}
\delta^* (\bm{q}) \stackrel{d \to \infty}{=} 1.
\label{G11.29}
\end{equation}
It effectively means that the momentum conservation
constraint may be ignored in $d = \infty$. Defining the position of
the interaction vertex in Fig.~\ref{fig1.4} by $\bm{R}_0 \equiv 0$,
\eqref{G11.29} means that in the lattice sum over $\bm{R}$ only the local term
$\bm{R} = 0$ contributes. This is, once again, the collapse-phenomenon discussed
above. Due to the irrelevance of momentum conservation an external
momentum $\bm{k}$ cannot enter into the internal structure of an irreducible
self-energy diagram; this makes the irreducible self-energy
 $\bm{k}$-independent in
$d = \infty$ (see \eqref{G11.23b}). Note, however, that the
total momentum of a particle must be conserved for the theory
to be meaningful.

Due to the simplifications caused by \eqref{G11.28} or \eqref{G11.23},
the most important obstacle for actual diagrammatic calculations in
finite dimensions $d \geq 1$, namely the integration over intermediate
momenta, is removed in $d = \infty$.
While in finite dimensions these integrations
lead to untractable technical problems, they become simple in
$d = \infty$, since one can replace them by one-dimensional
integrations over the DOS.

It should be noted that the limit $d \to \infty$ does not affect
the {\em dynamics} of the system at all. Time is always one-dimensional and
hence there is no ``collapse'' in the frequency variables. In spite of the
simplifications in position (or momentum) space the problem retains its
full dynamics in $d = \infty$.

\subsection{Interactions beyond the on-site interaction}
\label{sec:selfenergy}

\begin{figure}
\includegraphics[width=\textwidth]{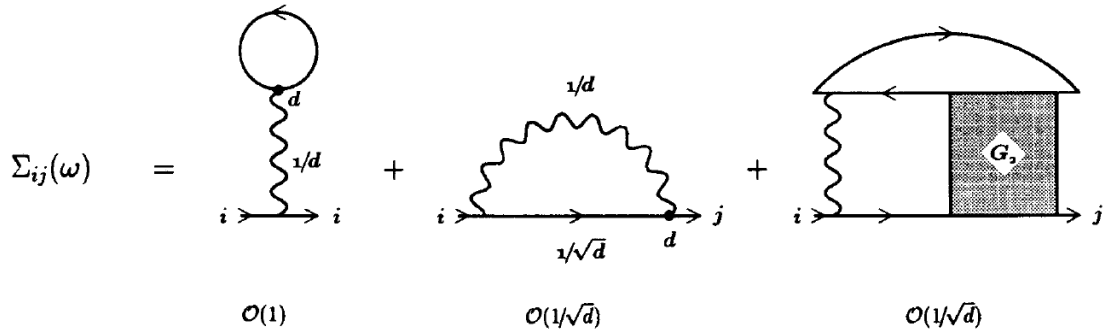}
\caption{Diagrams representing the irreducible self-energy.}
\label{fig1.5}
\end{figure}

In the case of more general interactions than the Hubbard interaction,
e.~g., nearest neighbor interactions such as
\begin{equation}
\hat{H}_{nn} = \sum_{\langle \bm{R}_i , \bm{R}_j \rangle} \; \sum_{\sigma \sigma'} \; V_{\sigma \sigma'} \hat{n}_{i \sigma}^{}
\hat{n}_{j \sigma'}^{}
\label{G11.30}
\end{equation}
the interaction constant has to be scaled, too, in the limit
$d \to \infty$. In the case of \eqref{G11.30}, which has the form
of a classical interaction, the ``classical'' scaling
\begin{equation}
V_{\sigma \sigma'} \to \frac{V_{\sigma \sigma'}^*}{Z}
\label{G11.31}
\end{equation}
is required. Of course, the propagator
still has the dependence \eqref{G11.22}. The self-energy has the
general diagrammatic form shown in Fig.~\ref{fig1.5}.

Due to \eqref{G11.31} all contributions, except for the Hartree-term, are
found to vanish in $d = \infty$. Hence nonlocal interactions only contribute
via their Hartree-contribution, which is purely static. This gives
the Hubbard interaction a unique role: of all interactions for fermionic lattice models only the
Hubbard interaction remains dynamical in the limit $d \to \infty$
\cite{MH89a}.

\subsubsection{The Hartree approximation revisited}

We are now able to answer the question whether the decoupling leading to the Hartree approximation \eqref{G14.2} also becomes exact
in the limit $d\rightarrow \infty$ as in the case of the Ising model.
%
Using \eqref{G14.4} the answer is simple: as
long as the on-site interaction in \eqref{G14.1} is present, the decoupling \eqref{G14.4} cannot  become exact in $d \to \infty$, because
for $i = j$ (and hence $\sigma = - \sigma'$) the r.h.s. of \eqref{G14.4} is
always of order unity. Namely,
 at a given site $\bm{R}_i$ the potential, in units of $U$, felt by an
electron with spin $\sigma$ is  either $0$ (if there is no
$-\sigma$ spin present)
or $1$ (if a $-\sigma$ spin is present).
These on-site fluctuation effects
are not described by the Hartree decoupling
approximation. This has been explicitly verified by  investigating  the thermodynamics of \eqref{G14.1} in $ d = \infty$
\cite{vanDongen91a}.
 On the other hand, if the on-site
interaction in \eqref{G14.1} did not exist (e.g., in the case of
spinless fermions) the Hartree approximation~\eqref{G14.4} would indeed become exact in the limit $d \to \infty$.

\subsection{One-particle and two-particle propagators}

Due to the $\bm{k}$-independence of the irreducible self-energy, \eqref{G11.23b}, the
one-particle propagator of an interacting lattice fermion system is given by
\begin{equation}
G_{\bm{k}, \sigma}^{} (\omega) =
\frac{1}{\omega - \epsilon_{\bm{k}} + \mu - \Sigma_{\sigma} (\omega)}.
\label{G11.32}
\end{equation}
Most importantly, the $\bm{k}$ dependence of $G_{\bm{k}} (\omega)$ comes
entirely from the energy dispersion
$\epsilon_{\bm{k}}$ of the {\em non}-interacting
particles. This means that for a homogeneous system with
 the propagator
\begin{equation}
G_{ij , \sigma} (\omega) = L^{-1} \; \sum_{\bm{k}} G_{\bm{k}, \sigma} (\omega)
e^{i \bm{k} \cdot (\bm{R}_i - \bm{R}_j) }
\label{G11.33}
\end{equation}
its local part, i.~e., $G_{ii , \sigma}$, can be calculated in closed form \cite{MH89b}
\begin{subequations}
\label{G11.34}
\begin{eqnarray}
G_{ii , \sigma} (\omega) & = & L^{-1} \sum\limits_{\bm{k}} G_{\bm{k}, \sigma} (\omega)
= \int\limits_{-\infty}^{\infty}  d E \frac{N_{\infty} (E)}{ \omega - E + \mu
- \Sigma_{\sigma} (\omega)}
\\[12pt]
\label{G11.34a}
& = & - i  \sqrt{\frac{\pi}{2}} \; \frac{1}{t^*} \;
e^{-z_{\sigma}^2} {\rm erfc} (- iz_{\sigma})
\\[12pt]
& \equiv & G_{\sigma}(\omega),
\label{G11.34b}
\end{eqnarray}
where $z_{\sigma} = (\omega + \mu - \Sigma_{\sigma} (\omega))/ \sqrt{2} t^*$, $N_{\infty} (E)$ is given by~\eqref{G11.12}, and ${\rm erfc(x)}$ is the error function. In the following we will limit our discussion to the paramagnetic phase and omit the spin index. The spectral function of the interacting system (often referred to as the DOS as in the non-interacting case) is then given by
\begin{equation}
A(\omega) = - \frac{1}{\pi} {\rm Im} G(\omega + i0^+);
\label{G11.34c}
\end{equation}
\end{subequations}
for $U=0$ one has $A(\omega)\equiv N(\omega)$.
In the limit $d \rightarrow \infty$ two quantities then play the most important role: the local propagator $G(\omega)$ and the self-energy $\Sigma (\omega)$.

\begin{figure}
\includegraphics[width=0.6\textwidth]{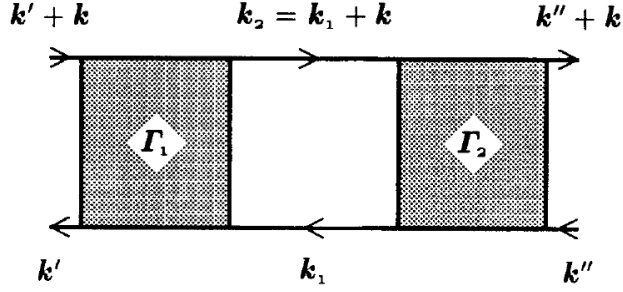}
\caption{Diagrammatic contribution to the two-particle propagator.}
\label{fig1.6}
\end{figure}

Concerning the two-particle propagator $G_2$ (or correlation functions, etc.)
the collapse phenomenon is a little different from
the one discussed below \eqref{G11.27}. Namely, it does not occur
for all external momenta \cite{MH89c, vanDongen89}.
A typical contribution to $G_2$ is shown in Fig.~\ref{fig1.6}, where
$\Gamma_1 , \Gamma_2$ are irreducible, momentum independent vertices.

The calculation of this contribution is analogous
to that in \eqref{G11.25}, and involves the
$\bm{k}$-sum over a product of two one-particle propagators, i.~e., over
$G_{\bm{k}_1 }^0 G_{\bm{k}_2}^0$ with $\bm{k}_2 = \bm{k}_1 + \bm{k} $.
 Introducing a generalized
DOS
\begin{equation}
N_{\bm{k}} (\epsilon_1 , \epsilon_2) = \sum_{\bm{k}_{1} , \bm{k}_{2}}
\delta (\epsilon_1 - \epsilon_{\bm{k}_1}) \delta (\epsilon_2
- \epsilon_{\bm{k}_2})
\delta^* (\bm{k}_2 - \bm{k}_1 - \bm{k} )
\label{G11.35}
\end{equation}
we find that for $d = \infty$ this quantity only factorizes into $N_{\infty}
(\epsilon_1) \; N_{\infty} (\epsilon_2 )$ if
$\bm{k} \neq 0$, \hbox{$\bm{k} \neq (\pi, \ldots , \pi)$}, i.~e., when
$\epsilon_{\bm{k}} \sim {\cal O} (1/\sqrt{d}).$
Hence the Bethe-Salpeter ladder collapses for all $\bm{k}$, except for
 these two special values.

\subsection{Weak-coupling correlation energy for
the Hubbard Model}

The correlation energy $E_c (U)$ is defined as the energy by which the
Hartree-Fock energy is lowered when genuine
correlations are included
\begin{equation}
E_c (U) : = E_{\rm exact} (U) - E_{\rm HF} (U).
\label{G11.36}
\end{equation}
For $d \to \infty$  second-order Goldstone perturbation theory in the weak interaction $U$ yields
the simple
expression ($n_{\uparrow} = n_{\downarrow} = \frac{1}{2})$ \cite{metzner89}
\begin{equation}
E_c^{(2)} (U) = - U^2 \; \int_{0}^{\infty} \; d \omega \; e^{2 \omega^2}
\; P^2 (E_{F} - \omega) P^2 (- E_{F} - \omega ),
\label{G11.38}
\end{equation}
where $P (x)$ is the Gaussian probability function. It should be stressed that for
general dimensions the calculation of $E_c^{(2)} $ involves $3d$ momentum
integrals over a singular integrand. Analytic calculations are thereby ruled out
and even numerical integration techniques
becomes very cumbersome. Indeed, the application
of Monte-Carlo integration techniques
becomes mandatory already for $d \stackrel{>}{\sim} 2$.
By contrast, the case $d = \infty$ is seen to be the simplest
of all dimensions, including $d = 1$,
since it only demands a one-dimensional integral. \\

\begin{figure}
\includegraphics[width=0.8\textwidth]{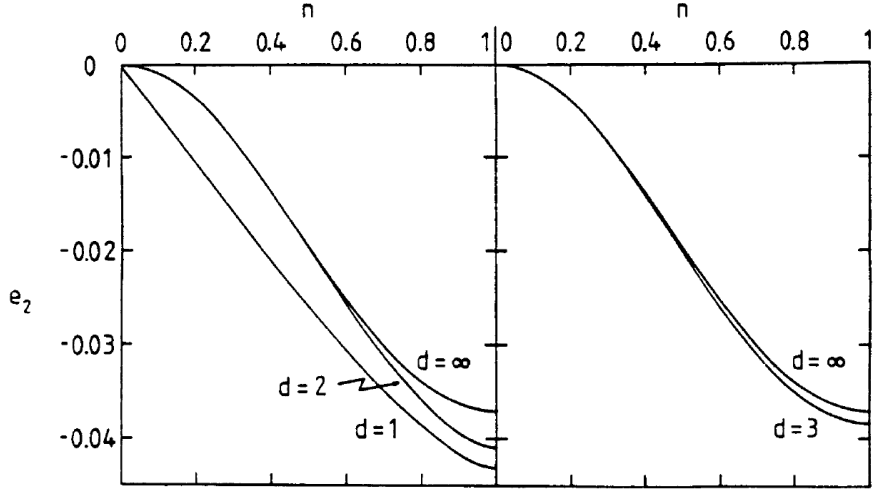}
\caption{Correlation energy
for the Hubbard model in second-order perturbation theory in $U$, \hbox{$e_2 = E_c^{(2)}/(2U^2/\mid\overline{\epsilon}_0\mid\,)$,}
vs. density $n$ for several lattice dimensions. Left: $d = 1,2, \infty$; right: $d = 3, \infty$. Here $\mid \overline{\epsilon}_0 \mid$
is the kinetic energy for $U=0$ and $n = 1$; after Ref. \cite{metzner89}.}
\label{fig1.7}
\end{figure}

In Fig.~\ref{fig1.7} $E_c^{(2)} /U^2$ is shown as a function of particle density $n$ for several
dimensions $d$. We see that the result for $d = 3$, which can only be
obtained by considerable effort, is very well approximated by the result
for $d = \infty$, which is easily calculated. The differences between $d = 1,2$ and $d = 3$
are mainly due to the singularities of the DOS in $d = 1,2$.


\subsection{Consequences of the $\bm{k}$-independence of the self-energy
}

We now discuss some more consequences of the
$\bm{k}$-independence of the self-energy
as derived by M\"{u}ller-Hartmann
\cite{MH89b}. Let us consider the Hubbard model
and concentrate on the paramagnetic phase.
That is, we do not wish to discuss the more complicated situation
with a broken symmetry for the moment.\footnote{On bipartite lattices
and for nearest neighbor hopping the Hubbard model has a
``perfect-nesting instability'' at half filling due to $\epsilon_{\bm{k}} =
- \epsilon_{\bm{k}  + \bm{Q}}$ with $\bm{Q} = (\pi, \ldots , \pi)$,
leading to an insulating state with antiferromagnetic correlations in $d\geq 2$. In this case
the system is not in a paramagnetic state even for arbitrarily small $U$.
However, by including a finite hopping amplitude to next-nearest
neighbors the paramagnetic phase becomes stable for small $U$.}
In the paramagnetic case at $T = 0$ the one-particle propagator \eqref{G11.32}
takes the form (again we do not write the spin index explicitly)
\begin{equation}
G_{\bm{k}} (\omega) = \frac{1}{\omega - \epsilon_{\bm{k}} + E_F - \Sigma (\omega)}.
\label{G11.39}
\end{equation}
In general, even when $\Sigma(\omega)$ is $\bm{k}$-dependent, the Fermi surface
is defined by the $\omega = 0$ limit of the denominator of \eqref{G11.39} as
\begin{subequations}
\label{G11.40}
\begin{equation}
\epsilon_{\bm{k}} + \Sigma_{\bm{k}} (0) = E_F.
\label{G11.40a}
\end{equation}
According to Luttinger and Ward \cite{LW60} the volume within the
Fermi surface is not changed by interactions, provided the effect of the
latter can be treated in infinite-order perturbation theory (hence no broken symmetry).
This is expressed by
\begin{equation}
n = \sum_{\bm{k} \sigma} \; \theta
[E_F - \epsilon_{\bm{k}} - \Sigma_{\bm{k}}
(0) ],
\label{G11.40b}
\end{equation}
where $n$ is the particle density and $\theta (x)$ is the step function.
In general, the $\bm{k}$-dependence of $\Sigma_{\bm{k}} (0)$
in \eqref{G11.40a}
implies that, in spite of \eqref{G11.40b}, the shape of
the Fermi surface of the interacting system will be quite different from that of the
non-interacting system (except for the rotationally invariant case
$\epsilon_{\bm{k}}=f(|\bm{k}|)$. For lattice fermion models in $d = \infty$,
where $\Sigma_{\bm{k}} (\omega) \equiv \Sigma (\omega)$,
\eqref{G11.40a} implies that the Fermi surface itself (and hence
the volume enclosed) is not changed by interactions.\footnote{In $d=\infty$ the notion of a Fermi surface of a lattice system is made complicated by the fact that in this limit the dispersion $\epsilon_{\bm{k}}$
is not a simple smooth function.} The Fermi energy is simply
shifted uniformly from its non-interacting value
$E_F^0$, i.~e.,  $E_F = E_F^0 + \Sigma (0)$, to keep $n$
in \eqref{G11.40b} constant.
From \eqref{G11.34a} we thus conclude that the $\omega  = 0$ value of the
local propagator, $G(0)$, and hence of the spectral function,
$A(0) = - \frac{1}{\pi}
{\rm Im} G(i0^+)$, is not changed
by interactions.\footnote{This behavior is well-known from the single-impurity Anderson model \cite{Hewson}.} Renormalizations of $N(0)$ can only come from a
$\bm{k}$-dependence of $\Sigma$, i.~e., if $\partial \Sigma /\partial \bm{k} \neq 0$.

For $\omega \to 0 $ the self-energy has the property
\begin{equation}
{\rm Im} \; \Sigma (\omega) \propto \omega^2
\label{G11.40c}
\end{equation}
which implies quasiparticle (Fermi liquid) behavior. The effective mass
\begin{equation}
\frac{m^*}{m} = \left. 1 - \frac{d \Sigma}{d \omega} \right|_{\omega = 0}
 = 1 + \frac{1}{\pi} \; \int_{-\infty}^{\infty} \; d \omega
\; \frac{{\rm Im} \Sigma (\omega + i0^-)}{\omega^2} \geq 1
\label{G11.40d}
\end{equation}
\end{subequations}%
is seen to be enhanced. In particular, the momentum distribution
\begin{equation}
n_{\bm{k}} = \frac{1}{\pi} \; \int_{-\infty}^{0} \; d \omega \; {\rm Im} G_{\bm{k}} (\omega)
\label{G11.41}
\end{equation}
has a discontinuity at the Fermi surface, given  by
$n_{k_{F}^-} - n_{k_F^+} = (m^*/m )^{-1}$, where $k_F^\pm = k_F \pm 0^+$.

\section{Dynamical mean-field theory for correlated lattice fermions}

Itinerant quantum mechanical models such as the Hubbard model
and its generalizations are much more
complicated than classical, Ising-type models.
Generally there do not even exist semiclassical approximations
for such models that might serve as a starting point for further
investigations. Under such circumstances the construction of
a  mean-field theory with  the comprehensive properties of the Weiss molecular
field theory for the Ising model will necessarily be much
more complicated, too. As discussed above there do exist well-known mean-field approximation
schemes, e.~g. Hartree-Fock, random-phase approximation, saddle-point
evaluations of path integrals, decoupling of operators. However,
these approximations do not provide mean-field theories
in the spirit of statistical mechanics, since
they are not able to provide a global description of a given
model (e.g., the phase diagram, thermodynamics, etc.)
in the  entire range of input parameters.

Here the limit of high spatial dimensions $d$ or coordination number $Z$  has again been extremely useful \cite{metzner89}. It provides the basis for the construction of a comprehensive mean-field theory for lattice fermions
 which is diagrammatically controlled and whose free energy has no unphysical singularities. The construction is based on the scaled Hamiltonian (26) and the simplifications in the many-body perturbation theory discussed in Sec. 3.2.
  There we saw that the local propagator $G(\omega)$, i.e., the amplitude for an electron to return to a lattice site, and the local but dynamical self-energy $\Sigma (\omega)$ are the most important quantities in such a theory.
Since the self-energy is a dynamical variable (in contrast to Hartree-Fock theory where it is merely a static potential) the resulting mean-field theory will also be dynamical
and can thus describe genuine correlation effects such as the Mott-Hubbard metal-insulator transition.

The self-consistency equations of this \emph{dynamical mean-field theory} (DMFT) for correlated lattice fermions can be derived in different ways. Nevertheless, all derivations make use of the fact that in the limit of high spatial dimensions  Hubbard-type models reduce to a ``dynamical single-site problem'', where the  $d$-dimensional lattice model is effectively described by the dynamics of the correlated fermions on a single site which is embedded in a ``bath'' provided by the other particles. In the following I will present two different derivations of the single-site action and the self-consistency equations of the DMFT.
I start with the approach by Jani\v{s} \cite{Janis91,Janis92b} who generalized the coherent potential approximation (CPA) for disordered systems.
 Then I discuss today's standard derivation developed independently by Georges and Kotliar \cite{Georges92} which is based on the mapping of the lattice problem onto a self-consistent single-impurity Anderson model; this approach was also employed by Jarrell \cite{Jarrell92}.
The DMFT equations derived within the CPA approach and the single-impurity approach, respectively, are identical. Nevertheless it is the Anderson-impurity formulation which was immediately adopted by the community since it makes contact with the theory of quantum impurities and Kondo problems; for a review see Ref. \cite{Fabrizio}. This is a well-understood branch of many-body physics  for whose solution efficient numerical codes had been developed already in the 1980's, in particular by making use of the quantum Monte-Carlo (QMC) method \cite{Hirsch86}.

Although the single-impurity based derivation of the DMFT is now the standard method I will present both approaches since it is always instructive to derive a theory in more than one way.

\subsection{Construction of the DMFT by generalizing the coherent potential approximation}
\label{sec:cpa}

\begin{figure}
\includegraphics[width=0.6\textwidth]{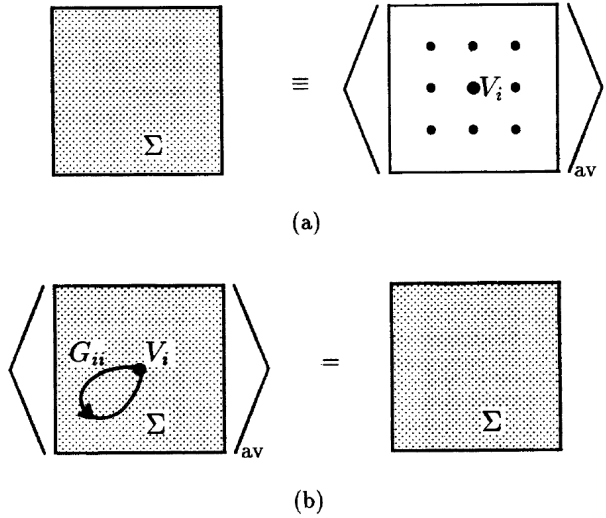}
\caption{(a) The random medium, described by local potentials $V_i$, is replaced
by an unknown, but exact, effective potential $\Sigma$.
(b) By demanding the average scattering from a single site with
potential $V_i - \Sigma$ to vanish,  $\Sigma$
is determined self-consistently; from Ref. \cite{Jerusalem}.}
\label{fig4.1}
\end{figure}

The \emph{coherent potential approximation} (CPA) is a well-known mean-field theory
which was originally developed in the context of disordered
systems \cite{Taylor67,Soven67,Yonezawa73,Elliott74}. To be specific let us consider
Anderson's  tight-binding Hamiltonian\footnote{Here the kinetic energy is already scaled according to \eqref{G11.11} to allow for the limit $Z \rightarrow \infty$ to be taken, in which case the CPA becomes exact (see Sec. 4.1.1.).} with local (``diagonal'') disorder
\cite{Anderson58}
\begin{equation}
\hat{H}_A = - \frac{t^*}{\sqrt{Z}} \; \sum_{\langle \bm{R}_i , \bm{R}_j \rangle} \; \hat{c}_i^+ \hat{c}_j
+ \sum_i \; V_i \hat{n}_i,
\label{G14.5}
\end{equation}
where $V_i$ is a random variable drawn from some distribution function
$P(V_i)$. The electrons described by \eqref{G14.5} do not
interact. Therefore we deal with the problem of a single particle moving
through a random medium. (Since there is no spin dependence one can suppress the spin index altogether and simply
work with a spinless  fermion). The problem is made complicated by the randomness. It requires one to
calculate the average of a physical quantity $X$ (which is a function of all site
energies $V_i$) with respect to $P(V_i)$. In particular, the \emph{arithmetic} average is obtained by
\begin{equation}
\langle X \rangle_{\rm av} := \prod_{\bm{R}_i} \; \int d V_i \; P(V_i) X
(V_1 , \ldots , V_L).
\label{G14.6}
\end{equation}
One may now proceed as follows:
\begin{enumerate}
\item The actual random medium, given by the local potentials $V_i$,
is thought to be replaced exactly  by an (unknown) {\em effective}
medium, described by a complex, frequency-dependent self-energy;
this defines the self-energy.
\item Since the effective medium is required to yield an
exact description of the random medium, we may remove
the medium at a site $\bm{R}_i$ ,
replace it by  an actual potential $V_i$
and then demand that, upon averaging, the scattering caused by the
perturbation of the medium due to $V_i$  vanishes identically. The self-consistency
condition expressed in the last step determines the previously unknown self-energy.
\end{enumerate}
Let $G_{ij} (z) \equiv G$ be the Green function of the
electron in the random medium, with $z$ as a complex frequency and $G_0$
as the unperturbed Green
function; we suppress site-indices for the moment \cite{Rickayzen80}.
The Lippmann-Schwinger equation for $G$ is given by
\begin{equation}
G = G_0 + G_0 VG \quad {\rm or} \quad G_0^{-1} G = 1 + VG.
\label{G14.7}
\end{equation}
We now introduce a self-energy
$\Sigma_{ij} (z) \equiv \Sigma$ into \eqref{G14.7}, which
plays the role of an unknown potential
\begin{equation}
(G_0^{-1} - \Sigma) G = 1 + (V - \Sigma) G,
\label{G14.8}
\end{equation}
and demand
\begin{equation}
\langle G \rangle_{\rm av} = (G_0^{-1} - \Sigma)^{-1}.
\label{G14.9}
\end{equation}
Multiplication of \eqref{G14.8} by $\langle G \rangle_{\rm av}$ yields
\begin{equation}
G =  \langle  G \rangle_{\rm av} + \langle G \rangle_{\rm av} (V - \Sigma) G.
\label{G14.10}
\end{equation}
Here $V - \Sigma$ is a new scattering potential,
 whose effect can be described by a $T$-matrix via
\begin{equation}
G = \langle G \rangle_{\rm av} + \langle G \rangle_{\rm av} T
\langle G \rangle_{\rm av},
\label{G14.11}
\end{equation}
where
\begin{equation} T = \frac{V - \Sigma}{1 - (V - \Sigma) \langle G \rangle_{\rm av}}.
\label{G14.12}
\end{equation}
Averaging of \eqref{G14.11} yields
\begin{equation}
\langle T \rangle_{\rm av} = 0,
\label{G14.13}
\end{equation}
which is a self-consistent equation for $\Sigma$. Eq. \eqref{G14.13} demands
that $\Sigma$ is determined in such a way that the scattering due to the
perturbation $V - \Sigma$ vanishes. If \eqref{G14.13} could be solved exactly
the entire problem would be solved. However, an exact solution
is usually not possible (an exception is the Lloyd  model where
$P(V_i)$ is given by a Lorentzian), so that an approximation
has to be made to proceed further. At this stage CPA \emph{assumes}
the self-energy to be site-diagonal
\begin{equation}
\Sigma_{ij} (\omega) = \Sigma (\omega) \delta_{ij},
\label{G14.14}
\end{equation}
i.~e., to be homogeneous. Eq. \eqref{G14.14} is equivalent to a
{\em single-site approximation} and corresponds to step 2 in
the construction of the CPA described
below \eqref{G14.6}. Since $\Sigma (\omega)$ is homogeneous
it is a $\bm{k}$ independent, but \emph{frequency} dependent
potential and thereby only adds to the frequency dependence
of $G_0^{-1}$, i.~e.,
the averaged propagator is simply given by the
unperturbed propagator with shifted frequency :
\begin{equation}
\langle G \rangle_{\rm av} = G_{ii}^0 (z - \Sigma (z)).
\label{G14.15}
\end{equation}
For  all sites $\bm{R}_i$ the condition
\eqref{G14.13} therefore reduces to $\langle T_i \rangle_{\rm av} = 0$, i.~e.,
\begin{equation}
\left\langle \; \frac{V_i - \Sigma (z)}{1 - (V_i - \Sigma (z) )
G_{ii}^0 (z - \Sigma (z))} \right\rangle_{\rm av} = 0,
\label{G14.16}
\end{equation}
where
\begin{equation}
G_{ii}^0 (z) = \int\limits_{-\infty}^{\infty} \; d \omega \; \frac{N (\omega)}{z - \omega}
\label{G14.17}
\end{equation}
is the local propagator, with $N(\omega)$ as the DOS of the unperturbed
system. Eq. \eqref{G14.16} implies that in the effective medium the average scattering
by a single site (``impurity'') vanishes. The single-site
aspect underlying the CPA may therefore be visualized as shown in Fig.~\ref{fig4.1} \cite{Yonezawa73}.

The CPA and its results have many attractive features:

\noindent (i) CPA is a non-perturbative,
but very simple and self-consistent theory;

\noindent (ii)  it may be considered the best single-site approximation
for the disorder problem
as can be inferred from the above
derivation;

\noindent (iii) it has the so-called
Herglotz-properties $\Sigma (z) = \Sigma^* (z^*)$ and
${\rm Im} (z-\Sigma (z)) \stackrel{>}{\scriptstyle{<}} $ for
$z \stackrel{>}{\scriptstyle{<}} 0$ \cite{MH73},
which implies that it has the correct analytic properties (positive DOS, etc);

\noindent (iv) it leads to very good qualitative and even quantitative
results for the one-particle properties of disordered systems. The
latter is true even in dimensions $d \stackrel{<}{\sim} 3$ and for parameter
values of the disorder strength and the impurity concentration where
CPA cannot
be linked to perturbation theory. These properties have made CPA the most widely used approximation
scheme for disordered systems.

\subsubsection{CPA and the limit $d \to \infty$}

The single-site aspect of the CPA outlined above and, in particular,
the property \eqref{G14.14}, indicate that
the CPA will become
exact in the limit of high coordination number $Z$. Indeed, by
investigating the moments of the electronic DOS  it was observed  that $1/Z$ appears as
a ``hidden'' small parameter, which governs the size of the corrections
to the CPA moments \cite{Schwartz}. However, by considering only the moments one cannot
draw conclusions about the validity of the
CPA itself.\footnote{In fact, in Ref. \cite{Schwartz} the
kinetic energy was not scaled according to $t \to t^*/\sqrt{Z}$, but
 the bandwidth was kept constant, corresponding to classical scaling ($t \to t^*/Z$). In this case the $Z \to \infty$
limit does not lead to CPA at all, but becomes trivial.}
Using the  scaling of $t$ in \eqref{G11.11}
one can show
\cite{vlaming92}
that \eqref{G14.14},
and hence CPA itself, indeed becomes exact for $Z \to \infty$,  irrespective
of the lattice structure. In other words, CPA solves the disorder
Hamiltonian \eqref{G14.5} exactly in the limit $Z \to \infty$. This finding
explains why for $Z < \infty$
the CPA can be so successful even for intermediate
values of the disorder strength and impurity concentration,
i.~e., when perturbation theory in these parameters is no longer  justified. In fact, we now see that there is an
additional small parameter, namely $1/Z$, which allows for a
perturbation expansion that is  independent of the values of the input parameters.
In view of the existence of a small parameter $1/Z$, CPA is seen to be a
controlled mean-field theory. It is therefore not so surprising that the
CPA often gives qualitatively and quantitatively correct results even
in dimensions $d \stackrel{<}{\sim} 3$.

\subsubsection{ Alternative derivation of CPA}
\label{sec:cpa_al}
\begin{figure}
\includegraphics[width=0.8\textwidth]{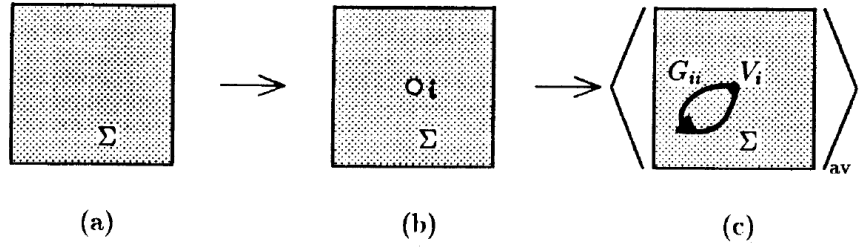}
\caption{Steps to construct the exact averaged free energy functional in $d = \infty$:
(a) Homogeneous effective medium,
(b) the medium is removed at site $\bm{R}_i$,
(c) the cavity in the medium at site $\bm{R}_i$ is filled with the actual potential
$V_i$. The probability amplitude for a particle from the medium to be at site
$\bm{R}_i$ (more precisely: the return amplitude) is given by the local propagator $G_{ii}$; from Ref. \cite{Jerusalem}.}
\label{fig4.2}
\end{figure}

I will now  show that the CPA can also be derived from a variational principle \cite{Janis92a}.
To this end the coherent potential, i.~e., the self-energy, will be determined
from the averaged  free-energy functional $\Omega$ of the
corresponding single-site
problem \cite{Janis89}. This
field-theoretical approach has the advantage
that it can be generalized to treat interacting
lattice models, such as Hubbard-type models and disorder models
on the same basis (see Sec.~\ref{sec:cpa_gen}).
Furthermore, the physical idea behind this single-site theory is
very transparent
and may be explained in terms of the following simple picture \cite{Janis92a}. To calculate the averaged free energy corresponding to a
single site $\bm{R}_i$ of the medium we have to determine the energy density
\begin{equation}
 \langle \Omega_i \rangle_{\rm av} \equiv
\frac {\langle \Omega \rangle_{\rm av}}{L},
\label{G14.18}
\end{equation}
where $L$ is the number of lattice sites. To this end we consider the second step of the CPA strategy outlined
below \eqref{G14.6} and drawn schematically in Fig.~\ref{fig4.2}:

\noindent (i) we start from the homogeneous, effective medium with free energy density
$\Omega_{\rm med}/L$;

\noindent (ii) we remove the medium at site $\bm{R}_i$,
i.~e., subtract a corresponding energy $\Omega_i$ ; and

\noindent (iii) replace it by a site with a bare potential $V_i$, i.~e., add a
corresponding averaged energy $\langle \Omega_i^{\rm bare}
\rangle_{\rm av}$; so we have
\begin{equation}
\frac {\langle \Omega \rangle_{\rm av}}{L}= \frac { \Omega_{\rm med}}{L}
- \Omega_i + \langle \Omega_i^{\rm bare} \rangle_{\rm av}.
\label{G14.19}
\end{equation}

To calculate the contributions in \eqref{G14.19} we make use of the
trace-log formula for the free energy density which in
the non-interacting case (superscript $0$) reads \cite{Negele88}
\begin{equation}
\frac {\Omega^0}{L} = - T \; {\rm tr} \sum_{\bm{k}} \; \ln
\bigl[ G_{\bm{k}}^0 (i \omega_n) \bigr]^{-1}.
\label{G14.20}
\end{equation}
Here $G_{\bm{k}}^0$ is given by \eqref{G11.24}, $\omega_n = (2n+1) \pi T$
are the Matsubara frequencies and the trace operation implies
\begin{equation}
{\rm tr} \widehat{=} \; \sum_{\sigma} \; \sum_{n = - \infty}^{\infty},
\label{G14.21a}
\end{equation}
where in the present problem the spin is unimportant. Using \eqref{G14.15}, i.~e.,
\begin{equation}
\langle G_{ii}^{} (z) \rangle_{\rm av} = G_{ii}^0 (z - \Sigma (z))
\label{G14.21b}
\end{equation}
we have
\begin{eqnarray}
\frac { \Omega_{\rm med}}{L} & = & - T \; {\rm tr} \; \sum_{\bm{k}} \; \ln \;
\Bigl[ G_{\bm{k}}^0 (i \omega_n - \Sigma (i \omega_n)) \Bigr]^{-1} \nonumber \\[10pt]
& = & - T \; {\rm tr} \int\limits_{-\infty}^{\infty} \; d E N(E) \ln
\; \bigl[ i \omega_n -\Sigma (i \omega_n) + \mu -E \bigr]
\label{G14.22}
\end{eqnarray}
and
\begin{equation}
\Omega_i = - T \; {\rm tr} \, \ln \;
\Bigl[ G_{ii}^0 (i \omega_n - \Sigma (i \omega_n)) \Bigr]^{-1}
= - T \; {\rm tr} \; \ln \; \bigl[ \langle G_{ii}^{} (i \omega_n)\rangle_{\rm av} \bigr]^{-1},
\label{G14.23}
\end{equation}
where $G_{ii}^0$ is the local, unperturbed propagator
needed to remove the medium at site $\bm{R}_i$. Finally,
$\langle \Omega_i^{\rm bare} \rangle_{\rm av}$ can be obtained from
\begin{equation}
\Omega_i^{\rm bare} = - T \; \ln {\cal Z}_i^{\rm bare},
\label{G14.24}
\end{equation}
where the local partition function is determined by the action $S_i^{\rm bare}$
as \cite{Negele88}
\begin{equation}
{\cal Z}_i^{\rm bare} = \int {D} c {D} c^* \;
\exp \big[- S_i^{\rm bare}
\{ c , c^* \} \big].
\label{G14.25}
\end{equation}
Here $c, c^*$ are  Grassmann (anti-commuting) variables
\cite{Popov87,Negele88}.
The action is given by
\begin{equation}
S_i^{\rm bare} =- {\rm tr} c_n^* [ \langle G_{ii}^{}(i \omega_n)\rangle_{\rm av}]^{-1}
c_n + {\rm tr}
c_n^{*} (V_i - \Sigma(i \omega_n)) c_n,
\label{G14.26}
\end{equation}
where $c_n\equiv c(i \omega_n)$, $c^*_n\equiv c^* (i \omega_n)$, and the two terms correspond to the kinetic and potential energy, respectively. We note that  the local potential at site
$\bm{R}_i$ is measured \emph{relative} to the surrounding effective
medium $\Sigma(i \omega_n)$. Since \eqref{G14.26} is the expression for a
non-interacting system (bilinear dependence on $c$ and $c^{*}$)
the functional intergral in \eqref{G14.25} is trivial
to perform and yields, together with the trace-log
formula  ${\rm det} A = \exp ({\rm tr} \ln A)$,
\begin{subequations}
\label{G14.27}
\begin{eqnarray}
\ln {\cal Z}_i^{\rm bare} & = & {\rm tr} \; \ln \left\{
[ \langle G_{ii}^{}(i \omega_n)\rangle_{\rm av}]^{-1}
 - V_i + \Sigma(i \omega_n) \right\}
\label{G14.27a} \\[10pt]
& = & {\rm tr} \ln ({\cal G}^{-1}(i \omega_n) - V_i),
\label{G14.27b}
\end{eqnarray}
where we introduced an {\em effective} local propagator  ${\cal G}$ by
\begin{equation}
{\cal G}^{-1} (i \omega_n)\equiv [ \langle G_{ii}^{}(i \omega_n)\rangle_{\rm av}]^{-1} + \Sigma(i \omega_n).
\label{G14.27c}
\end{equation}
\end{subequations}
The ${\cal G}(i \omega_n)$ propagator describes the coupling between the medium and the site
$\bm{R}_i$.

The averaged free energy, which is a functional of ${\cal G}^{-1} (\Sigma) $,
then follows as
\begin{eqnarray}
\langle \Omega \rangle_{\rm av}  =   - L \; T \; {\rm tr} \;
\Bigl\{ \int dE N(E) \ln \; [ \; i \omega_n  + \mu  - \Sigma (i \omega_n)- E \; ]
\nonumber \\[10pt]
 - \ln ({\cal G}^{-1}(i \omega_n) - \Sigma(i \omega_n) )
 +  \langle \ln  ({\cal G}^{-1}(i \omega_n) -V_i ) \rangle_{\rm av}
\Bigr\}.
\label{G14.28}
\end{eqnarray}
By taking the variational derivative of \eqref{G14.28} w.r.t. $\Sigma$, i.~e.,
using the stationarity condition
\begin{equation}
\frac{\delta \langle \Omega \rangle_{\rm av}}
{\delta {\cal G}^{-1} } = 0
\label{G14.29}
\end{equation}
we obtain
\begin{equation}
\frac{1}{{\cal G}^{-1} (i \omega_n) - \Sigma(i \omega_n)} = \left\langle
\frac{1}{{\cal G}^{-1}(i \omega_n) - V_i} \right\rangle_{\rm av}.
\label{G14.30}
\end{equation}
Together with \eqref{G14.21b} and \eqref{G14.27c} this equation is seen
to be identical to the self-consistent Eq.~\eqref{G14.16} for $\Sigma(i \omega_n)$.
Given a value ${\cal G}^{-1}(i \omega_n)$ we obtain $\Sigma(i \omega_n)$ from
\eqref{G14.30}, which determines a new value ${\cal G}^{-1}(i \omega_n) =
[ G_{ii}^{0} ((i \omega_n) - \Sigma(i \omega_n))]^{-1}  + \Sigma(i \omega_n)$ and so on.
Eq.~\eqref{G14.30} expresses particularly clearly the
single-site aspect of the CPA, as well as the role of $\Sigma(i \omega_n)$ as a homogeneous effective
potential that
describes the effect of the original random medium in the averaged system.

\subsubsection{Generalization of the CPA approach to interacting \hbox{systems}}
\label{sec:cpa_gen}

The CPA was extensively used in the 1970's to  investigate disordered systems. It was
also applied to interacting models, e.~g. the Hubbard model, by first
transforming the model (approximately) to a random alloy problem (``alloy
analogy'')~\cite{Elliott74}.

A new approach to the CPA, which makes use of field-theoretical
functional integral techniques in connection with
explicit diagrammatic perturbation theory,
was initiated by Jani\v{s} \cite{Janis86, Janis89}.
Thereby the range of
applicability of the CPA was extended to  interacting
lattice systems (spin systems and itinerant systems). In this generalized
single-site approach the free-energy functional can be derived in closed form.
The derivation is based on the general scheme for the construction of
\emph{conserving approximations} by Baym \cite{Baym62}, i.~e., the free
energy $\Omega \{ \Sigma \}$, which is a functional of $\Sigma$, is written as
\begin{equation}
\beta \Omega \{ \Sigma\} = \Phi \{ G \} - {\rm tr} (\Sigma G) - {\rm tr} \; \ln
[(G^0)^{-1} - \Sigma ].
\label{G14.31}
\end{equation}
Here $G_{\bm{k} \sigma}^0 (\omega)$ is
the free propagator, \eqref{G11.24}, and $G$ is the full
propagator which is determined by
\begin{subequations}
\label{G14.32}
\begin{equation}
\frac{\partial \Omega}{\partial \Sigma} = 0.
\label{G14.32a}
\end{equation}
The quantity $\Phi \{ G \}$ is obtained from a self-consistent perturbation
expansion in $G$ for the self-energy \cite{Baym62}, the latter being  defined
by
\begin{equation}
\frac{\delta \Phi}{ \delta G} = \Sigma.
\label{G14.32b}
\end{equation}
\end{subequations}
Hence $\Omega$ is a functional of $\Sigma$  only. To define $\Omega$
unambiguously one has to impose the boundary condition $\Phi = 0$ for
$\Sigma = 0$. The construction of
$\Omega \{ \Sigma \}$ amounts to the construction of the functional
\begin{equation}
\Lambda \{ \Sigma, G \} = \Phi \{ G \} - {\rm tr} (\Sigma G).
\label{G14.33}
\end{equation}
In a single-site theory, where the self-energy is purely local, $\Lambda$
 is fully determined by $\Sigma$ and the local part of $G$.
In this case the construction of $e^{-\beta \Lambda}$ reduces to a single-site problem, which
can be performed explicitly \cite{Janis86,Janis89}.

Due to the insight gained from the investigation of the large-$d$ limit
for fermionic lattice systems one can now conclude that
the generalized CPA approach \cite{Janis86,Janis89} becomes  exact
in $d = \infty$ \cite{Janis91}, just as the CPA for disordered systems and other single-site theories discussed
so far become exact in this limit. In particular,
this field-theoretical approach can be used to derive the exact free energy
for fermionic models in $d = \infty$ \cite{Janis91,Janis92b,Janis92a}. This leads to a comprehensive, controlled mean-field theory even for  interacting fermionic models which is conceptually
identical to the CPA for disordered systems or to the Weiss theory for the
Ising model. Of course, this theory is necessarily much more complicated in
detail than the previous mean-field theories since we now deal with a {\em dynamical}
single-site problem in a fermionic bath.

The physical idea behind the approach is the same as that described in the
last subsection in connection with the CPA. Let us
consider the motion of a particle on a lattice in $d = \infty$. The interaction with the other particles
affects the motion. This change is exactly described by a
yet unknown complex, dynamical field $\Sigma_{\sigma} (\omega)$.
Hence the
original system with its bare interactions has been exactly replaced by an
 effective medium; the latter is simply a system of  non-interacting, itinerant
electrons moving in a complex, homogeneous coherent potential $\Sigma_{\sigma} (\omega)$.

\subsubsection{Exact free energy functional for the Hubbard model in $d = \infty$ and the self-consistency equations}

We will now use the generalized CPA described above to construct an
exact expression for the free energy of the Hubbard model in $d = \infty$ \cite{Janis91,Janis92b,Janis92a}.
We proceed as in the case of disordered systems (see Sec.~\ref{sec:cpa_al} and Fig.~\ref{fig4.2}),
with $V_i$ replaced by $\hat{v}_{i \sigma} = U \hat{n}_{i, -\sigma}$, but
we do not have to perform any impurity average now. The
single-site free energy density $\Omega/L$ is given by
\begin{equation}
\frac { \Omega}{L} = \frac { \Omega_{\rm med}}{L} - \Omega_i +
\Omega_i^{\rm bare}.
\label{G14.34}
\end{equation}
Using the analog of \eqref{G14.21b}
\begin{equation}
G_{ii , \sigma}^{} (z) = G_{ii , \sigma}^{0} (z - \Sigma_{\sigma} (z))
\label{G14.35}
\end{equation}
the first two terms are given by (see \eqref{G14.21b}, \eqref{G14.23})
\begin{equation}
\frac { \Omega_{\rm med}}{L} = - T \; {\rm tr} \; \sum_{\bm{k}} \; \ln
\bigl[ G_{\bm{k}, \sigma}^{0} \bigl( i \omega_n - \Sigma_{\sigma} (i \omega_n )
\bigr) \bigr]^{-1}
\label{G14.36}
\end{equation}
\begin{equation}
\Omega_i = - T \; {\rm tr} \; \ln
[G_{ii , \sigma}^0 \bigl( i \omega_n - \Sigma_{\sigma} (i \omega_n)
\bigr) ]^{-1} = - T \; {\rm tr} \; \ln
[G_{ii , \sigma}^{} (i \omega_n)]^{-1}.
\label{G14.37}
\end{equation}
The contribution $\Omega_i^{\rm bare}$, obtained by replacing the medium on
site $\bm{R}_i$ by the actual, bare
interaction (here: the Hubbard interaction)
is again given by \eqref{G14.24} and \eqref{G14.25}, where
 the single-site action is now given by \cite{Janis91,Janis92a}
\begin{eqnarray}
S_i^{\rm bare} & = & -{\rm tr}\, c_{\sigma , n}^* [G_{ii , \sigma}(i \omega_n)]^{-1}
c_{\sigma , n} \nonumber \\[10pt]
& + & \left[ U \int\limits_0^{\beta} \; d \tau c_{\uparrow}^* (\tau)
c_{\uparrow}^{} (\tau) c_{\downarrow}^* (\tau) c_{\downarrow}^{} (\tau)
- {\rm tr} \,c_{\sigma , n}^* \Sigma_{\sigma} (i \omega_n)c_{\sigma , n}^{} \right].
\label{G14.38}
\end{eqnarray}
This expression has the same form as \eqref{G14.26}, but the
spin-dependence of the problem has now been taken into
account explicitly ($c_n \to c_{\sigma , n}$) and the local one-particle
{\em potential} $V_i$ has been replaced by the Hubbard {\em interaction}
between up and
down-spins.

In analogy with \eqref{G14.27c} we now introduce
an effective local propagator ${\cal G}_{\sigma} (i \omega_n)$ by
\begin{equation}
{\cal G}_{\sigma}^{-1} (i \omega_n)\equiv [G_{\sigma}(i \omega_n)]^{-1} + \Sigma_{\sigma}(i \omega_n),
\label{G14.39}
\end{equation}
where $G_{\sigma}(i \omega_n) \equiv G_{ii , \sigma} (i \omega_n)$.
The effective propagator ${\cal G}_{\sigma}(i \omega_n)$ again describes the coupling between
the medium and the interaction-site $\bm{R}_i$. Note, that sites do not communicate
with one another but
only via the effective medium. With \eqref{G14.39} the action \eqref{G14.38}
takes the form
\begin{equation}
S_i^{\rm bare}  \bigl\{ c_{\sigma} , c_{\sigma}^{*} ; {\cal G}_{\sigma}^{-1}
\bigr\} = -{\rm tr} \,c_{\sigma, n}^{*} \; {\cal G}_{\sigma}^{-1}(i \omega_n) c_{\sigma , n}
+ U \int_0^{\beta} d \tau c_{\uparrow}^* (\tau) c_{\uparrow} (\tau)
c_{\downarrow}^* (\tau) c_{\downarrow} (\tau).
\label{G14.40}
\end{equation}

In \eqref{G14.25} the partition function ${\cal Z}_i^{\rm bare}$ is given by an
integral over anticommuting Grassmann variables. It may be transformed into a conventional
functional integral over real, commuting variables by rewriting the Hubbard
interaction in \eqref{G14.40} using the Hubbard-Stratonovich transformation
\begin{eqnarray}
\exp \Biggl\{ - U \int\limits_{0}^{\beta} d \tau
c_{\uparrow}^* (\tau) c_{\uparrow} (\tau)
c_{\downarrow}^* (\tau) c_{\downarrow} (\tau ) \Biggr\}
 =  \int\limits {D} \eta {D} \xi \exp \Biggl\{
- \frac{1}{2 \beta} \int\limits_0^{\beta} d \tau \Biggl[ \eta^2 (\tau) + \xi^2 (\tau)
\nonumber \\[10pt]
 -  i \sqrt{2 U \beta} \bigl\{ \xi (\tau)
[ c_{\uparrow}^* (\tau) c_{\uparrow} (\tau) +
 c_{\downarrow}^* (\tau) c_{\downarrow} (\tau)  ]
 -  i \eta (\tau) [ c_{\uparrow}^* (\tau) c_{\uparrow} (\tau)
- c_{\downarrow}^* (\tau) c_{\downarrow} (\tau)  ] \bigr\} \Biggr] \Biggr\}.
\label{G14.41}
\end{eqnarray}
This is equivalent to the standard operator identity
\begin{equation}
\hat{n}_{i \uparrow} \hat{n}_{i \downarrow} = \frac{1}{4} [ (\hat{n}_{i \uparrow} + \hat{n}_{i \downarrow} )^2 - (\hat{n}_{i \uparrow} - \hat{n}_{i \downarrow})^2 ]
\label{G14.42}
\end{equation}
for the Hubbard interaction, where the two terms on the right-hand side correspond to
charge and spin fluctuations, respectively. In \eqref{G14.41} the fluctuations are
described by real fluctuating fields $\xi (\tau)$ and $\eta (\tau)$, respectively.
Now that the interaction problem has been rewritten in
terms of non-interacting particles in the
presence of infinitely many fluctuating fields, the integration over
the Grassmann variables in the expression for the partition
function can be performed explicitly,
yielding
\begin{equation}
{\cal Z}_i^{\rm bare} = \int {D} \eta {D} \xi \exp \; \big[
 - S_i^{\rm bare} \{ \eta, \xi; {\cal G}_{\sigma}^{-1} \} \big],
\label{G14.43}
\end{equation}
where now
\begin{equation}
\begin{array}{lll}
S_i^{\rm bare} \{ \eta , \xi; {\cal G}_{\sigma}^{-1} \} & = &  \frac{1}{2}
 \sum\limits_{\nu = - \infty}^{\infty} (\xi_{\nu}^2 + \eta_{\nu}^2 ) \\[10pt]
& - & {\rm tr} \ln \bigl[ \hat{{\cal G}}_{\sigma}^{-1} - \sqrt{\frac{U}{2 \beta}}
( \sigma \hat{\eta} + i \hat{\xi} ) \bigr]
\end{array}
\label{G14.44}
\end{equation}
with $(\hat{\xi})_{mn} = \xi_{m-n}, (\hat{\eta})_{mn} = \eta_{m-n}$
and
$(\hat{\cal G}_{\sigma}^{-1} )_{mn} = \delta_{mn} [ {\cal G}_{\sigma}
 (i \omega_n)]^{-1}$.

\begin{figure}
\includegraphics[width=1.0\textwidth]{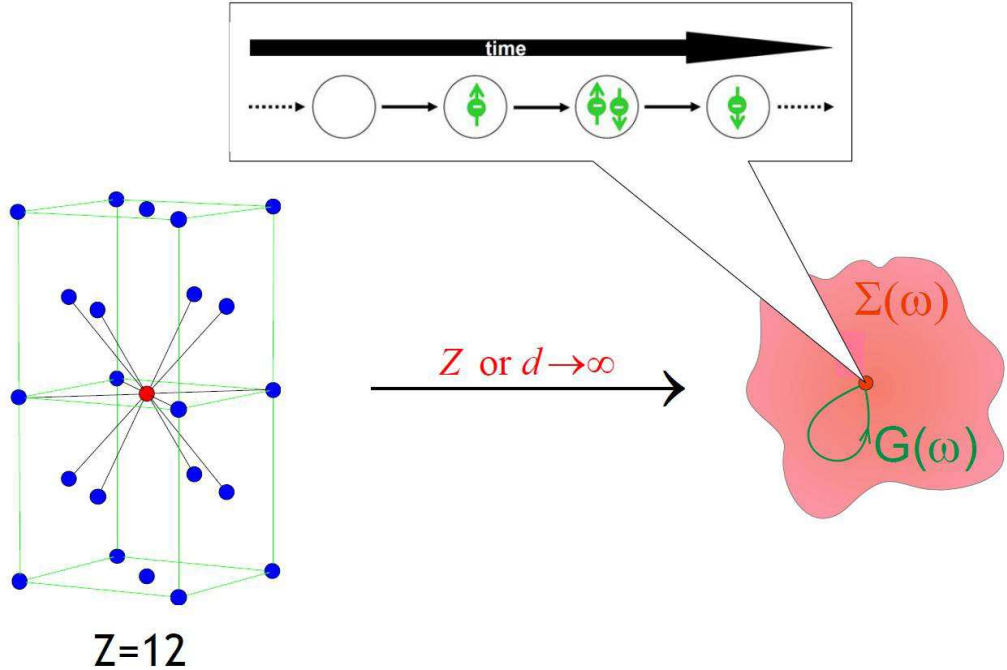}
\caption{In the limit $Z\rightarrow \infty $ the Hubbard model effectively reduces to a dynamical single-site problem, which may be viewed as a lattice site embedded in a dynamical mean field. Electrons may hop from the mean field onto this site and back, and interact on the site as in the original Hubbard model (see Fig. \ref{Hubbard_model}).
The local propagator $G(\omega)$ (i.e., the return amplitude) and the dynamical self-energy  $\Sigma (\omega)$ of the surrounding mean field play the main role in this limit. The quantum dynamics of the interacting electrons is still described exactly.
}
\label{DMFT}
\end{figure}

Making use of the relation \eqref{G14.20} and combining
the three contributions in \eqref{G14.27}, the total free energy is then found as
\begin{equation}
\hspace*{-12pt} \Omega = - L \; T \; {\rm tr} \; \Biggl\{ \int d \omega N(\omega) \ln
[i \omega_n + \mu - \Sigma_{\sigma} - \omega ] - \ln ({\cal G}_{\sigma}^{-1}
- \Sigma_{\sigma}) \Biggr\} - L \; T \; \ln {\cal Z}_i^{\rm bare}.
\label{G14.45}
\end{equation}
While $\Omega$ was originally a functional of $\Sigma_{\sigma}$
it is now understood as a functional of
$[ {\cal G}_{\sigma} (\Sigma_{\sigma})]^{-1}$.
The stationarity condition \eqref{G14.29} is then
\begin{equation}
\frac{\delta \Omega}{\delta {\cal G}_{\sigma}^{-1}} = 0,
\label{G14.46}
\end{equation}
which leads to
\begin{subequations}
\label{G14.47}
\begin{eqnarray}
\hspace*{-1cm}\frac{1}{[ {\cal G}_{\sigma}(i \omega_n )]^{-1}
-  \Sigma_{\sigma} (i \omega_n) }
& = & \frac{1}{{\cal Z}_i^{\rm bare}} \int
{D} c {D} c^*
(c_{\sigma , n} c_{\sigma , n}^* ) \exp \big[ -S_i^{\rm bare}
\{ c_{\sigma} , c_{\sigma}^* ; {\cal G}_{\sigma}^{-1} \} \big]
\label{G14.47a}
\\[10pt]
& = & \frac{1}{{\cal Z}_i^{\rm bare}} \int {D}\eta {D} \xi
\Biggl( \frac{1}{\hat{\cal G}_{\sigma}^{-1} - \sqrt{\frac{U}{2 \beta} }
(\sigma \hat{\eta} + i \hat{\xi} ) } \Biggr)_{nn} \nonumber \\[10pt]
& \times & \exp \big[ -S_i^{\rm bare}
\{ \eta , \xi ; {\cal G}_{\sigma}^{-1} \} \big]
\label{G14.47b}
\\[10pt]
& \equiv &  \Bigg\langle \Biggl( \frac{1}{\hat{{\cal G}}_{\sigma}^{-1} -
\sqrt{\frac{U}{2 \beta}} ( \sigma \hat{\eta} + i \hat{\xi}) } \Biggr)_{nn}
\Bigg\rangle_{\eta, \xi}
\label{G14.47c}
\end{eqnarray}
\end{subequations}%
with $\langle X \rangle_{\eta , \xi} = ( \int {D} \eta {D} \xi
X \exp [-S_i^{\rm bare}])/{\cal Z}_i^{\rm bare}$. The right-hand side of \eqref{G14.47a}  is nothing but the very definition
of the local propagator $G_{ii , \sigma} (i \omega_n)\equiv G_{\sigma} (i \omega_n)$ in terms of the action $S_i^{\rm bare}$.
Eqs.~\eqref{G14.47}, together with
\begin{equation}
[ {\cal G}_{\sigma} (z)]^{-1}  =
[G_{\sigma}^0 (z - \Sigma_{\sigma}(z) )]^{-1} + \Sigma_{\sigma}(z)
\label{G14.48}
\end{equation}
and the local action \eqref{G14.40} provide an exact, self-consistent set of equations for
$\Sigma_{\sigma}$ (or ${\cal G}_{\sigma}$) for the Hubbard model
in $d = \infty$ \cite{Janis91,Georges92,Janis92b}. In the paramagnetic phase  two quantities are seen to play the most important role: the local propagator $G_{\sigma} (\omega)$ and the self-energy $\Sigma (\omega)$; this is illustrated in Fig.~\ref{DMFT}.

Although the interaction problem is
mathematically much more complicated
 than the analogous expression \eqref{G11.30} for the disorder problem
without interactions it can, in principle, again be solved by iteration:
for given ${\cal G}_{\sigma}^{-1}$ we obtain $\Sigma_{\sigma}$ from \eqref{G14.47},
which yields a new ${\cal G}_{\sigma}^{-1}$ via \eqref{G14.48}, etc. The exact
local propagator is then provided by
$G_{ii , \sigma} = ({\cal G}_{\sigma}^{-1} - \Sigma_{\sigma} )^{-1}$.
As in the disorder case $G_{ii , \sigma}$ is completely expressed in terms of
 effective, averaged quantities.

The expression used in \eqref{G14.47c} brings out particularly clearly the
similarities and differences between the (on-site) interaction problem
and the analogous expression \eqref{G14.30} for the disorder case without interactions:

\noindent (i) on the left-hand side of \eqref{G14.47} the self-energy $\Sigma $ appears again as a
(homogeneous) effective medium, which is  obtained  exactly from the original
system by some averaging process;

\noindent (ii) however, this average is very different in the
two cases: in the disorder problem it involves an integration over the actual
disorder potentials $V_i$ with a  given disorder distribution
$P(V_i)$, while in \eqref{G14.47} it demands an integration over infinitely
many fluctuating (random) fields $\eta, \xi$, which simulate the actual interaction;

\noindent (iii) the latter integration leads to a highly non-trivial coupling
of the energies, i.~e., Matsubara frequencies $\omega_n$ (note, that this coupling
even exists in the  static limit, i.~e.,
for $\eta_{\nu} = \xi_{\nu} = 0$ for $\nu \neq 0$), while in the
disorder problem the corresponding Eq.~\eqref{G14.30} is diagonal
in the frequency. This shows clearly that,
although the interaction between electrons on different lattice sites has been reduced to an interaction of electrons with a
mean field, the \emph{dynamics} of the latter interaction is still
non-trivial. Once more we observe that the many-body
nature of the Hubbard model survives even in $d = \infty$, making an analytic
evaluation of the  local propagator $G_{\sigma}$ from
\eqref{G14.47}, \eqref{G14.48} generally
impossible.

\subsection{Construction of the DMFT as a self-consistent single-impurity Anderson model}
\label{sec:cavity}

\begin{figure}
\includegraphics[width=0.8\textwidth]{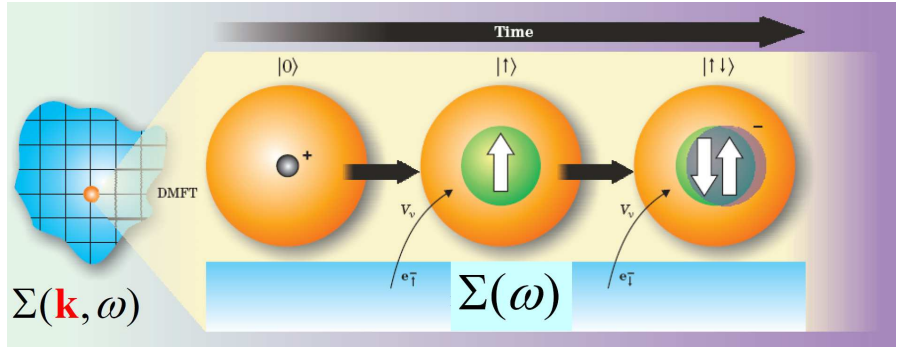}
\caption{The DMFT can be viewed as the mapping of the $d$-dimensional Hubbard model with self-energy $ \Sigma(\bm{k},\omega)$ onto a single site described by a single-impurity Anderson model which hybridizes with the bath provided by the other electrons \cite{georges96}. In this local problem the  self-energy $ \Sigma(\omega)$ is now $\bm{k}$ independent; after Ref. \cite{dmft_phys_today}.}
\label{DMFT_logo}
\end{figure}

Following the presentation of Georges, Kotliar, Krauth and Rozenberg \cite{georges96} the dynamical mean-field  equations will now be derived using the so-called \emph{cavity} method. This derivation starts by removing one lattice site together  with its bonds from the rest of the lattice. The remaining lattice, which now contains a cavity, is replaced by a particle bath which plays the role of the dynamical mean field (see Fig.~\ref{DMFT_logo}). So far the derivation and the underlying physical picture  coincides with that of the CPA approach described in the previous section. Now comes a new, physically motivated idea: the bath is coupled, via a \emph{hybridization}, to the cavity. The resulting problem then amounts to the solution of an effective single-impurity Anderson model where the degrees of freedom of the bath, represented by an appropriate hybridization function, have to be determined self-consistently.

To be specific, let us consider the partition function in the grand canonical ensemble
\begin{equation}
 {\mathcal Z} = \int \prod_{i \, \sigma} D c^{*} _{i \sigma} D c_{i \sigma} \exp[-S\{c^*_{i\sigma},c_{i\sigma}\}].
\end{equation}
The action $S\{c^*_{i\sigma},c_{i\sigma}\}$ for the Hubbard model is given by
\begin{eqnarray}
S\{c^*_{i\sigma},c_{i\sigma}\}=\int_0^{\beta} d \tau \left[ \sum_{i\sigma} c^{*}_{i \sigma} (\tau) (\frac{\partial}{\partial \tau} - \mu) c_{i \sigma} (\tau) + \sum_{ij \,\sigma} t_{ij} c^{*}_{i \sigma}(\tau)  c_{j \sigma} (\tau)  \right. \nonumber \\
\left. + \;\sum_{i}U c^{*}_{i\uparrow}(\tau)c_{i\uparrow}(\tau)c^{*}_{i\downarrow}(\tau)c_{i\downarrow}(\tau) \right ],
\end{eqnarray}
where again as in Sec.~\ref{sec:cpa_al}  we use Grassman variables $c^{*} _{i \sigma}$, $c _{i \sigma}$.
We split the action $S$ into three parts
\begin{equation}
S=S_{0}+\Delta S + S^{(0)},
\end{equation}
where $S_{0}$ is the part containing only variables on site $0$
\begin{equation}
S_0=\int_0^{\beta} d \tau \left[ \sum_{\sigma} c^{*}_{0 \sigma} (\tau) (\frac{\partial}{\partial \tau} - \mu) c_{0 \sigma} (\tau) + U c^{*}_{0\uparrow}(\tau)c_{0\uparrow}(\tau)c^{*}_{0\downarrow}(\tau)c_{0\downarrow}(\tau) \right ],
\end{equation}
$\Delta S$ contains the hoppings between site $0$ and other sites of the lattice $i\neq 0$
\begin{equation}
\Delta S=\int_0^{\beta} d \tau  \sum_{i \,\sigma} \left[ t_{i0} c^{*}_{i \sigma}(\tau)  c_{0 \sigma} (\tau) + t_{0i} c^{*}_{0 \sigma}(\tau)  c_{i \sigma} (\tau)\right ],
\end{equation}
and the rest, denoted by $S^{(0)}$, is the part of the action where the site $0$ and its bonds are removed, i.e., with $i,j\neq 0$
\begin{eqnarray}
S^{(0)}=\int_0^{\beta} d \tau \left[ \sum_{i\neq 0\,\sigma} c^{*}_{i \sigma} (\tau) (\frac{\partial}{\partial \tau} - \mu) c_{i \sigma} (\tau) + \sum_{ij \neq 0 \,\sigma} t_{ij} c^{*}_{i \sigma}(\tau)  c_{j \sigma} (\tau)  \right. \nonumber \\
\left.+ \; U \sum_{i\neq 0}
c^{*}_{i\uparrow}(\tau)c_{i\uparrow}(\tau)c^{*}_{i\downarrow}(\tau)c_{i\downarrow}(\tau) \right ].
\end{eqnarray}
Let us now rewrite the partition function $ {\mathcal Z}$ as
\begin{eqnarray}
 {\mathcal Z} = \int \prod_{ \sigma }D c^{*}_{0\sigma} D c_{0\sigma} \exp[-S_{0}\{ c^{*}_{0\sigma}, c_{0\sigma}\}]  \nonumber \\ \times\int \prod_{i\neq0\, \sigma } D c^{*}_{i\sigma} D c_{i\sigma} \exp[-S^{(0)}\{ c^{*}_{i\sigma}, c_{i\sigma}\}] \exp[-\Delta S \{ c^{*}_{0\sigma}, c_{0\sigma}, c^{*}_{i\sigma}, c_{i\sigma}\}]
\end{eqnarray}
and use the following ensemble average
\begin{equation}
\langle X\rangle_{(0)} \equiv \frac{1}{ {\mathcal Z}^{(0)}} \int \prod_{i\neq0\,\sigma } D c^{*}_{i\sigma} D c_{i\sigma} X\exp [-S^{(0)}\{ c^{*}_{i\sigma}, c_{i\sigma}\}]
\end{equation}
taken with respect to $S^{(0)}$ (the action where the site $i=0$ is excluded), with $ {\mathcal Z}^{(0)}$ being the corresponding partition function. Then the partition function reads
\begin{eqnarray}
 {\mathcal Z} =  {\mathcal Z}^{(0)}\int \prod_{\sigma }D c^{*}_{0\sigma} D c_{0\sigma} \exp[-S_{0}\{ c^{*}_{0\sigma}, c_{0\sigma}\}] \nonumber \\
 \times \; \langle \exp[-\Delta S \{ c^{*}_{0\sigma}, c_{0\sigma}, c^{*}_{i\sigma}, c_{i\sigma}\}]\rangle_{(0)}.
\end{eqnarray}
In the next step we expand the second exponent with respect to the
action $\Delta S$. As a result we obtain a formally infinite
series with all possible many-particle correlation functions, i.e.,
\begin{eqnarray}
 {\mathcal Z}=  {\mathcal Z}^{(0)}\int \prod_{\sigma }D c^{*}_{0\sigma} D c_{0\sigma} \exp[-S_{0}\{ c^{*}_{0\sigma}, c_{0\sigma}\}]  \left(1 -
\int_0^{\beta}d\tau \langle \Delta S(\tau)\rangle_{{(0)}}  \right. \nonumber \\
\left. + \;  \frac{1}{2!}
\int_0^{\beta}d\tau_1 \int_0^{\beta}d\tau_2 \langle \Delta S(\tau_1)\Delta
S(\tau_2) \rangle_{{(0)}}+ \cdot \cdot \cdot \right),
\end{eqnarray}
where we used $\Delta S \equiv \int_0^{\beta}d\tau  \Delta S(\tau)$. In the fermionic case only the correlation functions with equal number of $c$ and $c^{*}$ are non-zero. The lowest order term is second order and reads
\begin{eqnarray}
\frac{1}{2!}
\int_0^{\beta}d\tau_1 \int_0^{\beta}d\tau_2 \langle \Delta S(\tau_1)\Delta
S(\tau_2) \rangle_{{(0)}}  \nonumber \\
=\frac{1}{2!} \int_0^{\beta}d\tau_1 \int_0^{\beta}d\tau_2 {\sum_{\sigma}}{\sum_{j,k\neq0}}
\left[  \right. t_{j0}
t_{0k} \langle c^*_{j\sigma}(\tau_1) c_{k\sigma}(\tau_2) \rangle_{{(0)}} c_{0\sigma}(\tau_1) c^*_{0\sigma}(\tau_2)
\nonumber \\
+\; t_{0j}  t_{k0} \langle c_{j\sigma}(\tau_1) c^*_{k\sigma}(\tau_2)
\rangle_{{(0)}} c_{0\sigma}^*(\tau_1) c_{0\sigma}(\tau_2)  \left. \right] .  \label{Z_2}
\end{eqnarray}
Higher-order terms are obtained similarly. The above expression can be rewritten with the use of one-particle correlation function
\begin{eqnarray}
G_{jk\sigma}^{(0)}(\tau_1-\tau_2)= - \langle T_{\tau}
c_{j\sigma}(\tau_1)c_{k\sigma}^*(\tau_2)\rangle_{{(0)}}
\end{eqnarray}
and takes the form
\begin{eqnarray}
\frac{1}{2!}
\int_0^{\beta}d\tau_1 \int_0^{\beta}d\tau_2 \langle \Delta S(\tau_1)\Delta
S(\tau_2) \rangle_{{(0)}}\nonumber \\
 = - \int_0^{\beta}d\tau_1 \int_0^{\beta}d\tau_2 {\sum_{\sigma}}{\sum_{j,k\neq0}}
 t_{j0} t_{k0} G_{jk\sigma}^{(0)}(\tau_1-\tau_2)
c^*_{0\sigma}(\tau_1) c_{0\sigma}(\tau_2).
 \label{Z_2a}
\end{eqnarray}
Higher-order terms can be written in a similar way with the use of n-particle correlation functions.

A non-trivial limit $d \rightarrow \infty$ is obtained by scaling
the hopping amplitudes $t_{ij}$ as described
in Sec.~\ref{sec:scaling}. For example, in the
second-order contribution to the partition function, \eqref{Z_2a},
the hopping amplitudes must be scaled with $Z^{||\bm{R}_{0}-\bm{R}_{j}||/2}$ because
the one-particle correlation functions are  proportional to
$1/Z^{||\bm{R}_{0}-\bm{R}_{j}||/2}$ as discussed in Sec~\ref{sec:scaling}. In the calculation of higher-order
terms we find that all connected higher-order
terms vanish at least as $\mathcal{O}(1/Z)$. Consequently, in the
$Z\rightarrow \infty$ limit only the contribution
 $G_{jk\sigma}^{(0)}$, or disconnected
contributions made of products of  $G_{jk\sigma}^{(0)}$'s remain. Applying the linked-cluster theorem and collecting only connected contributions in the
exponential function one obtains the local action
\begin{eqnarray}
S_{\mathrm{loc}}=\left[\int_0^{\beta} d \tau  \sum_{\sigma} c^{*}_{0 \sigma} (\tau) (\frac{\partial}{\partial \tau} - \mu) c_{0 \sigma} (\tau) + U\int_0^{\beta} d \tau  c^{*}_{0\uparrow}(\tau)c_{0\uparrow}(\tau)c^{*}_{0\downarrow}(\tau)c_{0\downarrow}(\tau)\right.  \nonumber \\
\left. +  \int_0^{\beta}d\tau_1 \int_0^{\beta}d\tau_2 {\sum_{\sigma}}{\sum_{j,k\neq0}}
t^{*}_{j0} t^{*}_{k0} G_{jk\sigma}^{(0)}(\tau_1-\tau_2)
c^*_{0\sigma}(\tau_1) c_{0\sigma}(\tau_2)
 \right],
\end{eqnarray}
where the rescaled hoppings are denoted with a star.
Introducing the hybridization function
\begin{equation}
\Delta_{\sigma}(\tau_1-\tau_2)= - {\sum_{i,j\neq0}}t^{*}_{i0}t^{*}_{j0}
{G}_{ij\sigma}^{(0)}(\tau_1-\tau_2),
\label{hybridization_function}
\end{equation}
and employing the free (``Weiss'') mean-field propagator $\mathcal{G}_{\sigma}$
one can express the DMFT local action in the following form (here
the site index $i=0$ is omitted for simplicity)
\begin{eqnarray}
S_{\mathrm{loc}}=- \int_0^{\beta} d \tau_1  \int_0^{\beta} d \tau_2 \sum_{\sigma} c^{*}_{\sigma} (\tau_1)  \mathcal{G}^{-1}_{\sigma}(\tau_1-\tau_2) c _{\sigma}(\tau_2)   \nonumber \\
 + \;  U\int_0^{\beta} d \tau  c^{*}_{\uparrow}(\tau)c_{\uparrow}(\tau)c^{*}_{\downarrow}(\tau)c_{\downarrow}(\tau),
\label{local_action}
\end{eqnarray}
where
\begin{equation}
\mathcal{G}^{-1}_{\sigma}(\tau_1-\tau_2)=-\left( \frac{\partial}{\partial\tau_1}-\mu \right) \delta_{\tau_1\tau_2}-\Delta_{\sigma}(\tau_1-\tau_2).
\label{hybridization_function2}
\end{equation}

Finally, we need the relation between the Green
function ${G}^{(0) }_{ij\sigma}(\tau-\tau^{\prime})$ where the
site $i=0$ is removed and the full lattice Green function, i.e.,
\begin{equation}
{G}^{(0) }_{ij\sigma}={G}_{ij\sigma}-{G}_{i0\sigma}
{G}_{00\sigma}^{-1} {G}_{0j\sigma},
\label{lattice_local}
\end{equation}
which holds for a general lattice.

In order to obtain the full solution of the lattice problem it is convenient to express the relation between the local Green function ${G}_{00\sigma}\equiv{G}_{\sigma}$ and the dynamical (``Weiss'') mean field $\mathcal{G}^{-1}_{\sigma}$ in the form of a Dyson equation\footnote{It should be noted that, in principle, any one of the local functions $\Sigma_{\sigma}(i\omega_n)$, $\mathcal{G}^{-1}_{\sigma}(i\omega_n)$, or $\Delta_{\sigma}(i\omega_n)$ can be viewed as a ``dynamical mean field'' acting on particles on a site, since they all appear in the bilinear term of the local action (\ref{local_action}).}
\begin{eqnarray}
 [G_{\sigma}(i\omega_n)]^{-1}=\mathcal{G}^{-1}_{\sigma}(i\omega_n)- \Sigma_{\sigma }(i\omega_n)=i\omega_n+\mu-\Delta_{\sigma}(i\omega_n)- \Sigma_{\sigma }(i\omega_n).
\label{Dyson_eq}
\end{eqnarray}
Then the  lattice Green function (in ${\bm{k}}$-space) $G_{\bm{k}\,\sigma}(i\omega_n)$ is given by
\begin{eqnarray}
G_{\bm{k}\,\sigma}(i\omega_n)=\frac{1}{i\omega_n
-\epsilon_{\bm{k}}
+\mu  -\Sigma_{\sigma}(i\omega_n)}.  \label{dmft2}
\end{eqnarray}
After performing the so-called lattice Hilbert transform we recover the local Green function
\begin{eqnarray}
G_{\sigma}(i\omega_n)= \sum_{\bm{k}} G_{\bm{k}\,\sigma}(i\omega_n) = \sum_{\bm{k}} \frac{1}{i\omega_n
-\epsilon_{\bm{k}}
+\mu  -\Sigma_{\sigma}(i\omega_n)}.  \label{dmft3}
\label{hilbert_transform}
\end{eqnarray}
After analytic continuation to real frequencies the local (``$\bm{k}$ averaged'') propagator reads
\begin{eqnarray}
G_{\bm{k}\,\sigma}(\omega)=\frac{1}{\omega
-\epsilon_{\bm{k}}
+\mu  -\Sigma_{\sigma}(\omega)}.
\label{k-dependent-Prop}
\end{eqnarray}

It is very important to realize that although the DMFT corresponds to an effectively local problem, the propagator $G_{\bm{k}}(\omega)$ is a \emph{momentum-dependent} quantity. Namely, it depends on the momentum through the dispersion $\epsilon_{\bm{k}}$ of the non-interacting electrons, but there is no \emph{additional} momentum-dependence through the self-energy, since this quantity is strictly local within the DMFT.

The set of self-consistent equations \eqref{local_action}, \eqref{hybridization_function2}, \eqref{Dyson_eq}, \eqref{hilbert_transform} can be solved iteratively. In each step one solves the single-impurity problem given by the action ~\eqref{local_action}, then one finds the new self-energy from the Dyson equation~\eqref{Dyson_eq} and the new dynamical mean field from \eqref{hilbert_transform} and \eqref{Dyson_eq}. The single-impurity problem is still a complicated many-body interacting problem which cannot, in general, be solved exactly.

\subsection{Solution of the DMFT self-consistency equations
}


The dynamics of the full Hubbard model, \eqref{G11.7}, was found to remain complicated even in the
limit $d \to \infty$ because of the purely local nature
of the interaction. Hence an exact,
analytic evaluation of the self-consistent set of equations \eqref{G14.47},
\eqref{G14.48} for the local propagator $G_{\sigma}$ or the
effective propagator ${\cal G}_{\sigma} (i \omega_n)$ is
not possible.
Exact evaluations
are only feasible when  there is no coupling between the frequencies. This is the case, for example,
in the
Falicov-Kimball model \cite{fk_dmft},  which was solved analytically by Brandt and Mielsch \cite{Brandt} soon after the introduction of the $d \to \infty$ limit \cite{metzner89}.
A valuable semi-analytic approximation is provided by the so-called \emph{iterated perturbation theory}  (IPT) \cite{Georges92}, \cite{IPT}, \cite{georges96}.

Solutions of the general DMFT self-consistency equations require extensive numerical methods, in particular quantum Monte Carlo techniques \cite{Jarrell92},  \cite{Rozenberg92,Georges92b},  \cite{georges96}, the numerical renormalization group \cite{Bulla,NRG-RMP}, exact diagonalization \cite{Caffarel94,Si94,Rozenberg94a}, \cite{georges96}
and other techniques, whose discussion requires a separate series of lectures; here I refer the reader to the reviews quoted above.

It quickly turned out that the DMFT is a powerful tool for the investigation of electronic systems with strong correlations. It provides a non-perturbative and thermodynamically consistent approximation scheme for finite-dimensional systems which is particularly valuable for the study  of intermediate-coupling problems where perturbative techniques fail
\cite{freericks95}, \cite{georges96,Georges2003}, \cite{dmft_phys_today}, \cite{Maier-cluster,Jarrell-Salerno}.

In the remaining part of these lecture notes I shall discuss several applications of the DMFT to problems involving electronic correlations. In particular, I will address the Mott-Hubbard metal-insulator transition, and explain the connection of the DMFT with band-structure methods --- the LDA+DMFT scheme --- which is the first comprehensive framework for the \emph{ab initio} investigation of correlated electron materials.

\section{The Mott-Hubbard Metal-Insulator Transition}
\label{sec:mit}


The correlation induced transition between a paramagnetic metal and a
paramagnetic insulator, referred to as ``Mott-Hubbard metal-insulator
transition (MIT)'',  is  one of the most intriguing phenomena in
condensed matter physics~\cite{mott68,mott90,Gebhard}.
This transition is
a consequence of the competition
between the kinetic energy of the electrons and their local interaction $U$. Namely, the kinetic energy prefers the electrons to move (a wave effect) which leads to doubly occupied sites and thereby to interactions between the electrons (a particle effect). For large values of $U$ the doubly occupied sites become energetically very costly. The system may reduce its total energy by localizing the electrons. Hence the
Mott transition is a localization-delocalization transition, demonstrating the particle-wave duality of electrons.

Mott-Hubbard MITs
are, for example, found in transition metal oxides with
partially filled bands near the Fermi level. For such systems band theory
typically predicts metallic behavior. The most famous example is
V$_{2}$O$_{3}$ doped with Cr~\cite{McWhR,McWhetal,RMcWh}. In particular,
in (V$_{0.96}$Cr$_{0.04}$)$_{2}$O$_{3}$ the metal-insulator transition is of
first order below $T=380$\,K \cite{McWhetal}, with
discontinuities in the lattice parameters  and in the conductivity.
However, the two phases
remain isostructural.

Making use of the half-filled, single-band Hubbard model \eqref{G11.7}
  the Mott-Hubbard MIT was studied
intensively in the past \cite{HubbardI,RMcWh,mott68,mott90,Gebhard}. Important early results were obtained by Hubbard~\cite{Hub3} within a Green function decoupling scheme,
and by Brinkman and Rice~\cite{BR} who employed the Gutzwiller variational
method~\cite{Gutzwiller,Gutzwiller65}, both at zero temperature\footnote{The Gutzwiller variational method \cite{Gutzwiller,Gutzwiller65} consists of the choice of a simple projected variational wave function (``Gutzwiller wave function'') and a semi-classical evaluation of expectation values  in terms of this wave function (``Gutzwiller approximation''). As shown by Metzner and Vollhardt \cite{metzner88}, \cite{Jerusalem} the Gutzwiller approximation becomes exact in the limit $d\rightarrow \infty$. This result initiated the investigations of the Hubbard model in $d\rightarrow \infty$ \cite{metzner89}}. Hubbard's approach yields a
continuous splitting of the band into a lower and upper Hubbard band, but
cannot describe quasiparticle features. By contrast, the
Gutzwiller-Brinkman-Rice approach (for a review see Ref. \cite{Vollhardt84}) gives a good description of the low-energy, quasiparticle behavior, but cannot reproduce the upper and lower Hubbard bands. In the latter approach the MIT is signalled by the disappearance of the quasiparticle peak.

To solve this problem the DMFT has been extremely valuable since it provided detailed insights into the nature of the Mott-Hubbard MIT for all values of the interaction $U$ and temperature $T$ \cite{georges96,Bluemer-Diss}, \cite{dmft_phys_today}.

\subsection{DMFT and the three-peak structure of the spectral function}

\begin{figure}
\includegraphics[width=0.5\textwidth]{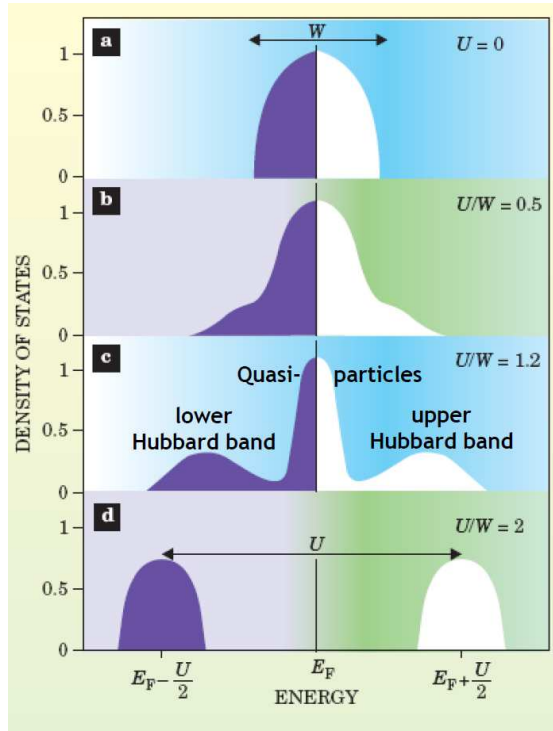}
\caption{Evolution of the spectral function (``density of states'') of the Hubbard model in the paramagnetic phase at half filling. a) non-interacting case, b) for weak interactions there is only little transfer of spectral weight away from the Fermi energy, c) for strong interactions a typical three-peak structure consisting of coherent quasiparticle excitations close to the Fermi energy and  incoherent lower and upper Hubbard bands  is clearly seen, d) above a critical interaction the quasiparticle peak vanishes and the system is insulating, with two well-separated Hubbard bands remaining; after Ref. \cite{dmft_phys_today}.}
\label{3_peak}
\end{figure}

\begin{figure}
\includegraphics[width=0.6\textwidth]{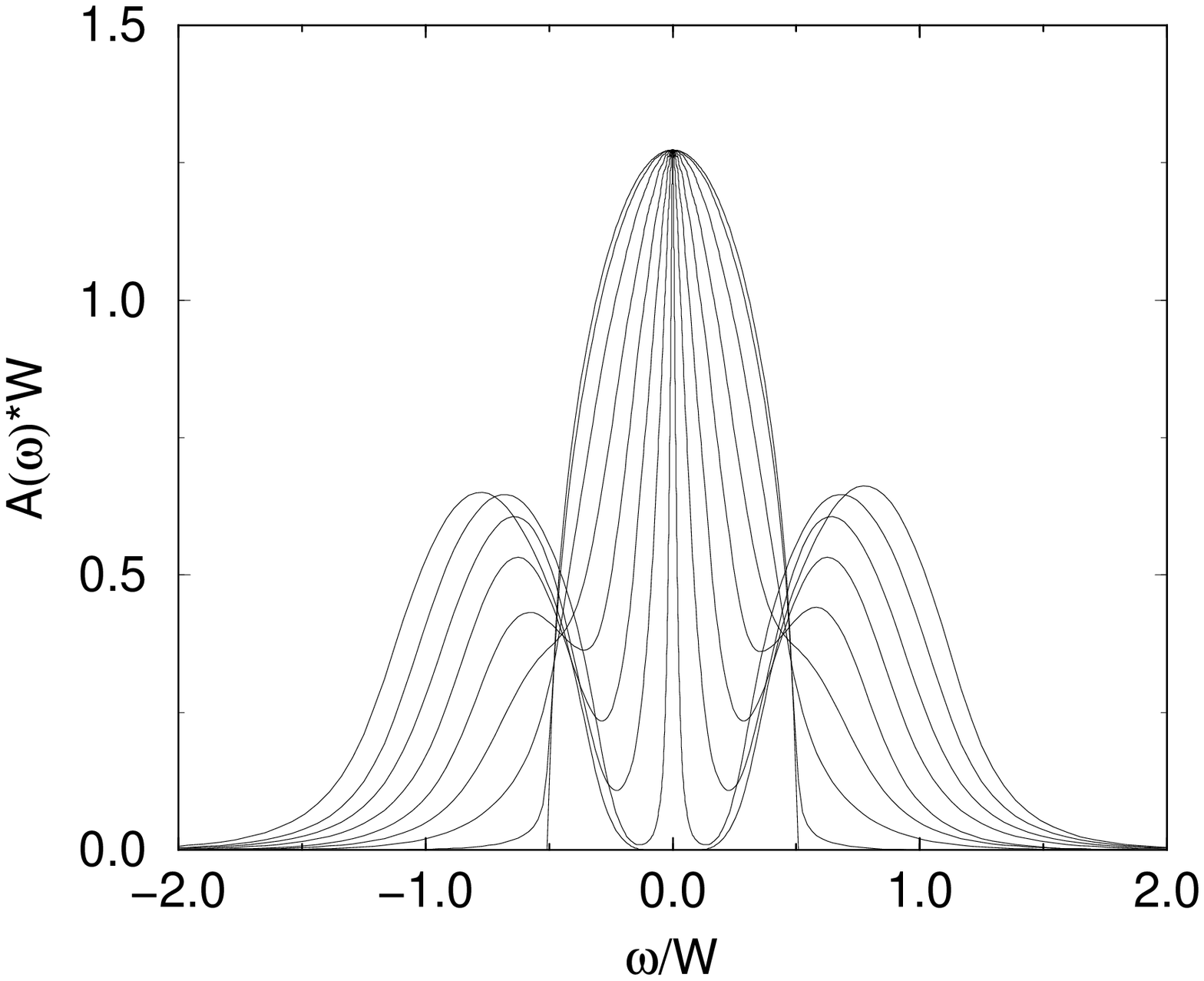}
\caption{Evolution of the $T=0$ spectral function of the
one-band Hubbard model with a semi-elliptic (``Bethe'') DOS for
interaction values $U/W=0,0.2,0.4,\ldots,1.6$ ($W$: band width)
calculated with the numerical renormalization group.
At the critical interaction $U_{\rm c2}/W\simeq1.47$ the metallic
solution disappears and the Mott gap opens; from Ref.~\cite{Vollhardt05}.}
\label{DOS_pinning}
\end{figure}

The Mott-Hubbard MIT is monitored by the spectral function $A(\omega)= - \frac{1}{\pi} {\rm Im} G(\omega + i0^+)$ of the correlated electrons\footnote{In the following we only consider the paramagnetic phase.}; here we follow the discussion of Refs.~\cite{Bulla01}, \cite{dmft_phys_today}, \cite{Vollhardt05}. The change of  $A(\omega)$ obtained within the DMFT for the one-band Hubbard model~\eqref{G11.7}
at $T=0$ and half filling $(n=1)$ as a function of the Coulomb repulsion $U$ (measured in units of the bandwidth $W$ of
non-interacting electrons) is shown in Figs.~\ref{3_peak} and~\ref{DOS_pinning}. While  Fig.~\ref{3_peak} is a schematic picture of the evolution of the spectrum when the interaction is increased, Fig.~\ref{DOS_pinning} shows actual numerical results obtained by the NRG~\cite{Bulla,Vollhardt05}.
Here magnetic order is assumed to
be suppressed (``frustrated'').

While at small $U$ the system can be described by coherent quasiparticles
whose DOS still resembles that of the free electrons, the spectrum in the Mott
insulator state  consists of two separate incoherent ``Hubbard
bands'' whose centers are separated approximately by the energy
$U$. The latter originate from atomic-like excitations at the
energies $\pm U/2$ broadened by the hopping of electrons away from the
atom. At intermediate values of $U$ the spectrum then has a
characteristic three-peak structure as in the single-impurity
Anderson model, which includes both the atomic features (i.e.,
Hubbard bands) and the narrow quasiparticle peak at low
excitation energies, near $\omega=0$. This
corresponds to a strongly correlated metal. The structure of the
spectrum (lower Hubbard band, quasiparticle peak, upper Hubbard
band) is quite insensitive to the specific form of the DOS of the
non-interacting electrons.

The width of the quasiparticle peak vanishes for $U\to U_{\rm c2}(T)$. The ``Luttinger
pinning'' at $\omega=0$ \cite{MH89b} is clearly observed.
On decreasing $U$, the transition from the insulator to the
metal occurs at a lower critical value $U_{\rm c1}$, where the gap
vanishes.

It is important to note that the three-peak spectrum
originates from a lattice model with only
\emph{one} type of electrons. This is in contrast to the
single--impurity Anderson model whose spectrum shows very similar
features, but is due to \emph{two} types of electrons, namely the
localized orbital at the impurity site and the free conduction
band. Therefore the screening of the magnetic moment which gives
rise to the Kondo effect in impurity systems has a different
origin in lattice systems. Namely, as explained by the DMFT, the same
electrons provide both the local moments and the electrons
which screen these moments \cite{georges96}.

\begin{figure}[t]
\includegraphics[width=0.7\textwidth]{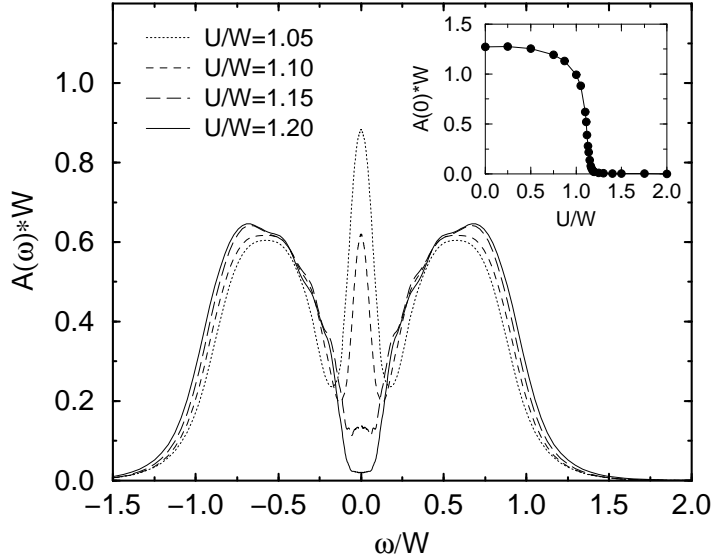}
\caption{Spectral function for the half-filled Hubbard model for
various values of $U$ at $T=0.0276~W$ in the crossover
region. The crossover from the metal to the insulator occurs via
a gradual suppression of the quasiparticle peak at $\omega\!=\!0$.
The inset shows the $U$ dependence of $A(\omega\! =\!0)$, in
particular the rapid decrease for $U\approx 1.1~W$; from Ref.~\cite{Bulla01}.} \label{fig:4.1}
\end{figure}

The evolution of the spectral function of the half-filled
frustrated Hubbard model at finite temperatures,
$T=0.0276~W$, is shown in Fig.~\ref{fig:4.1}. This temperature is
above the temperature of the critical point so that there is no
real transition but only a crossover from a metallic-like to an
insulating-like solution.
The height of the quasiparticle peak at the Fermi energy is no
longer fixed at its zero temperature value. This is due to a
finite value of the imaginary part of the self--energy.
The spectral weight of the quasiparticle peak is seen to be
gradually redistributed and shifted to the upper (lower) edge of
the lower (upper) Hubbard band. The inset of Fig.~\ref{fig:4.1}
shows the $U$-dependence of the value of the spectral function at
zero frequency $A(\omega\!=\!0)$. For higher values of $U$ the
spectral density at the Fermi level is still finite and vanishes
only in the limit $U\to\infty$ (or for $T\to 0$, provided that
$U>U_{\rm c2}(T=0)$).

For the insulating phase DMFT predicts the filling of the
Mott-Hubbard gap with increasing temperature. This is due to the
fact that the insulator and the metal are not distinct phases in
the crossover regime, implying that the insulator has a finite
spectral weight at the Fermi level. This behavior has
been detected experimentally by photoemission
experiments \cite{Mo04}.

\begin{figure}
\includegraphics[width=0.8\textwidth]{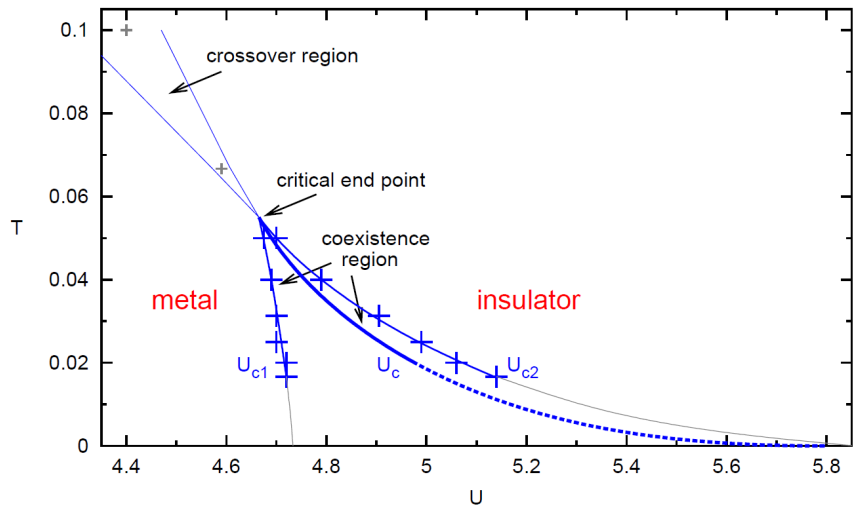}
\caption{High precision Mott-Hubbard MIT phase diagram showing the metallic phase and the insulating phase, respectively, at temperatures below the critical end point, as well as a coexistence region; from Ref. \cite{Bluemer-Diss}.}
\label{MIT_Bluemer}
\end{figure}

Altogether, the thermodynamic transition line $U_{\rm c}(T)$ corresponding to the Mott-Hubbard MIT  is found to be of first order at
finite temperatures, being associated with a hysteresis region in
the interaction range $U_{\rm c1}<U<U_{\rm c2}$ where $U_{\rm c
1}$ and $U_{\rm c 2}$ are the values at which the insulating and
metallic solution, respectively, vanishes \cite{georges96,Bulla},
\cite{Roz99}, \cite{Joo,Bulla01,Bluemer-Diss},
\cite{dmft_phys_today,Vollhardt05}. The high precision, state-of-the-art MIT phase diagram by Bl\"{u}mer \cite{Bluemer-Diss} is shown in Fig.~\ref{MIT_Bluemer}. The hysteresis region
terminates at a critical point.
For higher temperatures
the transition changes into a smooth crossover from
a bad metal to a bad insulator.

It is interesting to note that the slope of the phase transition line is negative down to $T=0$, which implies that for constant interaction $U$ the metallic phase can be reached from the insulator by decreasing the temperature $T$, i.e., by \emph{cooling}. This anomalous behavior (which corresponds to the Pomeranchuk effect \cite{VW90} in $^3$He, if we associate solid $^3$He with the insulator and liquid  $^3$He with the metal) can be easily understood from the Clausius-Clapeyron equation $dU/dT=\Delta S/\Delta D$. Here $\Delta S$ is the difference between the entropy  in the metal and in the insulator, and $\Delta D$ is the difference between the number of doubly occupied sites in the two phases. Within the single-site DMFT there is no exchange coupling $J$ between the spins of the electrons in the insulator, since the scaling \eqref{G11.11} implies $J \propto -t^{2}/U \propto 1/d \rightarrow 0$ for $d\rightarrow \infty$. Hence the entropy of the macroscopically degenerate insulating state is $S_{{\rm ins}}=k_{B}\ln2$ per electron down to $T=0$. This is larger than the entropy $S_{{\rm met}} \propto T$ per electron in the Landau Fermi-liquid describing the metal, i.e., $\Delta S = S_{{\rm met}} - S_{{\rm ins}} < 0$. At the same time the number of doubly occupied sites is lower in the insulator than in the metal, i.e., $\Delta D = D_{{\rm met}} - D_{{\rm ins}}>0$. The Clausius-Clapeyron equation then implies
that the phase-transition line $T$ vs. $U$ has a negative slope down to $T=0$. However, this is an artifact of the single-site DMFT. Namely, there will always exist an exchange coupling between the electrons leading to a vanishing entropy of the insulator at $T=0$. Since the entropy of the insulator vanishes faster than linearly with the temperature, the difference $\Delta S = S_{{\rm met}} - S_{{\rm ins}} $ eventually becomes positive, whereby the slope also becomes positive at lower temperatures\footnote{Here we assume for simplicity that the metal remains a Fermi liquid, and the insulator stays paramagnetic, down to the lowest temperatures. In fact, a Cooper pair instability will eventually occur in the metal, and the insulator will become long-range ordered, too. In this case the slope $dU/dT$ can change sign several times depending on the value of the entropy of the two phases  across the phase transition.}; this is indeed observed in cluster DMFT calculations \cite{Park08}. Since $\Delta S=0$ at $T=0$ the phase boundary must terminate  at $T=0$ with infinite slope.

\begin{figure}
\includegraphics[width=0.7\textwidth]{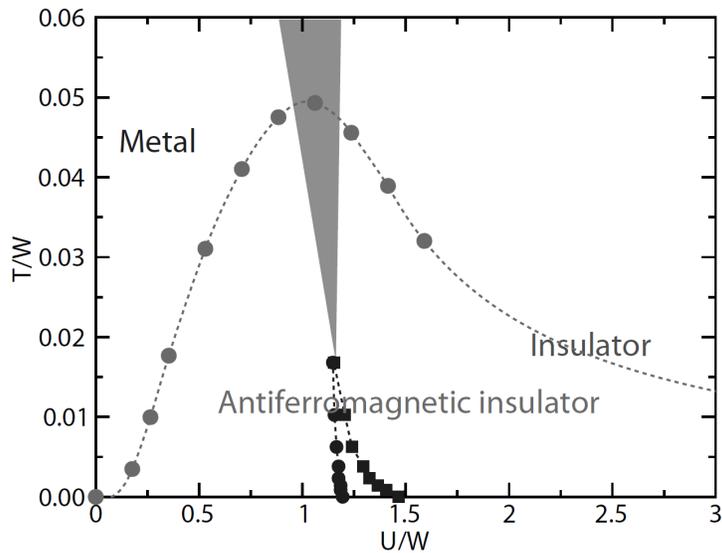}
\caption{On bipartite lattices and for half filling ($n=1$) the paramagnetic phase is unstable against antiferromagnetism. The metal-insulator transition is then completely hidden by the antiferromagnetic insulating phase; from Ref. \cite{AF_Pruschke}.}
\label{AF_Pruschke}
\end{figure}

At half filling and for bipartite lattices in dimensions $d>2$ (in $d=2$ only at $T=0$), the paramagnetic phase is unstable against antiferromagnetic long-range order. The metal-insulator transition is then completely hidden by the antiferromagnetic insulating phase, as shown in Fig.~\ref{AF_Pruschke}.

\section{Electronic correlations and disorder}

The properties of real materials are strongly influenced by the electronic
interaction and  randomness.
 In particular, Coulomb correlations and disorder are both driving forces behind
metal--insulator transitions (MITs) connected with the localization and
delocalization of particles.
While the Mott--Hubbard MIT is caused
by the electronic repulsion \cite{mott49,HubbardI,mott90}, the Anderson MIT is
due to coherent backscattering  of non-interacting particles from randomly
distributed impurities \cite{Anderson58,Lee85}.
The interplay between disorder and interaction effects gives rise to many fascinating phenomena \cite{Lee85,Altshuler85,Belitz94,Abrahams01,Kravchenko04,Byczuk10a}, some of which will now be discussed following  the presentation of Refs.~\cite{Byczuk05,Byczuk09,Byczuk10}.

The Mott-Hubbard MIT is characterized by the opening of a gap in the
density of states  at the Fermi level. By contrast, at the Anderson localization
transition the character of the spectrum at the Fermi level changes
from a continuous spectrum to a dense point spectrum. Both MITs can be characterized by a single
quantity, the local density of states (LDOS).
 Although the
LDOS is not an order parameter associated with a symmetry-breaking
phase transition, it discriminates between a metal and an insulator,
which is driven by correlations and  disorder.

\subsection{Arithmetic \emph{vs.} geometric averaging over the disorder}

The theoretical investigation of disordered systems requires the use of
probability distribution functions (PDFs) for the random quantities of
interest.
In physical or statistical problems one is usually interested in
``typical'' values of random quantities which are  mathematically determined by the most probable
value of the PDF, i.e., where the PDF becomes maximal.
In many cases the complete PDF is not known, i.e.,
only limited information about the system provided by certain averages
(moments or cumulants) is available.
In this situation  it is of great importance to choose the most informative
 average  of a random variable.
For example, if the PDF
of a random variable has a single peak and fast decaying tails this variable
is usually well estimated by its first moment, known as the \emph{arithmetic} average.
The arithmetic average of a function $F(\epsilon_i)$ is defined by (see also \eqref{G14.6})
\begin{equation}
F _{\mathrm{arith}}\equiv\langle F(\epsilon_i) \rangle_{\rm av} =\int d\epsilon
_{i} P (\epsilon _{i})F(\epsilon _{i}).
\end{equation}
 However, there are many examples,
e.g., from astronomy, the physics of glasses or networks, economy,
sociology, biology or geology, where the knowledge of the arithmetic average is insufficient
since the PDF is so broad that its characterization requires infinitely many
moments.
Such systems are said to be non-self-averaging.
One example is Anderson localization: when a disordered system is
near the Anderson MIT \cite{Anderson58}, most of the electronic quantities fluctuate strongly
and the corresponding PDFs possess long tails \cite{Mirlin94,Janssen98,Science10,Schubert10}.
At the Anderson MIT the corresponding moments might not even exist.
This is well illustrated by the local density of states (LDOS) of the
system. The arithmetic mean of this random one-particle quantity
does not resemble its  typical value at all.
In particular, it is non-critical at the Anderson transition \cite{Thouless74} and hence
cannot help to detect the localization transition. In this case the
\emph{geometric} mean \cite{lognormal,geometrical,geometrical2},
 \begin{equation}
F _{\mathrm{geom}}=\exp \left[ \langle \ln F(\epsilon_i)\rangle_{\rm
    av} \right],
\end{equation}
gives a much better
approximation of the most probable (``typical``)
value of the LDOS. It vanishes at a critical strength of the disorder and hence
provides  an explicit criterion for Anderson localization
\cite{Anderson58,Dobrosavljevic97,Dobrosavljevic03,Schubert03}, \cite{Byczuk05,Byczuk09,Byczuk10}.



\subsection{The Anderson-Hubbard model}

The fundamental electronic correlation model investigated here is
the Anderson-Hubbard model
\begin{equation}
\hat{H}=-t\sum_{ij,\sigma }\hat{c}_{i\sigma }^{+}
\hat{c}_{j\sigma}^{\phantom{+}}+
\sum_{i\sigma}
\epsilon_i n_{i\sigma} + U\sum_{i}\hat{n}_{i\uparrow
}\hat{n}_{i\downarrow }.  \label{one}
\end{equation}



The ionic energy
$\epsilon_i$  is a random, independent variable which
describes the local, quenched disorder affecting the motion of the
electrons. The disorder part is modeled by a corresponding PDF $P(\epsilon_i)$. For $P(\epsilon_i)=0$
the system is called \emph{pure}. Here we use  the continuous PDF
\begin{equation}
P(\epsilon _{i})=\frac{\Theta (\frac{\Delta}{2}-|\epsilon
_{i}|)}{\Delta},
\label{four}
\end{equation}
with $\Theta $ as the step function. The parameter $\Delta $
is a measure of the disorder strength.

\begin{figure}[tpb]
\includegraphics[clip,width=0.7\textwidth]{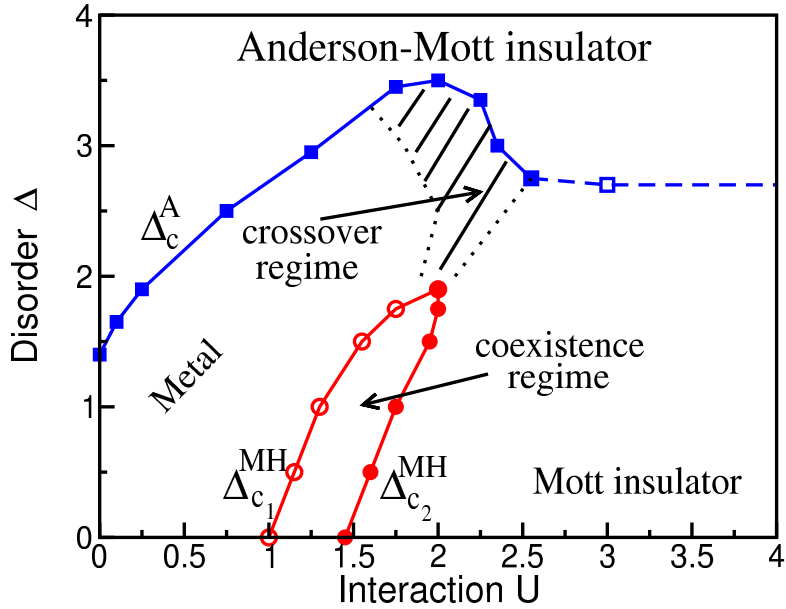}
\caption{Non-magnetic ground-state phase diagram of the
  Anderson-Hubbard  model at
half filling as calculated by DMFT with the geometrically averaged (``typical'')  LDOS; after Ref. \cite{Byczuk05}.}
\label{P-phase_diagram}
\end{figure}

A non-perturbative theoretical framework for the investigation of correlated
lattice electrons with a local interaction is again given  by the DMFT.
If in this approach the effect of local disorder is taken into account
through the arithmetic mean of the LDOS \cite{ulmke95} one obtains, in the
absence of interactions, the well-known coherent potential approximation
\cite{vlaming92} (see Sec. 4.1.), which does not describe the physics of  Anderson
localization. To overcome this deficiency Dobrosavljevi\'{c}  and collaborators formulated a variant of the DMFT where the
\emph{geometrically} averaged LDOS is computed from the solutions of the
self--consistent stochastic DMFT equations
\cite{Dobrosavljevic97} which is then incorporated into the self--consistency cycle \cite{Dobrosavljevic03}.
Thereby a mean--field theory of Anderson localization can be derived which
reproduces many of the expected features of the disorder--driven
MIT for non--interacting electrons  \cite{Dobrosavljevic03}.
This scheme  uses only one--particle quantities and is therefore
easily incorporated into the DMFT for disordered electrons in the presence of
phonons \cite{fehske}, or Coulomb correlations.
In particular, the DMFT with
geometrical averaging allows one to compute the phase diagram for the
Anderson-Hubbard model  with the continuous probability distribution
function (\ref{four}) at half filling
\cite{Byczuk05}. In this way we find that, although
 the metallic phase is enhanced for small and
intermediate values of the interaction and disorder, metallicity is
eventually destroyed upon further increase of the disorder. Surprisingly, the Mott and Anderson insulators
are found to be continuously connected.
The phase diagram for the non-magnetic ground state
is shown in Fig.~\ref{P-phase_diagram}.

Interacting quantum many-particle systems with disorder pose fundamental
challenges for theory and experiment not only in condensed matter physics
\cite{mott49,mott90,Lee85,Altshuler85,Belitz94,Abrahams01}, but most recently also
in the field of cold atoms in optical lattices
\cite{Lewenstein07,Bloch08,Billy08,Roati08,Aspect09,Lewenstein10}.
Indeed, ultracold gases have quickly developed into a fascinating new
laboratory for quantum many-body
physics \cite{Lewenstein07,Bloch08,Jaksch98,Greiner02,HTC,Mott1,Mott2}.
 A major advantage of cold atoms in optical lattices is the high degree
of controllability of the interaction and the disorder strength.
In particular, these quantum many-body systems will allow for the
first experimental investigation of the simultaneous presence of
strong interactions and strong disorder. This very interesting
parameter regime is not easily accessible in correlated electron
materials. Namely, at or close to half filling where interaction
effects become particularly pronounced, strong disorder implies
fluctuations (e.g., of local energies) of the order of the
band width, which usually leads to structural instabilities. These
limitations are absent in the case of cold atoms in optical lattices
where disorder can be tuned to become arbitrarily strong without
destroying the experimental setup. Since at half filling and in the
absence of frustration effects interacting fermions order
antiferromagnetically, several basic questions arise:

\noindent (i) How is a
non-interacting, Anderson localized system at half filling affected
by a local interaction between the particles?

\noindent (ii) How does an
antiferromagnetic insulator at half filling respond to disorder
which in the absence of interactions would lead to an Anderson
localized state?

\noindent (iii) Do Slater and Heisenberg antiferromagnets
behave differently in the presence of disorder?

\begin{figure}[tpb]
\includegraphics[clip,width=0.7\textwidth]{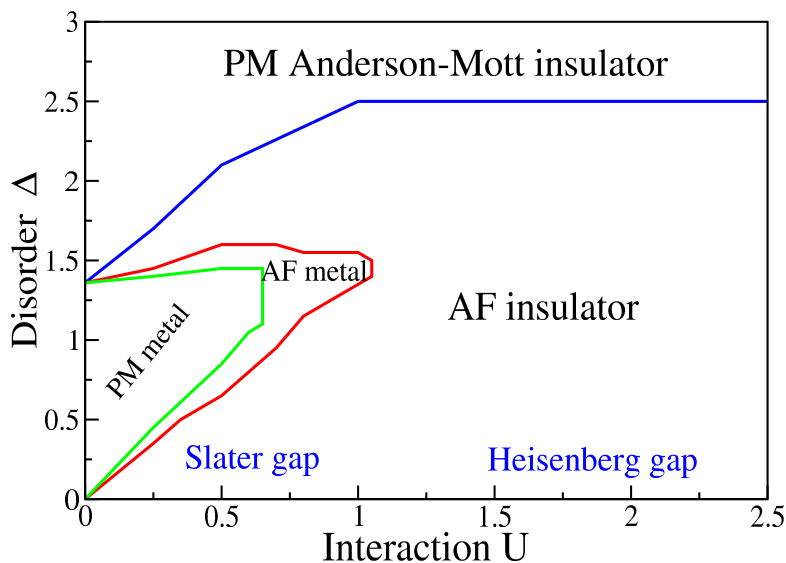}
\caption{Magnetic ground-state phase diagram of the
  Anderson-Hubbard  model at
half filling as calculated by DMFT with the geometrically averaged (``typical'') LDOS; after Ref. \cite{Byczuk09}. }
\label{AF-phase_diagram}
\end{figure}

\noindent In Ref. \cite{Byczuk09} answers to
 the above questions were obtained by calculating the zero temperature,
magnetic phase diagram of the disordered
Hubbard model at half filling using DMFT
with a geometric average over the disorder
and
allowing for a spin-dependence of the density of states (DOS). The results are collected in Fig.~\ref{AF-phase_diagram}.
 Depending on whether the interaction $U$ is weak
or strong the response of the system to disorder is found to be very
different. At strong interactions, $U/W\gtrsim 1$,
there exist only two phases, an AF insulating phase at weak disorder,
$\Delta /W\lesssim 2.5$, and a paramagnetic Anderson-Mott insulator at
strong disorder, $\Delta /W\gtrsim 2.5$. The transition between these
two phases is continuous.
By contrast, the non-magnetic phase diagram for weak
interactions, $U/W\lesssim 1,$ has a much richer structure
(Fig.~\ref{P-phase_diagram}). In particular, for weak disorder a
\textit{paramagnetic} metallic phase is found to be stable. It is separated from
the AF insulating phase at large $U$ by a narrow region of \textit{AF}
\textit{metallic} phase. The AF metallic phase is long-range ordered,
but there is no gap since the disorder leads to a redistribution of
spectral weight \cite{Byczuk09}.

\section{Theory of electronic correlations in materials}


\subsection{The LDA+DMFT approach}
\label{LDA+DMFT}

Although the Hubbard model is able to explain basic
features of the phase diagram of
correlated electrons it cannot explain the physics of real
materials in any detail. Clearly, realistic theories must take
into account the explicit electronic and lattice structure of the systems.

Until recently the electronic properties of solids were investigated by two
essentially separate communities, one using model Hamiltonians in
conjunction with many-body techniques, the other employing density
functional theory (DFT)~\cite{DFT1,DFT2}. DFT and its local density
approximation (LDA) have the advantage of being {\em ab initio} approaches
which do not require empirical parameters as input. Indeed, they are highly
successful techniques for the calculation of the electronic structure of
real materials~\cite{JonesGunn}. However, in practice DFT/LDA is seriously
restricted in its ability to describe strongly correlated materials where
the on-site Coulomb interaction is comparable with the band width. Here, the
model Hamiltonian approach is more general and powerful since there exist
systematic theoretical techniques to investigate the many-electron problem
with increasing accuracy. Nevertheless, the uncertainty in the choice of the
model parameters and the technical complexity of the correlation problem
itself prevent the model Hamiltonian approach from being
a flexible or reliable
enough tool for studying real materials. The two approaches are therefore
complementary. In view of the individual power of DFT/LDA and the
model Hamiltonian approach, respectively, it had always been clear that a combination of these techniques would be highly desirable for {\em ab initio} investigations of real materials, including,
e.g., $f$-electron systems and Mott insulators. One of the first successful
attempts in this direction was the LDA+U method~\cite{Anisimov91,Anisimov97}, which
combines LDA with a basically static, i.e., Hartree-Fock-like, mean-field
approximation for a multi-band Anderson lattice model
(with interacting and non-interacting orbitals). This method proved to
be a very useful tool in the study of long-range ordered, insulating states
of transition metals and rare-earth compounds. However, the paramagnetic
metallic phase of correlated electron systems such as high-temperature
superconductors and heavy-fermion systems clearly requires a treatment that
goes beyond a static mean-field approximation and includes dynamical
effects, e.g., the frequency dependence of the self-energy.

Here the recently developed LDA+DMFT method, a new computational scheme which
merges electronic band structure calculations and the dynamical
mean-field theory, has proved to be a breakthrough
\cite{Anisimov97a,Anisimov97aa,Nekrasov00,ldadmfta,psi-k,licht},
\cite{dmft_phys_today,Kotliar06,Held07,Katsnelson08,Kunes10}. Starting
from conventional band structure calculations in the local density
approximation (LDA) the correlations are taken into account by the
Hubbard interaction and a Hund's rule coupling term. The resulting
DMFT equations are solved numerically with a quantum Monte-Carlo
(QMC) algorithm. By construction, LDA+DMFT includes the correct
quasiparticle physics and the corresponding energetics. It also reproduces the LDA results in the limit of weak
Coulomb interaction $U$.
More importantly, LDA+DMFT correctly describes the
correlation induced dynamics near a Mott-Hubbard MIT and beyond.
Thus, LDA+DMFT is able to account for the physics at all values of the
Coulomb interaction and doping level.

 In the LDA+DMFT approach
\cite{Anisimov97a,psi-k,licht},
\cite{dmft_phys_today}
 the LDA band structure is expressed
by a one-particle Hamiltonian $\hat{H}_{\mathrm{LDA}}^{0}$, and is
then supplemented by the local Coulomb repulsion $U$ and Hund's
rule exchange $J$ (here we follow the presentation of Ref. \cite{Vollhardt05}). This leads to a material specific
generalization of the one-band model Hamiltonian

\begin{equation}
\hat{H} = \hat{H}_{\mathrm{LDA}}^{0}+{\ U}\sum_{m}\sum_{i}\hat{n}_{im\uparrow }\hat{n}_{i m\downarrow } \;+\;\sum_{i, m\neq {m'}, \sigma,{\sigma'}}\;(V-\delta _{\sigma {\sigma'}}J)\;\hat{n}_{im\sigma }\hat{n}_{im'\sigma'}.  \label{H}
\end{equation}%

Here $m$ and $m'$ enumerate the  three interacting $t_{2g}$
orbitals of the transition metal ion or the $4f$ orbitals in the
case of rare earth elements.
The interaction parameters are related by $%
V=U-2J$ which holds exactly for degenerate orbitals and is a good
approximation for the $t_{2g}$. The actual values for $U$ and $V$
can be obtained from an averaged Coulomb parameter $\bar U$ and
Hund's exchange $J$, which can be calculated by constrained LDA.

In the one-particle part of the Hamiltonian
\begin{equation}
\hat{H}_{\mathrm{LDA}}^{0} =  \hat{H}_{\rm{LDA}}
-{\sum_{i}}\sum_{m\sigma} \Delta\epsilon_d \,\hat{n}_{im\sigma}.
\end{equation}
the energy term containing $\Delta\epsilon_d$ is a shift of the
one-particle potential of the interacting orbitals. It cancels the
Coulomb contribution to the LDA results, and can also be calculated by
constrained LDA \cite{psi-k}.

Within the LDA+DMFT scheme the self-consistency condition
connecting the self-energy $\Sigma $ and the Green function $G$ at
frequency $\omega$ reads: \vspace{-0.5cm}

\begin{eqnarray}
G_{qm,q^{\prime }m^{\prime }}(\omega )=\!\frac{1}{V_{B}}\int{{\
d^{3}}{k}} \!&\left( \left[ \;\omega \bm{1}+\mu \bm{1}-H_{\mathrm{LDA}}^{0}
(\bm{k}%
) -\pmb{\Sigma}(\omega )\right]^{-1}\right)_{q  m,q^{\prime }m^{\prime
}} .&  \label{Dyson}
\end{eqnarray}

Here, $\bm{1}$ is the unit matrix, $\mu$ the chemical potential,  $H_{\mathrm{LDA}%
}^{0}(\bm{k})$ is the  orbital matrix  of the
 LDA Hamiltonian derived, for example, in a linearized
muffin-tin
orbital (LMTO) basis, $%
\pmb{\Sigma}(\omega)$ denotes the self-energy matrix which is nonzero
only between the interacting orbitals, and $[...]^{-1}$ implies
the inversion of the matrix with elements $n$ (=$qm$), $n^{\prime
}$(=$q^{\prime }m^{\prime }$), where $q$ and $m$ are the indices
of the atom in the primitive cell and of the orbital,
respectively. The integration extends over the Brillouin zone with
volume $V_{B}$ (we note that $\hat{H}_{\mathrm{LDA}}^{0}$ may include
additional non-interacting orbitals).

For cubic transition metal oxides
 Eq.\ \eqref{Dyson}
can be simplified  to
\begin{eqnarray}
G(\omega)&\!=\!&G^{0}(\omega-\Sigma (\omega))=\int d\epsilon
\frac{N^{0}(\epsilon )}{\omega-\Sigma (\omega)-\epsilon}
\label{intg}
\end{eqnarray}
if the degenerate $t_{2g}$ orbitals crossing the Fermi level are
well separated from the other orbitals~\cite{psi-k}. For non-cubic
systems the degeneracy is lifted. In this case we employ Eq.\
\eqref{intg} as an approximation, using different $\Sigma_m
(\omega)$, $N^{0}_m(\epsilon )$ and $G_m(\omega)$ for the three
non-degenerate
 $t_{2g}$ orbitals.

The Hamiltonian \eqref{H} is solved within the DMFT using standard
quantum Monte-Carlo (QMC) techniques~\cite{Hirsch86} to solve the self-consistency
equations. From the imaginary time QMC Green function
we calculate the physical (real frequency) spectral function with
the maximum entropy method~\cite{MEM}.

\subsection{Single-Particle Spectrum of Correlated Electrons in Materials}

Transition metal oxides are an ideal laboratory for the study of
electronic correlations in solids. Among these materials, cubic
perovskites have the simplest crystal structure and thus may be
viewed as a starting point for understanding the electronic
properties of more complex systems. Typically, the $3d$ states in
those materials form comparatively narrow bands with width
$W\!\!\sim \!2\!-\!3\,$~eV, which leads to strong Coulomb
correlations between the electrons. Particularly simple are
transition metal oxides with a 3$d^{1}$ configuration since, among
others, they do not show a complicated multiplet structure.

Photoemission spectra provide a direct experimental tool to study
the electronic structure and spectral properties of electronically
correlated materials. Intensive experimental investigations of
spectral and transport properties of strongly correlated 3$d^{1}$
transition metal oxides started with investigations by Fujimori
\textit{et al.}~\cite{fujimori}. These authors observed a
pronounced lower Hubbard band in the photoemission spectra (PES)
which cannot be explained by conventional band structure theory.


\subsubsection{Excursion: Detection of electronic correlations in solids by
photoemission spectroscopy}

\begin{figure}
\includegraphics[width=\textwidth]{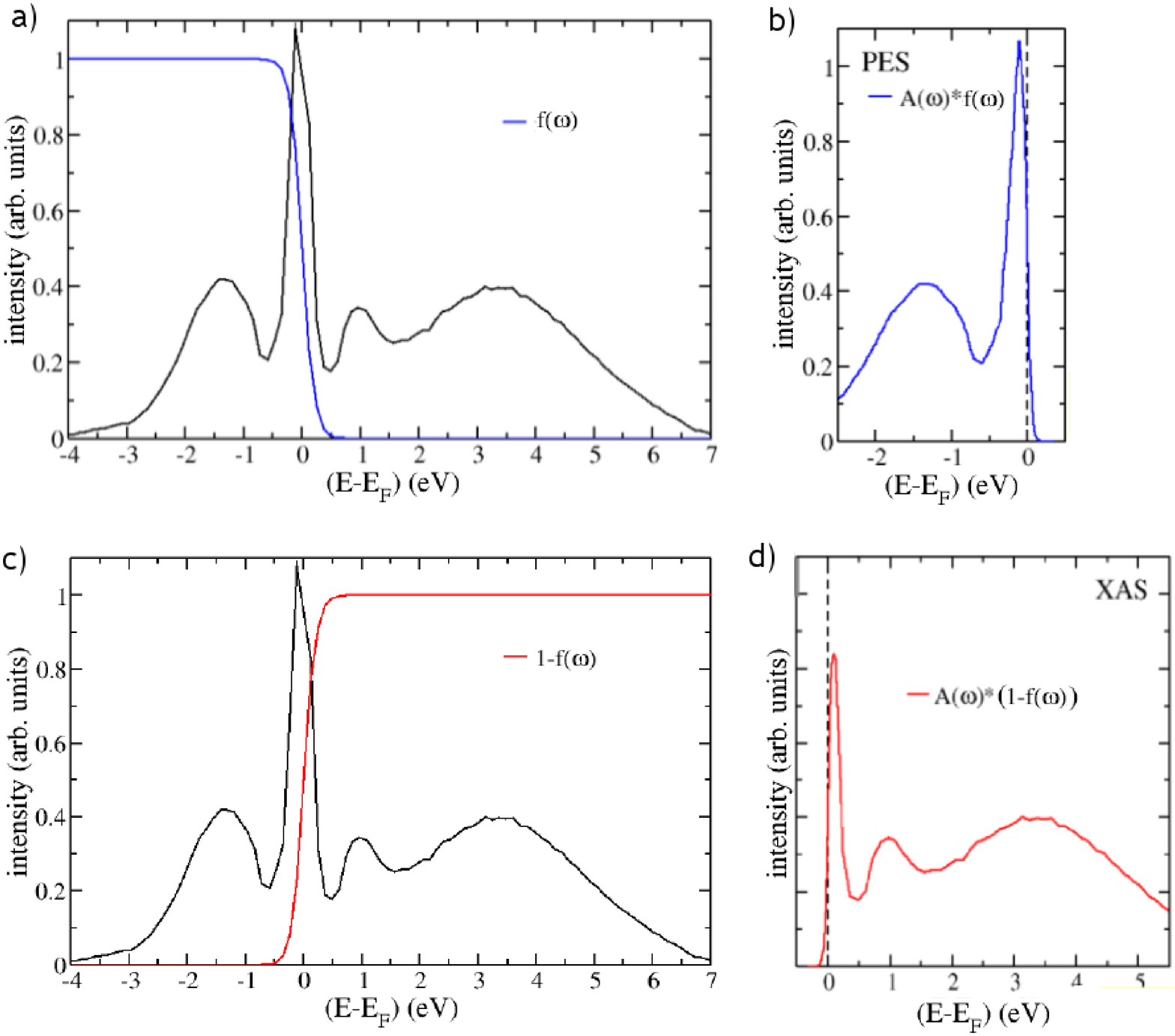}
\caption{Using photoemission spectroscopy (PES) the occupied electronic states can be measured. This corresponds to a multiplication of the (unknown) spectral function of a material with the Fermi function $f(\omega)$ as shown in panel a).  PES thus only measures the lower part of the full spectral function as shown in panel b). Inverse photoemission spectroscopy (IPES) or techniques like x-ray absorption spectroscopy (XAS) measure the unoccupied states. This corresponds to a multiplication of the spectral function with $1-f (\omega)$ as shown in panel c). Hence IPES or XAS measures the upper part of the full spectral function as shown in panel d).}
\label{fig1}
\end{figure}

In photoemission spectroscopy (PES) a photon of a given energy is used
to emit an electron whose properties (energy, angular distribution) are
measured in a detector. Angular resolved PES is referred to as ARPES.
This technique measures {\em occupied} electronic states. This means that only the states described by
the full spectral function of a material multiplied by the
Fermi function $f(\omega)$  are measured (see Fig.~\ref{fig1}a), leading to the typical result shown
in Fig.~\ref{fig1}b. By contrast, inverse photoemission spectroscopy (IPES)
measures the {\em unoccupied} electronic states. IPES is harder to perform
and not as accurate as PES. But in many situations information about the
unoccupied states is also available by X-ray absorption spectroscopy
(XAS). In this case only the states described by the spectral function of a material
multiplied by $1-f (\omega)$ (Fig.~\ref{fig1}c) are measured, leading to a typical result shown
in Fig.~\ref{fig1}d.

Spectroscopic techniques are routinely used to investigate correlated
electronic systems. They provide very valuable information about the
system since they measure the spectral function of a material which can
be calculated theoretically.
In particular, photoemission techniques allow one to detect the
correlation induced shift of spectral weight discussed in Sec.~\ref{sec:mit}.

In the following we will employ the LDA+DMFT approach to compute the
$\bm{k}$-integrated electronic spectra of two correlated
materials, the 3$d^{1}$ system (Sr,Ca)VO$_3$
and the charge-transfer insulator NiO.


\subsubsection{Sr$_x$Ca$_{1-x}$VO$_3$}

SrVO$_{3}$ and CaVO$_{3}$ are simple transition metal compounds
with a 3$d^{1}$ configuration (here we follow the presentation of Sekiyama \emph{et al.}~\cite{Sekiyama03} and Nekrasov \emph{et al.}~\cite{Nekrasov05}). The main effect of the substitution
of Sr ions by the isovalent, but smaller, Ca ions is to decrease
the V-O-V angle from $\theta = 180^{\circ }$ in SrVO$_{3}$ to
$\theta \approx 162^{\circ }$ in the orthorhombically distorted
structure of CaVO$_{3}$. However, this rather strong bond bending
results only in a 4\% decrease of the one-particle bandwidth $W$
and thus in a correspondingly small increase of the ratio $U/W$ as
one moves from SrVO$_{3}$ to CaVO$_{3}$.

\begin{figure}
\includegraphics[width=0.7\textwidth]{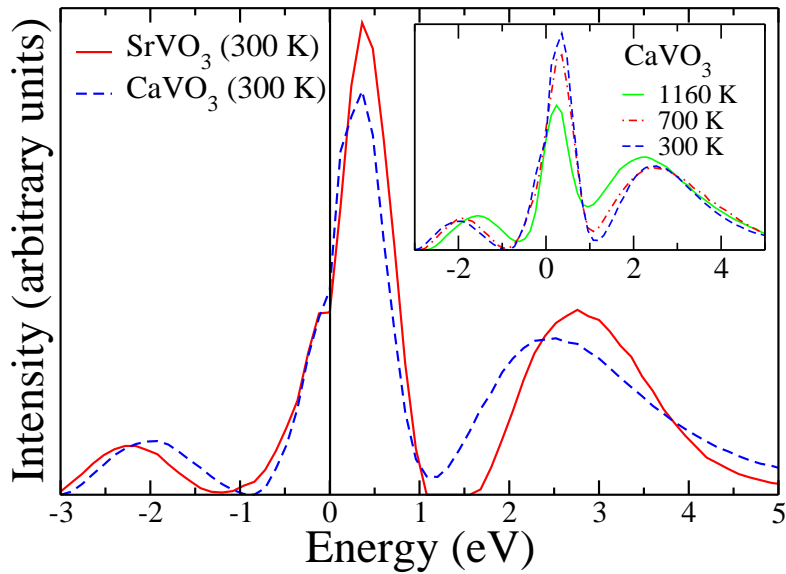}
\caption{LDA+DMFT(QMC) spectrum of SrVO$_{3}$ (solid line) and
CaVO$_{3}$ (dashed line) calculated at T=300 K; inset: effect of
temperature in the case of CaVO$_{3}$; after Ref. \cite{Sekiyama03}.}
\label{fig_ldadmft}
\end{figure}

LDA+DMFT(QMC) spectra of SrVO$_{3}$ and CaVO$_{3}$ were calculated
by Sekiyama \textit{et al.} \cite{Sekiyama03} by starting from the
respective LDA DOS of the two materials; they are shown in Fig.~\ref{fig_ldadmft}.
These spectra show genuine correlation effects, i.e., the
formation of lower Hubbard bands at about  1.5 eV and upper
Hubbard bands at about 2.5 eV, with well-pronounced quasiparticle
peaks at the Fermi energy. Therefore both SrVO$_{3}$ and
CaVO$_{3}$ are strongly correlated metals.
The DOS of the two systems shown in Fig.~\ref{fig_ldadmft} are
quite similar. In fact, SrVO$_{3}$ is slightly less correlated
than CaVO$_{3}$, in accord with their different LDA bandwidths.
The inset of Fig.~\ref{fig_ldadmft} shows that the effect of
temperature on the spectrum is small for $T \lesssim 700$~K.
Spectra of SrVO$_{3}$ and CaVO$_{3}$ were also calculated
independently by Pavarini {\em et al.}\cite{Pavarini}.

\begin{figure}
\includegraphics[width=0.9\textwidth]{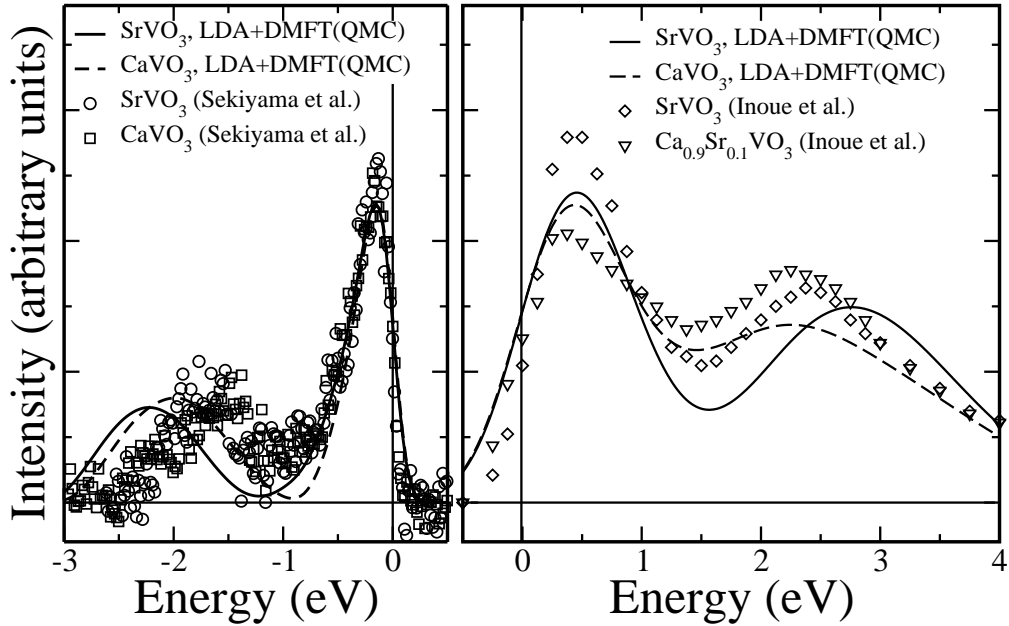}
\caption{Comparison of the calculated, parameter-free
LDA+DMFT(QMC) spectra of SrVO$_{3}$ (solid line) and CaVO$_{3}$
(dashed line) with experiment. Left: Bulk-sensitive
high-resolution PES (SrVO$_{3}$: circles; CaVO$_{3}$: rectangles). Right: 1s XAS for
SrVO$_{3}$ (diamonds) and Ca$_{0.9}$Sr$_{0.1}$VO$_{3}$
(triangles)~\protect\cite{Inoue94}. Horizontal line: experimental
subtraction of the background intensity; after Ref.~\cite{Nekrasov05}.}
\label{fig_XPS}
\end{figure}

Since the three $t_{2g}$ orbitals of this simple 3$d^{1}$ material
are (almost) degenerate the spectral function has the same
three--peak structure as that of the one-band Hubbard model shown
in Fig.~\ref{fig:4.1}. The temperature induced decrease of the
quasiparticle peak height is also clearly seen. As noted in Sec.~\ref{sec:mit} the actual form of the spectrum no longer resembles the input
(LDA) DOS, i.e., it essentially depends only on the first three
energy moments of the LDA DOS (electron density, average energy,
band width).
In the left panel of Fig.~\ref{fig_XPS} the
LDA+DMFT(QMC) spectra at 300K are compared with experimental high-resolution bulk
PES.
For this purpose  the theoretical spectra were multiplied with the
Fermi function at the experimental temperature (20$\,$K) and Gauss
broadened with the experimental resolution of $0.1\,$eV
\cite{Sekiyama03}. The quasiparticle peaks in theory and experiment
are seen to be in very good agreement. In particular, their height
and width are almost identical for both SrVO$_{3}$ and CaVO$_{3}$.
The difference in the positions of the lower Hubbard bands may be
partly due to (i) the subtraction of the (estimated) oxygen
contribution which might also remove some $3d$ spectral weight
below $-2$~eV, and (ii) uncertainties in the {\em{ab initio}}
calculation of the local Coulomb interaction strength. In the right panel of Fig.~\ref{fig_XPS}
comparison is made with XAS data of Inoue \textit{et
al.}~\cite{Inoue94}. Core-hole life time effects were considered
by Lorentz broadening the spectrum with 0.2~eV~\cite{Krause79},
multiplying with the inverse Fermi function (80K), and then Gauss
broadening with the experimental resolution of
$0.36\,$eV~\cite{Inoue03}. Again, the overall agreement of the
weights and positions of the quasiparticle and upper $t_{2g}$
Hubbard band is good, including the tendencies when going from
SrVO$_{3}$ to CaVO$_{3}$ (Ca$_{0.9}$Sr$_{0.1}$VO$_{3}$ in the
experiment). For  CaVO$_{3}$ the weight of the quasiparticle peak
is somewhat lower than in the experiment. In contrast to one-band
Hubbard model calculations, the material specific results
reproduce the strong asymmetry around the Fermi energy w.r.t.
weights and bandwidths. The results also give a different
interpretation of the XAS than in Ref. \cite{Inoue94} where the
maximum at about $2.5\,$eV was attributed to an $e_g$ band and not
to the $t_{2g}$ upper Hubbard band. The slight differences in the
quasiparticle peaks (see Fig.~\ref{fig_ldadmft}) lead to different
effective masses, namely $m^*/m\!=\!2.1$ for SrVO$_{3}$ and
$m^*/m\!=\!2.4$ for CaVO$_{3}$. These theoretical values agree
with $m^{\ast }/m\!=\!2-3$ for SrVO$_{3}$ and CaVO$_{3}$ as
obtained from de Haas-van Alphen experiments and thermodynamics
\cite{Inoue02}.

The experimentally determined spectra of SrVO$_{3}$ and CaVO$_{3}$ and the good agreement with parameter-free LDA+DMFT calculations confirm the existence of a pronounced three-peak structure in a correlated bulk material. Although the DMFT had predicted such a behavior for the Hubbard model (see Sec. 5.1.) it was not clear whether the DMFT result would really be able to describe real materials in three dimensions. Now it has been confirmed that the three-peak structure not only occurs in single-impurity Anderson models but also in three-dimensional correlated bulk matter.

\subsubsection{NiO}


Already in 1937, at the outset of modern solid state physics, de Boer and Verwey \cite{deboer} drew attention
to the surprising properties of materials with incompletely filled 3$d$-bands, such as NiO. This
observation prompted  Mott and Peierls \cite{mott37} to discuss the interaction between the electrons.
Ever since transition metal oxides (TMOs) were investigated intensively (here we follow the presentation of Kune\v{s} \emph{et al.}\cite{kun07a}).
This interest further increased when it was discovered that
TMOs display an amazing multitude of ordering and electron correlation phenomena, including
high temperature superconductivity, colossal magnetoresistance and Mott metal-insulator transitions \cite{tokura}.
In the late 1950's MnO and NiO were taken as the textbook examples of antiferromagnets.
However, when the importance of local Coulomb correlations in the transition metal $d$-shell
was realized TMOs were considered candidates for Mott insulators \cite{mott37}.
In the mid 1980's Zaanen, Sawatzky and Allen (ZSA) introduced their classification of
TMOs and related compounds into Mott-Hubbard and charge-transfer (CT) systems \cite{zaa85}.
In the early TMOs the ligand $p$-band is located well below the transition metal $d$-band
and thus plays a minor role in the low energy dynamics. Such a case, called
Mott-Hubbard system in the ZSA scheme, is well described by a multi-band Hubbard model.
On the other hand, the late TMOs belong to the CT type where the $p$-band is situated between
the interaction split $d$-bands. A more general Hamiltonian where
the $p$-states are explicitly included is then needed, which can be viewed as a combination
of multi-band Hubbard and Anderson lattice models.
A major impulse for detailed investigations of CT systems, and especially of their hole doped regime,
came with the discovery of high temperature superconductivity in cuprate perovskites.
While the standard three-band Hamiltonian for cuprates \cite{emery} contains only one $d$-orbital
per lattice site, the description of cubic transition metal monoxides, the prominent member of which
is NiO, requires the full set of $d$-orbitals. The latter are of interest not only
for fundamental research, but play an important role also in fields such as geophysics \cite{cohen}.
Furthermore, recent progress in high pressure experiments \cite{yoo}
made the insulator-to-metal transition in some TMOs accessible in the laboratory,
providing yet another stimulus for theoretical investigations.



NiO is a type II antiferromagnet ($T_N=523$ K) with
a magnetic moment of almost 2$\mu_B$ and a large gap surviving well above $T_N$.
The standard LDA band theory predicts NiO to be a metal \cite{mat72},
or an antiferromagnetic insulator \cite{ter84} if spin polarization is allowed.
A severe underestimation of the gap and the magnetic moment
suggests, however, that the Slater antiferromagnetic state
obtained within LDA does not describe the true nature
of NiO. On the other hand exact diagonalization studies on small clusters
were quite successful in describing the single- and two-particle
spectra \cite{fuj84}, showing that the local Coulomb interactions are important.
This made it clear that an explicit treatment of Coulomb interactions within the
$3d$ shell is needed, and methods such as
LDA+U \cite{ani91}, self-interaction correction \cite{sva91}, or GW \cite{ary95} were applied.
The static, orbitally dependent self-energy of LDA+U
enforces a separation of the occupied and unoccupied $d$-bands and thus opens a gap comparable
to experiment. This in turn leads to a significant improvement of the description of static properties
such as the local moment or the lattice dynamics \cite{savrasov}.
However, the LDA+U method is limited to an ordered state and does not
yield the electronic excitations and the effect of doping correctly. A systematic inclusion of dynamical correlations is only made possible by the
DMFT.
\begin{figure}
\includegraphics[width=0.8\textwidth]{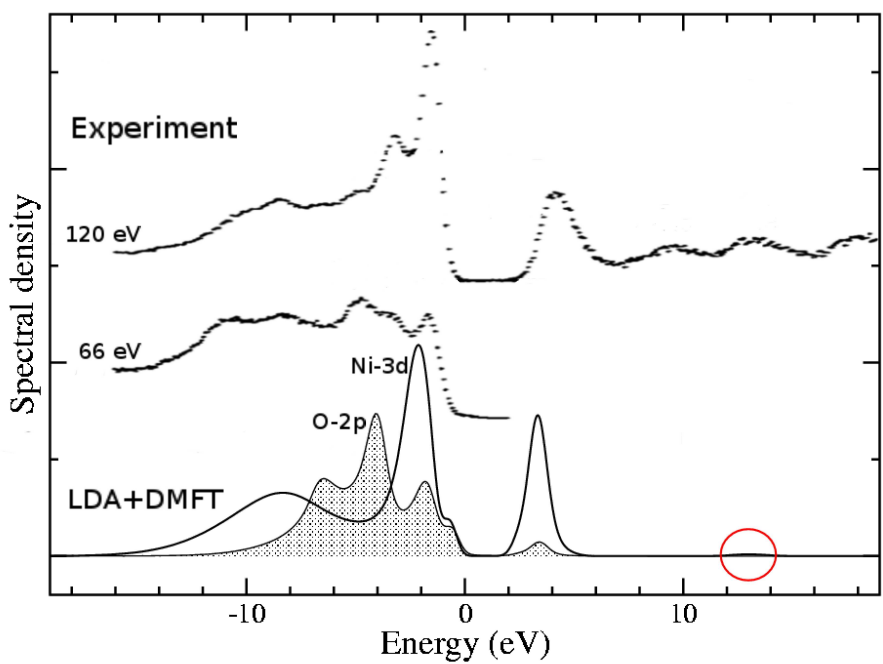}
\caption{\label{fig:total} Theoretical Ni-$d$ (solid line) and O-$p$ (shaded) resolved spectral densities compared
to photoemission and inverse photoemission data obtained at
120 eV and 66 eV photon frequencies after Ref. \cite{saw84}. Gaussian broadening of 0.6 eV full width at
half maximum corresponding to the experimental resolution was applied to the theoretical curves.
The circle marks position of the $d^{10}\underline{L}$ excitation; from Ref. \cite{kun07a}.}
\end{figure}

In Fig.~\ref{fig:total} the calculated spectral densities \cite{kun07a} (resolved into Ni $3d$
and O $2p$ contributions)
are compared to photoemission and inverse photoemission data \cite{saw84}.
Using the full $p-d$ Hamiltonian it is possible to cover the entire
valence and conduction bands spectra. Features corresponding to $4s$ and $4p$
bands at 10 eV and 13 eV, respectively, are not included in the theoretical spectrum.
The relative intensity of the $2p$ contribution increases with decreasing photon energy \cite{eastman}.
Therefore the 120 eV spectrum is dominated by Ni $3d$ emission, while at 66 eV photon energy
the O $2p$ contribution peaked around -4 eV is resolved (for a detailed
orbital decomposition see Ref. \cite{eastman}).
The theoretical spectrum very well reproduces the experimental features, including
the size of the gap, the $d$ character of the conduction band, the broad $d$ peak at -9 eV,
the position of the $p$-band, and the strong $d$ contribution at the top
of the valence band.
While the gap and the Hubbard subbands can be described already with the
static theory (LDA+U) \cite{ani91}, a dynamical treatment is apparently needed
to capture the substantial redistribution of spectral weight between the incoherent (-9 eV)
and resonant (-2 eV) features in the $d$ spectrum.



We see that by including the ligand $p$ states and the on-site Coulomb interaction within the same framework one is able
to provide a full description of the valence band spectrum and, in particular, of the distribution
of spectral weight between the lower Hubbard band and the resonant peak at the top of the valence band.
Good agreement with the available photoemission and inverse-photoemission data is found without need
for adjustable parameters \cite{kun07a}. In a similar way ARPES data can also be explained in detail \cite{kun07b}.


\subsection{Correlation induced structural transformations}


In materials with correlated electrons the interaction between spin,
charge, orbital, and lattice degrees of freedom leads to a wealth of
ordering phenomena and complex phases \cite{tokura}. The diverse
properties of such systems and their great sensitivity with respect
to changes of external parameters such as temperature, pressure,
magnetic field or doping also make them highly attractive for
technological applications \cite{tokura}. In particular, orbital
degeneracy is an important and often inevitable cause for this
complexity \cite{KK}. A fascinating example is the cooperative
Jahn-Teller (JT) effect --- the spontaneous lifting of the
degeneracy of an orbital state --- leading to an occupation of
particular orbitals (``orbital ordering'') and, simultaneously, to a
structural relaxation with symmetry reduction.



Applications of  LDA+DMFT so far mainly
employed linearized and higher order muffin-tin orbital (L(N)MTO)
methods \cite{LNMTO} and concentrated on the study of correlation
effects within the electronic system for a given ionic lattice (here we follow the presentation of Leonov \emph{et al.} \cite{Leonov08,Leonov10}). On
the other hand, the interaction of the electrons with the ions also
affects the lattice structure. LDA+DMFT investigations of
particularly drastic examples, e.g., the volume collapse in paramagnetic
Ce \cite{HM01,AB06} and Pu \cite{SKA01+, SKA01+1,SKA01+2}, and the magnetic moment
collapse in MnO \cite{MnO}, incorporated the lattice by calculating
the total energy of the correlated material as a function of the
atomic volume. However, for investigations going beyond equilibrium-volume calculations, e.g., of the cooperative JT effect and other
subtle structural relaxation effects, the L(N)MTO method is not
suitable since it cannot determine atomic displacements reliably.
This is partly
due to the fact that the atomic-sphere approximation used in the
L(N)MTO scheme, with a spherical potential inside the atomic sphere,
completely neglects multipole contributions to the electrostatic
energy originating from the distorted charge density distribution
around the atoms.



Recently Leonov \emph{et al.} \cite{Leonov08,Leonov10} formulated
 a computational scheme which now allows one to
calculate lattice-relaxation effects caused by electronic
correlations. To this end the GGA+DMFT --- a merger of the ``generalized gradient approximation'' (GGA) and
DMFT --- was formulated within a plane-wave pseudopotential approach
\cite{TL08,LG06}. Thereby the limitations of the L(N)MTO
scheme in the direct calculation of total energies are overcome. In
particular, this new method can be applied to determine the orbital order
and the cooperative JT distortion in the paramagnetic phase of the
prototypical JT system KCuF$_{3}$.

KCuF$_{3}$ is long known to be a prototypical material with a
cooperative JT distortion \cite{KK} where the electronic degrees of
freedom are the driving force behind the orbital order
\cite{KK,LA95,MK02}. Indeed, the relatively high (tetragonal)
symmetry makes KCuF$_{3}$ one of the simplest systems to study. In
particular, only a single internal structure parameter, the shift of
the in-plane fluorine atom from the Cu-Cu bond center, is needed to
describe the lattice distortion.



\begin{figure}[tbp]
\centerline{\includegraphics[width=0.6\textwidth,clip]{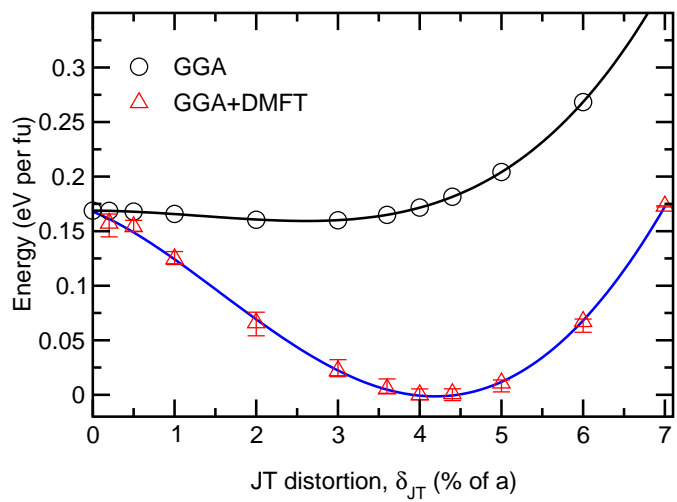}}
\caption{ Comparison of the total energies of
paramagnetic KCuF$_3$ computed by GGA and GGA+DMFT(QMC) as a
function of the JT distortion. Error bars indicate the statistical
error of the DMFT(QMC) calculations; from Ref. \cite{Leonov08}. } \label{fig:energy}
\end{figure}

The total energies as a function of the JT distortion
$\delta_{JT}= \frac{1}{2}(d_l-d_s)/(d_l+d_s)$,
where $d_l$ and $d_s$ denote the long and short Cu-F bond distances,
 obtained by the GGA and GGA+DMFT, respectively, are compared in
Fig.~\ref{fig:energy}. It should be noted that the GGA not only predicts a
\textit{metallic} solution, but its total energy is seen to be
almost constant for $0<\delta_{JT}\lesssim 4\%$. Both features are
in contradiction to experiment since the extremely shallow minimum
at $\delta_{JT}\simeq2.5\%$ would imply that KCuF$_3$ has no JT
distortion for $T\gtrsim$ 100 K. By contrast, the inclusion of the
electronic correlations among the partially filled Cu $e_g$ states
in the GGA+DMFT approach leads to a very substantial lowering of the
total energy by $\sim$ 175 meV per formula unit (fu). This implies
that the strong JT distortion persists up to the melting temperature
($>1000$ K), in agreement with experiment. The minimum of the
GGA+DMFT total energy is located at the value $\delta_{JT}= 4.2\%$
which is also in excellent agreement with the experimental value of
4.4\% \cite{BM90}. This clearly shows that the JT distortion in
paramagnetic KCuF$_3$ is caused by electronic correlations.



The GGA+DMFT scheme introduced in Refs. \cite{Leonov08,Leonov10} opens the way for fully
microscopic investigations of the structural properties of strongly
correlated electron materials such as lattice instabilities observed
at correlation induced metal-insulator transitions.


\section{Kinks in the dispersion of strongly correlated electron systems}


The dispersion relation $E_{\bm{k}}$ indicates at which energy and crystal
  momentum one-particle excitations can occur in a solid. The coupling between
    the excitations may lead to abrupt changes in the slope of the
    dispersion, referred to as ``kinks''.   Such kinks thus carry important information about
    interactions in a many-body system (here we follow the presentation of Byczuk \emph{et al.} \cite{byczuk07,byczuk07a}).



In systems with a strong electron-phonon coupling kinks in the
  electronic dispersion at 40-60 meV below the Fermi level are well
  known (''Kohn anomaly''). Therefore the kinks detected at 40-70 meV below the Fermi level in
    the electronic
    dispersion of high-temperature superconductors are taken as
    evidence for phonon \cite{Lanzara2002,Shen2002} or
    spin-fluctuation based \cite{He2001,Hwang2004} pairing mechanisms.
    Kinks in the electronic dispersion at binding energies ranging from 30 to
    800 meV are also found in various other metals
    \cite{Hengsberger1999,Rotenberg2000,Schafer2004,Aiura2004,Yoshida2005}
    posing questions about their origins.


 Starting from the unexpected finding of kinks in the momentum-resolved spectral functions of SrVO$_3$ calculated by LDA+DMFT \cite{Nekrasov07}, Byczuk \emph{et al.} \cite{byczuk07}
recently discovered
  a  novel, purely electronic mechanism yielding kinks    in the
electron dispersion. This mechanism does not
require a  coupling of two different excitations as in previously
known cases. The theory applies to strongly correlated metals whose
spectral function shows well separated Hubbard subbands and central
peak as, for example, in transition metal-oxides.
For a microscopic description of these electronic kinks the single-band
  Hubbard model \eqref{G11.7}  with
  particle-hole symmetry was investigated by the DMFT at
  $T=0$ \cite{byczuk07} as will be described below.

%
%

The effective dispersion relation $E_{\bm{k}}$ of the
  one-particle excitation is determined by the singularities of
  $G(\bm{k},\omega) =(\omega+\mu-\epsilon_{{ \bf k}}-\Sigma(\bm{    k},\omega))^{-1}$, which give rise to peaks in the spectral
  function $A(\bm{k},\omega)=-{\rm Im}G(\bm{k},\omega)/\pi$.
Here
  $\omega$ is the frequency, $\mu$ the chemical potential,
  $\epsilon_{\bm{k}}$ the bare dispersion relation, and
  $\Sigma(\bm{k},\omega)$ is the self-energy.
 If
  the damping given by the imaginary part of $\Sigma(\bm{k},\omega)$
  is not too large, the effective dispersion $E_{\bm{k}}$ is thus determined by
\begin{equation}
E_{\bm{k}}+\mu-\epsilon_{\bm{k}}-{\rm Re}\Sigma(\bm{k},E_{\bm{k}})=0.
\end{equation}
  Any kinks in $E_{\bm{k}}$ that do not originate from
  $\epsilon_{\bm{k}}$ must therefore be due to  changes in the slope
  of  ${\rm Re}\Sigma(\bm{k},\omega)$.

  The DMFT self-consistency equations are now used to express
  $\Sigma(\bm{k},\omega)=\Sigma(\omega)$ as
  $\Sigma(\omega)=\omega+\mu-1/G(\omega)-\Delta(G(\omega))$, where
  $G(\omega)=\int G(\bm{k},\omega)\,d\bm{k}$ is the $\bm k$-averaged local Green
  function and $\Delta(G)$ is an
  energy-dependent hybridization function, written here as a
  function of $G(\omega)$.

  \begin{figure}[tbp]
\includegraphics[clip,width=110mm]{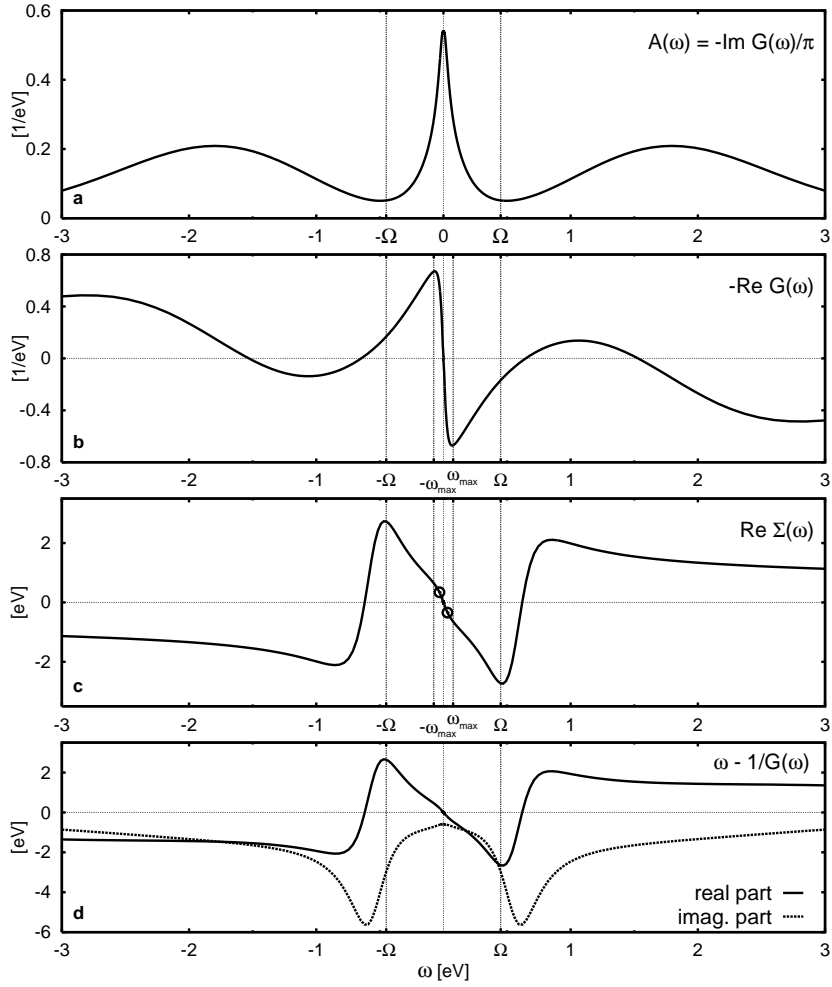}\vspace*{-5mm}
    \caption{Local propagator and self-energy for a strongly
      correlated system. (a) Correlation-induced three-peak spectral
      function
  $A(\omega)=-{\rm Im}G(\omega)/\pi$ with dips at $\pm\Omega=0.45$
  eV.  (b) Corresponding real part of the propagator,
  $-{\rm Re}G(\omega)$, with minimum and maximum at
  $\pm\omega_{{\rm max}}$ inside the central spectral peak.
  (c) Real part of the self-energy with kinks at
  $\pm\omega_*$ (circles), located at the points of maximum
  curvature of ${\rm Re}G(\omega)$,
  ($\omega_*=0.4\omega_{{\rm max}}=0.03$ eV).  (d)
  $\omega-1/G(\omega)$ contributes to the self-energy. In general
  ${\rm Re}[\omega-1/G(\omega)]$ (full line) is linear in
  $|\omega|<\Omega$.  The other contribution to the self-energy is
  $-\Delta(G(\omega))\approx-(m_2-m_1^2)G(\omega)$ (to lowest order in
  the moments $m_i$ of $\epsilon_{\bm{k}}$; here $m_2-m_1^2$=0.5
  eV$^2$).  Therefore the nonlinearity of $-{\rm Re}[G(\omega)]$ at
  $\pm\omega_*$ determines the location of kinks; after Ref. \cite{byczuk07}. }
    \label{kinks}
  \end{figure}

\begin{figure}
\includegraphics[width=1.2\textwidth]{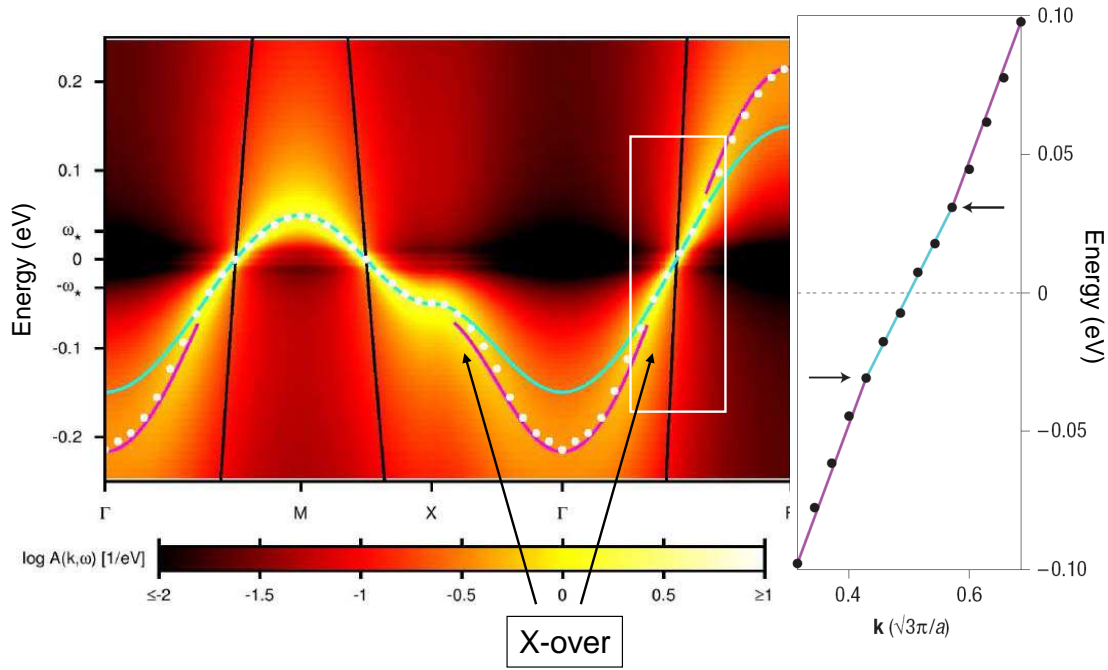}
\caption{Kinks in the dispersion relation $E_{\bm{k}}$ of the Hubbard model on a
  cubic lattice with interaction $U$=3.5 eV,
  bandwidth $W\approx$ 3.46 eV, $n=1$, implying a Fermi-liquid renormalization factor $Z_{\rm FL}$=0.086.  The intensity plot represents
      the spectral function $A(\bm{k},\omega)$.  Close to the Fermi
      energy the effective dispersion (white dots) follows the
      renormalized band structure
      $E_{\bm{k}}=Z_{\text{FL}}\epsilon_{\bm{k}}$ (light line).  For
      $|\omega|>\omega_\star$ the dispersion has the same shape but
      with a different renormalization,
      $E_{\bm{k}}=Z_{\text{CP}}\epsilon_{\bm{k}}-c\;\text{sgn}(E_{\bm{k}})$
      (dark line). Here $\omega_\star$=0.03 eV, $Z_{\text{CP}}=0.135$,
      and $c=0.018$ eV are all calculated from
      $Z_{\text{FL}}$ and $\epsilon_{\bm{k}}$ (black line).  A
      subinterval of $\Gamma$-R (white frame) is plotted on the right,
      showing kinks at $\pm\omega_\star$ (arrows); after Ref. \cite{byczuk07}.}\label{kinks2}
\end{figure}

  Kinks in ${\rm Re}\Sigma(\omega)$ are found to appear at a new small
  energy scale which emerges quite generally for a three-peak spectral
  function $A(\omega)$, see Fig.~\ref{kinks}.  Kramers-Kronig relations imply that
  ${\rm Re}[G(\omega)]$ is small near the dips of $A(\omega)$,
  located at $\pm\Omega$. Therefore ${\rm Re}[G(\omega)]$ has a
  maximum and a minimum at $\pm\omega_{{\rm max}}$ \emph{inside the
    central spectral peak} (Fig.~\ref{kinks}b). This directly leads
  to kinks in
  ${\rm Re}\Sigma(\omega)$ for the following reason.
  There are two contributions to $\Sigma(\omega)$:
  $\omega+\mu-1/G(\omega)$ and $-\Delta(G(\omega))$.  While
  ${\rm Re}[\omega+\mu-1/G(\omega)]$ is \emph{linear} in the large
  energy window $|\omega|<\Omega$ (Fig.~\ref{kinks}d), the
  term $-{\rm Re}[\Delta(G(\omega))]$ is approximately proportional
  to $-{\rm Re}[G(\omega)]$ (at least to first order in a moment
  expansion), and thus remains linear only in a much narrower energy
  window $|\omega|<\omega_{{\rm max}}$.  The sum of
  these two contributions produces pronounced kinks in the real part
  of the self-energy at $\pm\omega_*$, where
  $\omega_*=0.41 \omega_{{\rm max}}$ is the energy where
  ${\rm Re}[G(\omega)]$ has maximum curvature (marked by
  circles in Fig.~\ref{kinks}c).  The Fermi-liquid (FL) regime with
  slope $\partial {\rm Re}\Sigma(\omega)/\partial\omega =
  1-1/Z_{{\rm FL}}$ thus extends only throughout a small part of the
  central peak ($|\omega |<\omega_\star$). At intermediate
  energies ($\omega_\star<|\omega|<\Omega$) the slope
  is then given by $\partial{\rm Re}\Sigma(\omega)/\partial\omega =
  1-1/Z_{{\rm CP}}$. The kinks at $\pm\omega_*$ mark the
  crossover between these two slopes.
  As a consequence there is also a kink at $\omega_*$ in the effective band structure
  $E_{\bm{k}}$ (Fig.~\ref{kinks2}).

  The FL regime terminates at the kink energy scale $\omega_\star$, which
  cannot be determined within FL theory itself.  One of the most surprising results of the investigation
  is the fact that it
is possible to describe properties of the system outside the Landau-Fermi-liquid regime fully
analytically. Namely, one can express the
  quantities $\omega_\star$ and $Z_{\bm{CP}}$
in terms of $Z_{{\rm FL}}$ and the bare density of states
  alone.
  Explicitly, one finds $\omega_\star=0.41Z_{{\rm FL}}D$, where $D$
    is an energy scale of the noninteracting system, e.g., $D$ is approximately given by
  half the bandwidth \cite{byczuk07}.

  The energy scale $\omega_*$ involves only the bare band
  structure which can be obtained, for example, from band structure
  calculations, and the
  FL renormalization
  $Z_{{\rm FL}}=1/(1-\partial
  {\rm Re}\Sigma(0)/\partial\omega)\equiv m/m^*$ known from, e.g.,
  specific heat measurements or many-body calculations.
  It should be noted that since phonons are not involved in this mechanism,
  $\omega_\star$ shows no isotope effect.  For strongly interacting
  systems, in particular close to a metal-insulator transition,
  $\omega_\star$ can become quite small, e.g., smaller
  than the Debye energy.

The theory described above explains the kinks  in the slope of the dispersion as a direct consequence of the electronic interaction~\cite{byczuk07}. The same mechanism may also lead to kinks in the low-temperature electronic specific heat~\cite{Toschi09}. The kinks have also been
  linked to maxima in the spin susceptibility~\cite{Uhrig09}. Of
  course, \emph{additional} kinks in the electronic dispersion may also arise from the coupling of electrons to bosonic degrees of freedom, such as
  phonons or spin fluctuations.  Interestingly, recent experiments~\cite{Nickel09} have found evidence for kinks in
  Ni(110), which may be due to the electronic
  mechanism presented here.


\section{Summary and Outlook}


Due to the intensive international research over the last two decades the DMFT has quickly developed into a powerful method for the investigation of electronic systems with strong correlations. It provides a comprehensive, non-perturbative and thermodynamically consistent approximation scheme for the investigation of finite-dimensional systems (in particular for dimension $d=3$), and is particularly useful for the study of problems where perturbative approaches are inapplicable. For this reason the DMFT has now become the standard mean-field theory for fermionic correlation problems,  including cold atoms in optical lattices \cite{Rapp}, \cite{Snoek,Mott2}. The study of models  in non-equilibrium within a suitable generalization of the DMFT has become yet another fascinating new research area \cite{Schmidt02,Turkowski05,Freericks06b,Eckstein2008a,Tran08x,Freericks08,Tsuji08,Freericks09,Eckstein2008b,Eckstein2008c,Eckstein2009}.

Until a few years ago research into correlated electron systems concentrated on homogeneous bulk systems. DMFT investigations of systems with internal or external inhomogeneities such as thin films and multi-layered nanostructures are still very new ~\cite{Potthoff99,Freericks-book,Takizawa06,Chen07,Byczuk08x,Helmes08}. They are particularly important in view of the novel types of functionalities of such systems, which may have important applications in electronic devices. Here the DMFT and its generalizations will certainly be very useful.

In particular, the development  of the \emph{ab initio} band-structure calculation technique referred to as LDA+DMFT has proved to be a breakthrough in the investigation of electronically correlated materials. It has already provided important insights into the spectral and magnetic properties of correlated electron materials, e.g., transition metals and their oxides \cite{psi-k,licht}, \cite{dmft_phys_today,Kotliar06,Held07,Katsnelson08}. Clearly, this approach has a great potential for further developments. Indeed, it is not hard to foresee that the LDA+DMFT framework will eventually develop into a comprehensive \emph{ab initio} approach which is able to describe, and even predict, the properties of complex correlated materials.

\section*{Acknowledgments}

These lecture notes gave me the opportunity to discuss concepts and results obtained in collaboration with numerous coauthors over more than 20 years. I wish to express my deep gratitude to all of them, with particular thanks to (in the chronological order of our first coauthorship): Walter Metzner, Florian Gebhard, Peter van Dongen,  Vaclav Jani\v{s}, Ruud Vlaming, G\"{o}tz Uhrig, Karsten Held, Marcus Kollar, Nils Bl\"{u}mer, Walter Hofstetter, Thomas Pruschke, Vladimir Anisimov, Ralf Bulla, Igor Nekrasov, Theo Costi, Volker Eyert, Georg Keller,  Krzysztof Byczuk, Jim Allen, Shigemasa Suga, Martin Ulmke, Gabi Kotliar, Martin Eckstein, Ivan Leonov, and Jan Kune\v{s}.
I am also much obliged to Anna Kauch for her assistance in the preparation of these lecture notes and to Krzysztof Byczuk and Marcus Kollar for helpful discussions.
This project was supported in part by the Deutsche Forschungsgemeinschaft through SFB~484 and TRR~80.








\hyphenation{Post-Script Sprin-ger}

\end{document}